\documentclass[twocolumn,pra,superscriptaddress]{revtex4}
\usepackage{amssymb}
\usepackage{amsmath}
\usepackage{graphicx,bm}
\usepackage{epstopdf}
\usepackage{subfigure}
\usepackage{natbib}
\usepackage{epsfig}
\usepackage{amsfonts}
\usepackage{mathrsfs}
\usepackage{CJK}
\usepackage[toc,page,title,titletoc,header]{appendix}
\usepackage[usenames, dvipsnames, x11names]{xcolor}

\usepackage[colorlinks]{hyperref}
\hypersetup{
 colorlinks=true,
 citecolor=red,
 linkcolor=blue,
 urlcolor=cyan}

\usepackage{comment}

\definecolor{gray}{rgb}{0.5, 0.5, 0.5}

\newtheorem{theorem}{Theorem}[section]

\newtheorem{lemma}[theorem]{Lemma}

\def \HH{\mathcal{H}}
\def \HHPT{\mathcal{H}_\mathcal{PT}}
\def \MM{\mathcal{M}}
\def \HHtb{\hat{\mathcal{H}}_{\rm tb}}
\def \hc{\hat{c}}
\def \PT{\mathcal{PT}}

\begin{document}
\title{Non-Hermitian dynamics without dissipation in quantum systems}

\author{Yu-Xin Wang}
\affiliation{Institute  for  Molecular  Engineering,  University  of  Chicago, 5640  South  Ellis  Avenue,  Chicago,  Illinois  60637,  U.S.A.}

\author{A. A. Clerk}
\affiliation{Institute  for  Molecular  Engineering,  University  of  Chicago, 5640  South  Ellis  Avenue,  Chicago,  Illinois  60637,  U.S.A.}

\date{\today}

\begin{abstract}
Models based on non-Hermitian Hamiltonians can exhibit a range of surprising and potentially useful phenomena.  Physical realizations typically involve couplings to sources of incoherent gain and loss; this is problematic in quantum settings, because of the unavoidable fluctuations associated with this dissipation.  Here, we present several routes for obtaining unconditional non-Hermitian dynamics in non-dissipative quantum systems.  We exploit the fact that quadratic bosonic Hamiltonians that do not conserve particle number give rise to non-Hermitian dynamical matrices.  We discuss the nature of these mappings from non-Hermitian to Hermitian Hamiltonians, and explore applications to quantum sensing, entanglement dynamics and topological band theory.  The systems we discuss could be realized in a variety of photonic and phononic platforms using the ubiquitous resource of parametric driving. 
\end{abstract}

\maketitle

\section{Introduction}

Systems whose dynamics are governed by a non-Hermitian Hamiltonian exhibit a wealth of unique phenomena, and have been the subject of considerable recent theoretical and experimental interest \cite{Christodoulides2018}.  Non-Hermitian dynamics is typically realized by the structured introduction of both loss and gain, and is usually studied in the context of coupled-mode systems or tight-binding lattices with linear dynamics.  Such systems can exhibit the spontaneous breaking of parity-time ($\mathcal{PT}$) symmetry, as well as exceptional points in parameter space, where Hamiltonian eigenvalues coalesce.  A variety of phenomena in such non-Hermitian systems have been studied, including quasi-adiabatic evolution and chiral mode switching \cite{Moiseyev2011,Uzdin2011,Moiseyev2013,Killingbeck2013,Moiseyev2013b,Viennot2014,Milburn2015, Harris2016, Rotter2016,Kepesidis2016,Zhang2018,Read2018}, directional invisibility \cite{Christodoulides2011}, the possibility of enhanced parameter sensing \cite{Wiersig2014,Wiersig2016,Yang2017,HKLau2018,LJiang2018}, and even applications to robust wireless power transfer \cite{Assawaworrarit2017}.

While the majority of work on non-Hermitian $\mathcal{PT}$-symmetric systems has been in classical settings, it is natural to ask whether their rich properties can also be exploited in quantum systems.  A natural stumbling block is that in the quantum context, the gain and loss typically used to implement non-Hermitian dynamics invariably introduces noise into the system; as explored in several studies \cite{HKLau2018,LJiang2018,Kepesidis2016}, this limits the utility of many non-Hermitian effects in quantum systems.  
While in principle such bath-induced noise effects could be avoided using measurement and postselection \cite{Ashida2017,Murch2019}, this is difficult if not infeasible in many setups.

In this paper, we present and analyze an alternative method for realizing effective non-Hermitian dynamics in a quantum setting that requires {\it no} couplings to external dissipative baths, and requires no measurement-induced conditioning.  The basic idea is to exploit the unitary physics of squeezing (and anti-squeezing) in parametrically-driven quantum bosonic systems.  As is well known, this coherent form of driving can lead to dynamics that exhibits exponential growth and/or decay in time.  We show that in a wide range of contexts, this can be made to parallel the exponential growth and decay associated with incoherent gain/loss processes, allowing a route for the noiseless implementation of non-Hermitian dynamics.  At a formal level, we utilize the unitary correspondence between the non-Hermitian dynamical coupled mode equations of interest, and the Heisenberg equations of motion in our Hermitian bosonic system.  
We provide a detailed analysis of how this idea can be implemented both in simple two-mode systems (with and without $\mathcal{PT}$ symmetry), as well as in more complicated multi-mode lattice systems.  We also use this general mapping to explore a variety of non-Hermitian phenomena (e.g.~chiral mode switching, exceptional-point sensing) in a dissipation-free quantum setting.  We close by  showing how these mappings can also be useful when considering topological band structure in non-Hermitian systems.

\begin{figure}[t]
  \includegraphics[width=80mm]{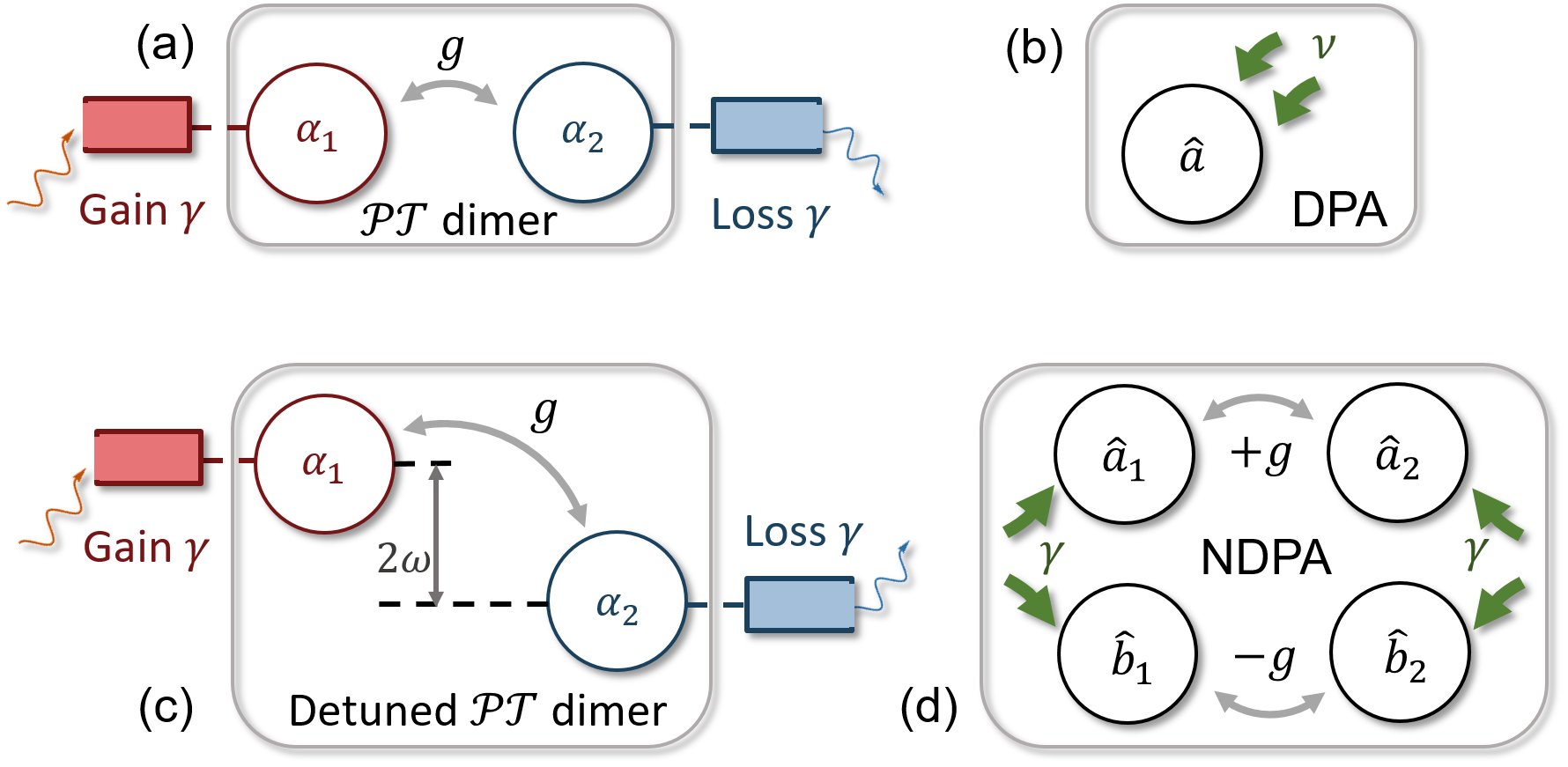}
  \caption{Schematics depicting non-Hermitian two mode systems and equivalent Hermitian driven bosonic setups. (a) Standard two-mode $\mathcal{PT}$ dimer with balanced gain and loss.  This system is unitarily equivalent to the system in (b):  a single-mode bosonic degenerate parametric amplifier(DPA) with drive amplitude $\nu = \gamma$.  (c)  Detuned gain-loss dimer, where an energy detuning between modes breaks $\PT$ symmetry; this is equivalent to the system in (d), a four-mode non-degenerate parametric amplifier setup.  As discussed in the text, these mappings can be extended to multi-mode and lattice systems.   }
    \label{schemaref}
\end{figure}

We stress that the non-Hermitian nature of dynamical matrices in quadratic bosonic systems has long been realized.
Recent work has utilized this to establish mappings between specific 1D models \cite{McDonald2018,Lieu2018}, as well as a means for applying non-Hermitian symmetry classes to bosonic problems \cite{Lieu2018,Ueda2018}.
Our focus is quite different:  we discuss general methods that enable one to realize a given non-Hermitian Hamiltonian of interest without dissipation using a parametrically-driven (but Hermitian) bosonic system, paying special attention to $\PT$-symmetric systems.

\section{Dissipationless non-Hermitian two-mode dynamics}
We start by reviewing the basic dynamics of a non-Hermitian two-mode $\mathcal{PT}$-symmetric dimer, and show how this can be directly mapped to the {\it unitary} squeezing dynamics generated by a dissipation-free parametric amplifier model.  We then extend this discussion to two-mode non-Hermitian systems where $\mathcal{PT}$ symmetry is broken, and show that a similar mapping to a quantum system is still possible.

\subsection{Review: $\mathcal{PT}$ Dimer}
\label{sec:PTDimerReview}

A standard two-mode $\mathcal{PT}$ dimer consists of two tunnel-coupled modes (amplitudes $\alpha_1(t), \alpha_2(t)$) where mode $1$ ($2$) is subject to gain (loss), with the gain and loss rates set equal to $\gamma$ (see Fig.~\ref{schemaref}).  The equations of motion are
\begin{equation}
    i \frac{d}{dt} 
        \left( \begin{array}{c}
            \alpha_1(t) \\
            \alpha_2(t)
            \end{array}
        \right) = 
    \mathcal{H}_{\mathcal{PT}} 
            \left( \begin{array}{c}
            \alpha_1(t) \\
            \alpha_2(t)
                    \end{array}
        \right),
        \label{eq:PTEOM}
\end{equation}
where the $2 \times 2$ effective non-Hermitian Hamiltonian is
\begin{equation}
	\label{eqPToH}
	{{ { \mathcal{H}}}_{\mathcal{PT}}} = \left( {\begin{array}{*{20}{c}}
  	{ + i\frac{\gamma }{2}}&g \\ 
  	g  &{ - i\frac{\gamma }{2}} 
	\end{array}} \right)= i{ \frac{\gamma }{2}} {\sigma _z} + g{\sigma _x}.
\end{equation} 
$g$ is the tunneling amplitude (which we take without loss of generality to be real and positive), and $\sigma_z,\sigma_x$ are standard Pauli matrices.  We will use the calligraphic symbol $\mathcal{H}$ throughout to denote non-Hermitian Hamiltonians.
Defining the time-reversal operation $\mathcal{T}$ as complex conjugation, 
and defining parity $\mathcal{P}$ as the interchange of the two modes, we see that ${\mathcal{H}_\mathcal{PT}}$ is invariant under $\mathcal{PT}$.

The eigenvalues of ${\mathcal{H}_\mathcal{PT}}$ are given by
\begin{equation}
    {\lambda _ \pm } =  \pm \sqrt {{g^2} - { \left( {\frac{\gamma }{2}}  \right) ^2}} . 
    \label{eqPT2mspec}
\end{equation} 
At the critical point $g = \gamma/2 \equiv g_c$, the $\mathcal{PT}$-symmetric Hamiltonian $\mathcal{H}_\mathcal{PT}$ is defective, corresponding to a (second order) exceptional point in parameter space.  For $g < g_c$, the mode eigenvalues become complex, implying exponential growth and decay in the time domain; this transition is often referred to as the ``spontaneous breaking of   $\mathcal{PT}$ symmetry".

Consider next a more general non-Hermitian 2 mode problem, where the dynamics are again given by Eq.~(\ref{eq:PTEOM}) with 
$\HHPT$ replaced by a more general effective Hamiltonian: 
\begin{equation}
	\mathcal{H} = \left( {\vec c + i\vec d} \right) \cdot \vec \sigma. 
	\label{eq:PTCond}
\end{equation}
Here ${\vec c}$, $\vec d$ are real vectors, and we have ignored any constant-matrix part of $\HH$ (as this has a trivial effect on dynamics).  This general non-Hermitian Hamiltonian is unitarily equivalent to $\HHPT$ (and has eigenvalues of the form in Eq.~\eqref{eqPT2mspec}) whenever its Hermitian and anti-Hermitian parts are orthogonal, i.e.
\begin{equation}\label{eqPTreq}
	\vec c \cdot \vec d=0.
\end{equation}

While the preceding discussion is all classical, one might think that the extension to a quantum setting is trivial:  just replace mode amplitudes $\alpha_1(t),\alpha_2(t)$ in Eq.~(\ref{eq:PTEOM}) by canonical bosonic Heisenberg-picture annihilation operators $\hat{a}_1(t),\hat{a}_2(t)$, and interpret the result as a set of coupled Heisenberg equations of motion.  This in general is not a valid evolution, as the dynamics will not preserve canonical commutation relations, i.e.~at all times $[\hat{a}_j(t),\hat{a}^\dagger_{j'}(t)] = \delta_{jj'}$.  This is perhaps most evident in the simple case where $g=0$, and one has simple exponential growth (decay) of mode 1 (2).  

At a physical level, this inconsistency arises because the gain/loss terms that give rise to the non-Hermitian part of $\mathcal{H}$ arise from couplings to dissipative environments.  In addition to providing gain/loss, these baths will also drive the system with noise, and this noise cannot be neglected in the quantum case.  By adding appropriate inhomogeneous quantum noise terms to the RHS of Eq.~(\ref{eq:PTEOM}), one can then obtain an allowed quantum evolution (i.e.~commutation relations are preserved in time).  A systematic procedure for constructing quantum noise terms consistent with a given non-Hermitian Hamiltonian was presented recently in Ref.~\cite{HKLau2018}.

\subsection{Mapping to a degenerate parametric amplifier}
\label{sec:DPAMapping}

Our goal is to realize the effective non-Hermitian dynamics of Eq.~(\ref{eq:PTEOM}) in a quantum system {\it without} having dissipation and the corresponding driving noise.  To that end, we consider the quantum Hamiltonian of a single bosonic mode $\hat{a}$ that is parametrically driven (i.e.~subject to two-photon driving).  In an appropriate rotating frame, the Hamiltonian is:
\begin{equation}
    \label{eqPAHsing}
    {{\hat H}_\mathrm{DPA }} = \delta {{\hat a}^\dag }\hat a  +{\frac{\nu}{2}} \left(i { {{\hat a}^{\dag 2} }  + h.c.} \right),
\end{equation}
where $\delta$ represents a detuning term, and $\nu$ is the magnitude of the parametric drive.  This is the basic Hamiltonian of a degenerate parametric amplifier (DPA).  Without loss of generality, we work in a gauge where $\nu$ is real and positive in what follows.

Despite having only a single mode, the dynamics has a 2 component structure, as the parametric drive couples $\hat{a}$ and $\hat{a}^\dagger$.  Defining a vector of operators
\begin{equation}
    \left| {\hat a} \right\rangle  = { ( {{\hat a},{{\hat a}^\dag }}  )^T},
\end{equation}
the Heisenberg equations of motions can be written as
\begin{equation}
    i{\partial _t}\left| {\hat a} \right\rangle  = 
    \MM_{\mathrm{DPA}} \left| {\hat a} \right\rangle,
    \label{eq:DPAHeisEOM}
\end{equation} 
where the dynamical matrix $\MM_{\mathrm{DPA}}$ is 
\begin{equation}\label{eqeff2DH}
    \MM_{\mathrm{DPA}} = \left( {\begin{array}{*{20}{c}}
    {{\delta   }}& {i\nu } \\ 
    {   {i\nu  }}&{ - {\delta }  } 
\end{array}} \right) =\delta \sigma_z + i\nu \sigma_x.
\end{equation}

We see immediately that the structure of the Heisenberg EOM for this Hermitian problem mirrors that of the effective non-Hermitian dimer system in Eq.~(\ref{eq:PTEOM}), with the dynamical matrix 
$\MM_{\mathrm{DPA}}$ playing the role of an effective non-Hermitian Hamiltonian $\HH$.  Further, $\MM_{\mathrm{DPA}}$ satisfies the condition in Eq.~(\ref{eq:PTCond}), implying that it is unitarily equivalent to $\HHPT$ in Eq.~(\ref{eqPToH}) (with $\delta=g$ and $\nu=\gamma/2$).  This effective $\PT$ symmetry cannot be broken in our single mode problem (as Eq.~(\ref{eqPAHsing}) is the most general single-mode, quadratic, Hermitian and bosonic Hamiltonian).  

Not surprisingly, the eigenvalues of $\MM_{\mathrm{DPA}}$ have exactly the same structure as the $\PT$ dimer:  
\begin{equation}
    {\lambda_{\mathrm{DPA},\pm} } =  \pm \sqrt {{\delta^2} - {{\nu}^2}} . \label{eqPA2ml}
\end{equation} 
It follows that the parametric drive $\nu$ plays the role of the incoherent gain/loss in $\HHPT$, whereas the detuning $\delta$ plays the role of the tunnel coupling.    As we show in the next subsection, this allows us to directly map the physics of the threshold transition in our DPA system to the ``spontaneous" breaking of $\mathcal{PT}$-symmetry that occurs in $\HHPT$ when $\gamma/2$ is made larger than $g$.  In particular, the DPA dynamical matrix exhibits an exceptional point when $\nu = \delta$, i.e. at the parametric oscillation threshold.

Before exploring this connection, we return to the problem of commutation relations:  why doesn't the non-Hermitian nature of the dynamical matrix (and the possibility of exponential growth / decay) in time cause issues in our DPA system?  The easiest way of seeing this is to explicitly make the unitary transformation that maps the dynamical matrix $\MM_\mathrm{DPA}$ in Eq.~(\ref{eqeff2DH}) to the gain-loss form of $\HHPT$ in Eq.~(\ref{eqPToH}). 
With this transformation, the Heisenberg equations of motion in Eq.~(\ref{eq:DPAHeisEOM}) now take the form:
\begin{equation}
    \label{eq2mefqua}
    i{\partial _t}
        \left( \begin{matrix}
            {\hat{q}}  \\
            { i \hat{p}}  \\
        \end{matrix} \right) 
    = 
        \left( \begin{array}{cc}
            i \nu   & \delta \\
            \delta  &   - i \nu
        \end{array} \right)
        \left( \begin{matrix}
   {\hat{q}}  \\
   {i \hat{p}}  \\
\end{matrix} \right)
\end{equation} 
where $\hat{q}\equiv (\hat{a} + \hat{a}^\dagger)/\sqrt{2}$,
$\hat{p}\equiv  i  (\hat{a}^\dagger - \hat{a})/\sqrt{2}$ are canonical (Hermitian) quadrature operators.
With this transformation, we see that the dynamical matrix for the DPA in the quadrature basis has explicitly the same form as the non-Hermitian Hamiltonian $\HHPT$ in Eq.~(\ref{eqPToH}) describing the gain-loss $\mathcal{PT}$ dimer.  We can also see why there is no longer any issue with commutation relations: the exponential growth that occurs when $|\delta| < \nu$ causes $\hat{q}$ to grow exponentially in time, and $\hat{p}$ to decay exponential in time, {\it at the same rate}.  This is just standard, unitary squeezing dynamics.  This exponential growth preserves the canonical 
$[\hat{q},\hat{p}] = i$ commutation relation at all times.   

We thus see that by exploiting the squeezing / anti-squeezing, we can replicate the dynamics of the non-Hermitian $\mathcal{PT}$ dimer in Eq.~(\ref{eq:PTEOM}).  Of course, in this mapping we have just a single mode, and hence only a single complex degree of freedom (whereas in Eq.~(\ref{eq:PTEOM}), there are two complex degrees of freedom).  In Eq.~\eqref{eq2mefqua}, this manifests itself in the fact that the relative phase between the two amplitudes {\it must} be $i$.  By using a two-mode, non-degenerate parametric amplifier system, this phase constraint can be lifted; this will be discussed in Sec.~\ref{subsec:StandardPTChain}.  
Sec.~\ref{sec:TwoModeQMFS} presents yet another approach allowing even more flexibility.

Before proceeding, we briefly pause to note another connection between the $\mathcal{PT}$ dimer Hamiltonian $\HHPT$ and the DPA dynamical matrix $\MM_{\mathrm{DPA}}$: they are are both pseudo-Hermitian matrices. By definition, a pseudo-Hermitian matrix is isospectral with its Hermitian conjugate, so that  \begin{equation}
\HH^\dag = \eta \HH \eta^{-1},
\end{equation}
where $\eta$ is Hermitian and invertible \cite{JMP20021}.  It is easy to see that the dynamical matrix of a generic multi-mode bosonic parametric amplifier system is pseudo-Hermitian; this was recently explicitly pointed out by Lieu \cite{Lieu2018} (see also Appendix \ref{AppSecpsH}).  This connection is however of limited use for our problem: while a given $\mathcal{PT}$-symmetric Hamiltonian is always pseudo-Hermitian, it is not necessarily unitarily equivalent to the dynamical matrix of some quantum bosonic system having the same number of modes (see Appendix~\ref{appsec:MapCEx}).

\subsection{``Phase transitions", exceptional points and conserved quantities}
\label{sganaSEC}

A consequence of the above mapping is that the so-called $\mathcal{PT}$ symmetry breaking phase transition in $\HHPT$ is equivalent to the threshold transition in a parametric amplifier.   Recall that $\HHPT$ exhibits a transition in the eigenvectors and eigenvalues as a function of $g$; this is referred to as the  ``spontaneous breaking of $\mathcal{PT}$ symmetry" \cite{Bender2013}.  The transition occurs at $g = \gamma/2 \equiv g_c$, i.e.~the point at which $\HHPT$ has an exceptional point.  
When $g >g_c$, one is in the $\mathcal{PT}$-unbroken phase.  $\HHPT$ has purely real eigenvalues, and its   right eigenvectors  ${{\vec r}_ \pm }$ are delocalized (i.e.~their amplitudes in each mode are equal) as
\begin{equation}
{{\vec r}_ \pm } = \frac{1}{{\sqrt 2 }}\left( {1, - i{e^{ \pm i\alpha }}} \right) ^T,
\end{equation}
with $\alpha  = \arccos \left( {\gamma /2g} \right)$.
In contrast, when $g$ is reduced below $g_c$, one is in the $\mathcal{PT}$-broken phase.  $\HHPT$ has purely imaginary eigenvalues, and the eigenvectors now exhibit localization.

The above behaviour is equivalent to the threshold behaviour of a detuned DPA.  For $|\delta| > \nu$, the parametric drive is too non-resonant to cause any instability, and the system has purely oscillatory dynamics (as it would if $\nu = 0$).  In contrast, when $|\delta|$ is reduced below $\nu$, one crosses the threshold for parametric instability.  One now has exponential decay and growth, which (as discussed) corresponds to the squeezing / anti-squeezing of canonically conjugate quadratures.  The effective localization of the eigenvectors in this regime corresponds to the fact that the amplified quadrature is predominantly $\hat{q}$, while the squeezed quadrature is predominantly $\hat{p}$.

Finally, consider the case where one tunes $\delta = \nu$ and is thus exactly at the EP.  The DPA Hamiltonian in this case is:
\begin{equation}
    \hat{H}_{\rm DPA} = 
    \frac{\nu}{2}
    \left(
    e^{-i \pi/4}
    \hat{a} +
    e^{i \pi/4}\hat{a}^\dagger \right)^2
    \equiv \nu \hat{Q}^2.
\end{equation}
The Hermitian quadrature operator $\hat{Q}$ is a conserved quantity, and thus $\hat{H}_{\rm DPA}$ is  said to possess a quantum non-demolition (QND) structure. This structure is directly responsible for the lack of any oscillatory dynamics.  The co-existence of exceptional points and conserved QND quadrature operators is not just limited to this simple example: it is a generic feature in particle non-conserving bosonic Hamiltonians.  For example, in Appendix \ref{app:HigherOrderEP}, we discuss a 3-mode system that can be tuned to a third-order EP; this coincides with it having two conserved QND quadrature operators.

\subsection{Mapping for more general two mode non-Hermitian Hamiltonian}
\label{sec:TwoModeQMFS}

We now discuss a more general approach for realizing non-Hermitian two-mode dynamics in dissipation-free quantum systems.  Unlike the mapping to a DPA discussed in Sec.~\ref{sec:DPAMapping}, this alternate method does not require a $\mathcal{PT}$-symmetric non-Hermitian Hamiltonian $\HH$, and does not place restrictions on the phases of mode amplitudes.  Our approach adapts the concept of quantum-mechanics free subsystems (QMFS) introduced by Tsang and Caves \cite{Caves2012}:  by introducing extra bosonic modes, one can have a {\it commuting} set of operators with arbitrary (possibly non-Hermitian) dynamics.  As all relevant operators commute, there is no need to add noise terms.    While QMFS are conventionally discussed and utilized for quantum back-action evasion \cite{Caves2012,Clerk2013,Clerk2015,Polzik2017,Polzik2018}, we show here that they are also a powerful tool for realizing effective non-Hermitian quantum dynamics in a dissipationless setting.  

Consider a two-mode non-Hermitian system where $\mathcal{PT}$ is explicitly broken by the addition of a detuning term $\omega$:
\begin{equation}
\label{eqnH2m} 
    {\mathcal{H}}_\omega = \left( {\omega   +i \frac{ \gamma}{2} } \right) {\sigma _z} + g   {\sigma _x}.
\end{equation}
This Hamiltonian is not unitarily equivalent to a $\mathcal{PT}$ system (c.f.~Eq.~\eqref{eqPTreq}), and thus its dynamics cannot be realized by a DPA using the mapping of Sec.~\ref{sec:DPAMapping}.  

As usual, the goal is to have a quantum system whose Heisenberg equations of motion are governed by $\mathcal{H}_\omega$ without any extra added quantum noise terms, i.e.  
\begin{equation}
    i \frac{d}{dt} 
        \left( \begin{array}{c}
            \hat{z}_1(t) \\
            \hat{z}_2(t)
            \end{array}
        \right) = 
    \mathcal{H}_{\omega} 
            \left( \begin{array}{c}
            \hat{z}_1(t) \\
            \hat{z}_2(t)
                    \end{array}
        \right).
        \label{eq:BrokenEOM}
\end{equation}
The operators $\hat{z}_j$ should play the analogous role of the mode amplitudes in the classical coupled-mode equations Eq.~(\ref{eqPToH}), and hence should encode two complex degrees of freedom.  As discussed, the obvious choice where $\hat{z}_j$ represent canonical annihilation operators of two bosonic modes does not work:  the resulting dynamics would not in general preserve canonical commutation relations.  

Clearly, a simple solution would be to use operators $\hat{z}_j$ where for all $j,j'$
\begin{equation}
    [\hat{z}_j, \hat{z}^\dagger_{j'}] = [\hat{z}_j, \hat{z}_{j'}] = 0.
    \label{eq:Commutes}
\end{equation}
As all operators commute, there would be no additional quantum constraints on Eq.~(\ref{eq:BrokenEOM}).   Throughout this paper, we will use the term {\textit{pseudo-modes}} to denote a set of fully commuting operators $\hat{z}_j, \hat{z}^\dagger_j$ that obey some desired non-Hermitian dynamics. While these pseudo-mode operators are not canonical bosonic annihilation / creation operators, they can play the role of mode amplitudes in the classical coupled mode theory. 

For our two-mode problem, we can construct appropriate pseudo-modes by considering a system of {\it four} canonical bosonic modes, with annihilation operators $\hat{a}_1, \hat{a}_2, \hat{b}_1, \hat{b}_2$.  Each mode can be written in terms of Hermitian quadrature operators ($j=1,2$):
\begin{eqnarray}
    \hat{a}_j & = \frac{1}{\sqrt{2}} \left( \hat{x}_{a,j} + i \hat{p}_{a,j} \right), \\
    \hat{b}_j & = \frac{1}{\sqrt{2}} \left( \hat{x}_{a,j} + i \hat{p}_{a,j} \right) .
\end{eqnarray}
One could now also define collective quadrature operators in the standard manner:
\begin{subequations}\label{eq:ColQuad4mode}
\begin{align}
    \hat{x}_{\pm ,j} & = \frac{1}{\sqrt{2}} \left( \hat{x}_{a,j} \pm  \hat{x}_{b,j} \right), \\
    \hat{p}_{\pm ,j} & = \frac{1}{\sqrt{2}} \left( \hat{p}_{a,j} \pm  \hat{p}_{b,j} \right) .
\end{align}
\end{subequations}
These satisfy standard canonical commutation relations, namely 
$\left[ {{{\hat x}_{ \pm , j }},{{\hat p}_{ \pm ,j' }}} \right]=i{{\delta }_{jj'}}$, $\left[ {{{\hat x}_{ \pm ,j}},{{\hat x}_{ \mp ,j'}}} \right]=\left[ {{{\hat p}_{ \pm ,j}},{{\hat p}_{ \mp ,j'}}} \right]=\left[ {{{\hat x}_{ \pm ,j}},{{\hat p}_{ \mp ,j'}}} \right]=0$.  Note that all $+$ collective quadrature operators commute with all $-$ operators.

We can now construct non-Hermitian pseudo-mode operators $\hat{z}_j$ with the desired properties by building them out of a fully commuting set of four collective quadrature operators.  While there are many possible choices, we will use:
\begin{equation}
    \label{eq:PseudoMode2m}
 \hat{z}_j = \hat{x}_{+ ,j} + i \hat{p}_{- ,j}= {{\hat a}_j} +{\hat b_j^\dag }.
\end{equation}
Eq.~(\ref{eq:Commutes}) is thus satisfied:  we have two complex degrees of freedom where there are no constraints from commutation relations.

All that remains is to construct a physical (Hermitian) Hamiltonian where the four collective quadratures of interest are dynamically coupled as per Eq.~(\ref{eq:BrokenEOM}).  
This can be accomplished using
\begin{align}
    &{{\hat H}_{\omega \mathrm{PA}}}  = \omega   \left( {\hat a_1^\dag {{\hat a}_1} - \hat a_2^\dag {{\hat a}_2} + \hat b_2^\dag {{\hat b}_2} - \hat b_1^\dag {{\hat b}_1}} \right) \nonumber\\
 + &\left[ { g  \left(\hat a_1^\dag {{\hat a}_2} -  \hat b_1^\dag {{\hat b}_2} \right) + i \frac{\gamma }{2}  \left(\hat a_1^\dag \hat b_1^\dag  -   \hat a_2^\dag \hat b_2^\dag \right) + h.c. } \right]. 
 \label{eqnH2mpa}
\end{align}
This represents a system of two tunnel-coupled non-degenerate parametric amplifiers.
One can verify that the Heisenberg equations of motion for collective quadratures generated by the {\it Hermitian} Hamiltonian ${{\hat H}_{\omega \mathrm{PA}}} $ correspond to Eq.~(\ref{eq:BrokenEOM}), with the pseudo-modes defined in Eq.~(\ref{eq:PseudoMode2m}).  We thus have our desired mapping.

Note that with this choice, the collective quadratures that do not appear in the definition of $\hat{z}_j$ can be used to construct another pair of pseudo-modes:
\begin{equation}
    \hat{ \tilde{z}}_j =\hat{x}_{- ,j} + i \hat{p}_{+ ,j}.
    \label{eq:PseudoModeConj2m}
\end{equation}
The dynamics does not couple $\hat{z}$ and $\hat{\tilde{z}}$ operators; using Eq.~(\ref{eqnH2mpa}), the latter satisfy: 
\begin{equation}
    i \frac{d}{d t } 
        \left( \begin{array}{c}
            \hat{ \tilde z}_1( t ) \\
            \hat{ \tilde z}_2( t )
            \end{array}
        \right) = 
    \HH^\dag_{\omega} 
            \left( \begin{array}{c}
            \hat{ \tilde  z}_1( t ) \\
            \hat{ \tilde z}_2( t )
                    \end{array}
        \right).
        \label{eq:BrokenEOM2mConj}
\end{equation}
Thus, in doubling the degrees of freedom, we have constructed two sets of commuting ``pseudo-mode" operators; the first set evolves according to $\HH_\omega$, the second to $\HH^\dagger_\omega$.  

It is instructive to also consider the structure of the Heisenberg equations of motion when written in terms of the true canonical mode annihilation operators; the desired non-Hermitian structure is present there as well.
Letting $\left| {\hat v}_2  \right\rangle$ denote the four-vector of operators ${ ( {{\hat a_1},{\hat a_2},{\hat b_1^\dag},{\hat b_2^\dag} }  )^T}$,  the Heisenberg equations of motion generated by $\hat{H}_{\omega \mathrm{PA}}$ have the general form 
\begin{equation}
    i \frac{d}{d t } 
       \left| {\hat v}_2  \right\rangle  = 
   {{\MM}_{\omega \mathrm{PA}}}
            \left| {\hat v}_2  \right\rangle.
        \label{eq:DynMat4m}
\end{equation}
Here ${{\MM}_{\omega \mathrm{PA}}}$ is the system's mode-basis dynamical matrix; it is 
unitarily equivalent to a $\mathcal{PT}$-symmetric matrix:
\begin{eqnarray}
&\HH_{\omega \mathcal{PT}}=\left( {\begin{array}{*{20}{c}}
  { \HH_{\omega}}&0 \\ 
  0&{ \HH^*_{\omega}} 
\end{array}} \right)= {\mathcal{U}_4} {{\MM}_{\omega \mathrm{PA}}} {\mathcal{U}_4^\dag },\label{eq:UDMtoNH4m}\\
&{\mathcal{U}_4} = \frac{1}{{\sqrt 2 }}\left( {\begin{array}{*{20}{c}}
  {{\mathbb{I}_2}}&{-{\mathbb{I}_2}} \\ 
  {{\mathbb{I}_2}}&{{\mathbb{I}_2}} 
\end{array}} \right).
\end{eqnarray}
This provides another way to interpret our mapping:  by doubling the degrees of freedom and introducing a mirror system of the detuned $\PT$ dimer $\HH_{\omega}$ in Eq.~(\ref{eqnH2m}) which evolves under $\HH^*_{\omega}=\HH^\dag_{\omega}$, we effectively restore $\mathcal{PT}$ symmetry for the entire, composite system, allowing a mapping to a parametrically-driven bosonic Hamiltonian.

We end this section by stressing that our construction using four modes is not limited to the particular non-Hermitian Hamiltonian $\HH_{\omega}$, but can be used to realize the dynamics of {\it any} non-Hermitian two-mode Hamiltonian $\HH$.  One again represents the quasi-mode operators $\hat{z}_1$ and $\hat{z}_2$ using Eqs.~(\ref{eq:PseudoMode2m}).  One obtains the desired dynamics in Eq.~(\ref{eq:BrokenEOM}) (with $\HH_{\omega}$ replaced by $\HH$) if the Hermitian Hamiltonian describing the four mode system is taken to be:
\begin{align}
    {\hat H}_{\mathrm{QMFS}}& = \frac{1}{2}\sum\limits^2_{j,j'=1}  
    \left[ \left(\HH + \HH^\dag\right)_{jj'}   
        \left({\hat a_j^\dag {{\hat a}_{j'}} }-{\hat b_j {{\hat b}_{j'}^\dag} }\right)
        \right.\nonumber\\
    & \left.+     
    \left(\HH - \HH^\dag\right)_{jj'} 
    \left({\hat a_j^\dag {{\hat b}_{j'}^\dag} }- {\hat a_{j'}  {{\hat b}_j } }\right) \right].\label{EqHQMFS2m}
\end{align}
We see that the particle-number conserving terms are associated with the Hermitian part of $\HH$, whereas the non-Hermitian parts of $\HH$ are associated with particle-nonconserving two-photon driving terms.


\section{Dissipationless non-Hermitian lattice dynamics}
\label{sec:Lattices}

We now show that the approaches in the previous section for realizing effective non-Hermitian dynamics in driven, dissipation-free quantum bosonic systems can be generalized to a multi-mode lattice setting.
We will focus on approach where the number of modes in the original non-Hermtian system and the bosonic system are identical; this will be accomplished by using non-degenerate parametric driving (where pairs of photons are added to distinct modes).

\subsection{Standard non-Hermitian $\mathcal{PT}$-symmetric tight-binding chain}
\label{subsec:StandardPTChain}

\label{sec:MultiModePTtoPA}

We start with a simple, but paradigmatic case:  a one-dimensional, nearest-neighbour tight-binding chain with on-site gain/loss terms that respects $\mathcal{PT}$ symmetry.  We refer to this as a ``standard" $\mathcal{PT}$ tight-binding chain.  Non-Hermitian lattice models of this form have been the subject of many recent studies
(see, e.g., \cite{Christodoulides2011,Wiersig2014,Yang2017,Szameit2017,Barnett2013,Lieu2018,Lieu2018a}).  We show that it is possible to realize identical dynamics in a Hermitian driven bosonic system, without any need to introduce dissipation {\it or} double the number of degrees of freedom.  We also show that this approach can be generalized to a wider class of models.

We consider a 1D lattice of coupled modes having $2N$ sites, labelled (from left to right) by $j \in \{-N,-N+1,...,-1,1,...,N-1,N\}$.  We will also (as is common) describe our non-Hermitian Hamiltonian using second-quantized notation, with $\hat{c}_j$ being the mode annihilation operator on site $j$.    The non-Hermitian lattice Hamiltonian then has the form:  
\begin{align}
    \HHtb
    & =
    \sum_{j=-N+1}^{-1} \left( t_j  \hc_{j}^\dagger  \hc_{j-1} + h.c. \right) +
    \left( t_0  \hc_{1}^\dagger  \hc_{-1} + h.c. \right)  \nonumber \\
    & +
    \sum_{j=1}^{N-1} \left( t_j  \hc_{j+1}^\dagger  \hc_{j} + h.c. \right)
    + i \sum\limits_{j} {\frac{\gamma _j}{2} { \hc_j^\dagger {{ \hc}_j}}} . \label{eqPTmulc} 
\end{align}
The first three terms represents Hermitian hopping on the lattice, with hopping strength $t_j$ on each bond (which we take to be real without loss of generality). The last, non-Hermitian term describes on-site gain/loss on each site, with a corresponding rate $\gamma_j/2$. 

We now constrain this model by insisting that it be $\mathcal{PT}$-symmetric.  $\mathcal{P}$ is defined as the real-space operation which maps $\hat{c}_j$ to $\hat{c}_{-j}$, and
$\mathcal{T}$ is defined as before as simple complex conjugation of the Hamiltonian matrix.  $\mathcal{PT}$ symmetry thus requires: 
\begin{subequations}
\begin{align}
t_j=&t_{ -j},\\
 -\gamma _j=&\gamma _{-j}.
\end{align}
\end{subequations}
Note that the class of models of this form includes the widely-studied non-Hermitian $\mathcal{PT}$-symmetric Su-Schrieffer-Heeger (SSH) model \cite{SSH1979,Szameit2017,Barnett2013,OL2013}.  This would correspond to a dimerized structure for the hoppings and loss:
${t_j} = t + {\left(  -  \right)^j}t'$ and ${\gamma _j} = {\left(  -  \right)^j}{\gamma _0}$.

\begin{figure}
  \includegraphics[width=60mm]{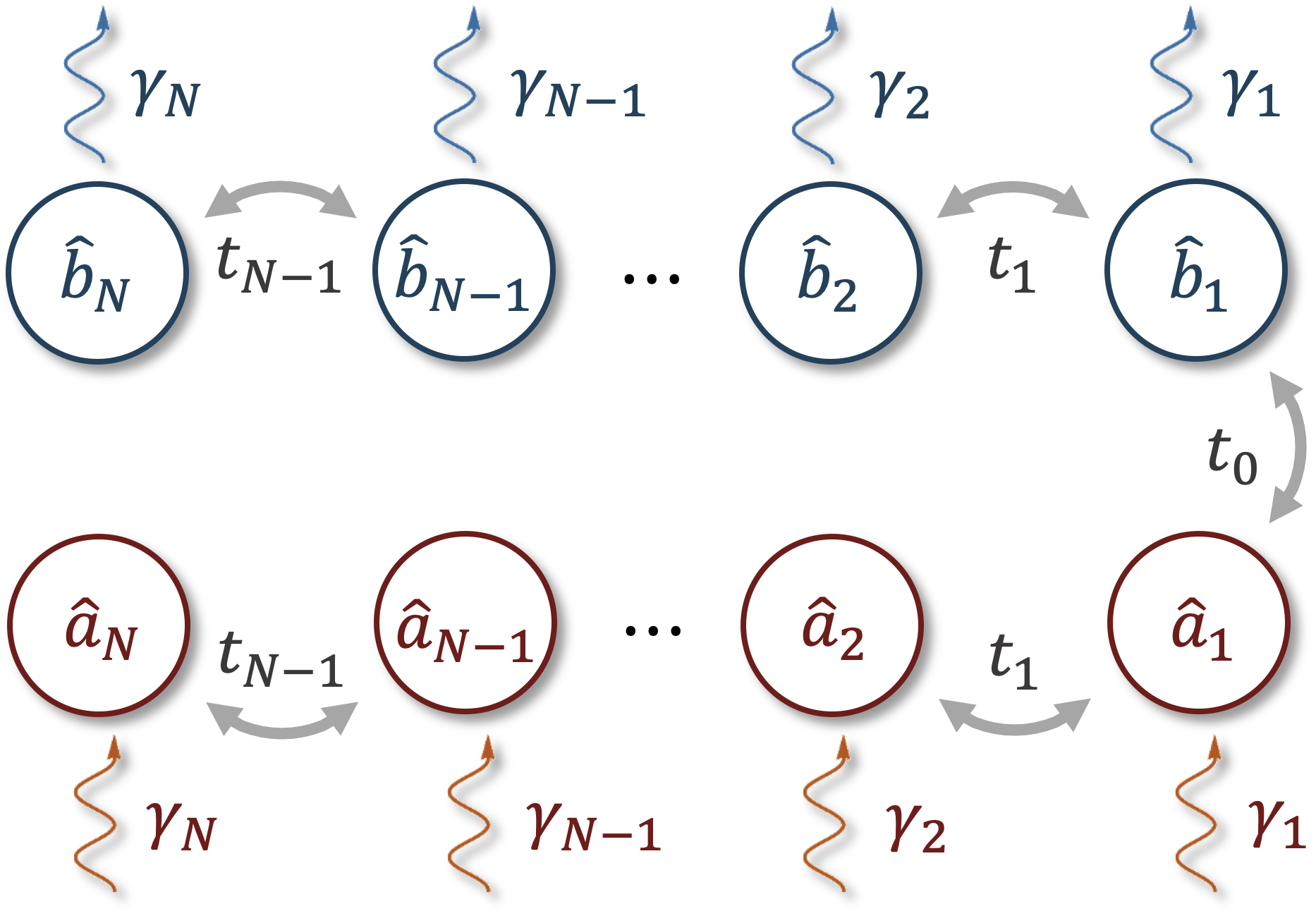}
  \caption{Illustration of a $2N$-mode nearest-neighbor tight-binding $\mathcal{PT}$-symmetric system, whose Hamiltonian $\HHtb$ is given in Eq.~\eqref{eqPTmul}. Insisting on $\PT$ symmetry, and relabelling ${{ \hat{a}}_{j}} \to \hc_{-j}$ and $ {{ \hat{b}}_{j}} \to \hc_j$, the figure also represents the tight-binding Hamiltonian $\HHtb$ in Eq.~\eqref{eqPTmulc}. There always exists unitary correspondence between a system of this form and a Hermitian parametrically-driven bosonic system having an identical number of modes (see discussion in main text).}
  \label{nnPTfrf}
\end{figure}

It will be useful to re-write the Hamiltonian in a more structured form by relabelling the mode operators via
\begin{equation}
    \hc_{-j} \to {{ \hat{a}}_{j}}, \quad \hc_j \to {{ \hat{b}}_{j}} \quad ( j = 1,2, \cdots ,N ).
\end{equation}
As depicted in Fig.~\ref{nnPTfrf}, the Hamiltonian becomes
\begin{align} 
    \HHtb & = 
    \sum\limits_{j,j'=1}^N \left[ {\Omega _{jj'}}\left( {{\hat a}_j^\dag {{\hat a}_{j'}} + {\hat b}_j^\dag { {\hat b}_{j'}}} \right)\right. \nonumber\\
    + & i{\Gamma _{jj'}}\left( {{\hat a}_j^\dag {{\hat a}_{j'}} -  {\hat b}_j^\dag { {\hat b}_{j'}}} \right) + {J_{jj'}}  {\hat a}_j^\dag { {\hat b}_{j'}}+{{\tilde J}_{jj'}}\left. {\hat b}_j^\dag { {\hat a}_{j'}} \right], \label{eqPTmul} 
 \end{align} 
where $\Omega,\Gamma, J$ and ${\tilde J}$ are all $N \times N$ coefficient matrices with entries given by
\begin{subequations}\label{eq:1DtbCoeffMat}
\begin{eqnarray}
{\Omega _{ll'}} & = & {t_l}{\delta _{l',l + 1}} + {t_{l'}}{\delta _{l',l - 1}}  ,\label{eq:1DtbOmega}\\ 
{\Gamma _{ll'}} & = & {\gamma _l}{\delta _{l,l'}}/2 , \\ 
{J_{ll'}} & = & {t_0}{\delta _{l,1}}{\delta _{l',1}}= {{\tilde J}_{ll'}}.
\end{eqnarray}
\end{subequations}
In this new basis, $\mathcal{P}$ is simply the operation which interchanges ${{ a}_{j}}$ and ${{  b}_{j}}$
($j = 1,2, \cdots ,N$). 

We now proceed in analogy to our treatment of the $\mathcal{PT}$ dimer in Sec.~\ref{sec:DPAMapping}.  We first obtain the equations of motion for the ${{ a}_{j}}$ and ${{  b}_{j}}$ modes amplitudes ${{ \alpha}_{j}}$ and ${{ \beta}_{j}}$, generated by $\HHtb$ as
\begin{equation}
i{\partial _t }\left( {\begin{array}{*{20}{c}}
  {{\vec {\alpha }} } \\ 
  {{\vec {\beta}   }} 
\end{array}} \right)  = \HH_{\mathrm{tb}} \left( {\begin{array}{*{20}{c}}
  {{\vec {\alpha }} } \\ 
  {{\vec {\beta}   }} 
\end{array}} \right),
\end{equation}
with the resulting non-Hermitian dynamical matrix given by 
\begin{equation}
 \HH_{\mathrm{tb}} =\left( {\begin{array}{*{20}{c}}
  {{\Omega + i \Gamma}}&{{  J}} \\ 
  { {\tilde J} }&{{\Omega - i \Gamma}} 
\end{array}} \right),
\end{equation}
with ${\tilde J} =J$. 

The block structure 
of the matrix $\HH_{\mathrm{tb}}$ 
(corresponding to ${{ a}_{j}}$/${{  b}_{j}}$ modes) 
allows us to make a simple rotation ${\mathcal{U}_\mathrm{tb}}$ which moves the non-Hermitian gain/loss terms to the off-diagonal blocks: 
\begin{eqnarray}
 &   {\MM_\mathrm{tb}} ={\mathcal{U}_\mathrm{tb}} \HH_{\mathrm{tb}} {\mathcal{U}_\mathrm{tb}^\dag }=\left( {\begin{array}{*{20}{c}}
  {\Omega  + J}& { i \Gamma }  \\ 
  { i \Gamma }&{\Omega  - J} 
\end{array}} \right),\\
&{\mathcal{U}_\mathrm{tb}} = \frac{1}{{\sqrt 2 }}\left( {\begin{array}{*{20}{c}}
  {{\mathbb{I}_N}}&{{\mathbb{I}_N}} \\ 
  {{\mathbb{I}_N}}&{-{\mathbb{I}_N}} 
\end{array}} \right).
\end{eqnarray}
In analogy to the two-mode problem in Sec.~\ref{sec:DPAMapping}, the rotated matrix ${\MM_\mathrm{tb}}$ now has the form of a dynamical matrix of a parametrically driven bosonic system with $2N$ sites.  However, unlike the mapping in Sec.~\ref{sec:DPAMapping}, the relevant system here involves non-degenerate parametric drives (i.e.~two-photon driving terms that involve distinct modes).  The dynamical matrix ${\MM_\mathrm{tb}}$ above corresponds to the {\it Hermitian} bosonic Hamiltonian
\begin{align}
    {\hat H}_{\mathrm{NDPA}} &= 
    \sum\limits_{j,j'}^N  \left[ {\Omega _{jj'}}\left( {\hat a_j^\dag {{\hat a}_{j'}} - \hat b_j^\dag {{\hat b}_{j'}}} \right)  \right.  \nonumber\\
    +&{J_{jj'}}\left( {\hat a_j^\dag {{\hat a}_{j'}} + \hat b_j^\dag {{\hat b}_{j'}}} \right) + i{\Gamma _{jj'}} \left.\left( {\hat a_j^\dag \hat b_{j'}^\dag  - {{\hat b}_j}{{\hat a}_{j'}}} \right) \right].\label{eqpamulR}
 \end{align} 

To be explicit, the Heisenberg equations of motion corresponding to this Hermitian Hamiltonian can be compactly written as
\begin{subequations}
\begin{align}
i{\partial _t}\left| {\hat v}_N \right\rangle  &=  {\MM_\mathrm{tb}}  \left| {\hat v}_N \right\rangle,
\end{align} 
\end{subequations} 
where $\left| {\hat v}_N \right\rangle  = { ( {{\hat a_1},{\hat a_2},\cdots,{\hat a_N},{\hat b_1^\dag},{\hat b_2^\dag},\cdots,{\hat b_N^\dag} }  )^T}$.  Thus, we see that the dynamics of the general $\mathcal{PT}$-symmetric non-Hermitian gain-loss lattice model in Eq.~\eqref{eqPTmul}  can be realized by the non-dissipative, Hermitian quantum Hamiltonian in Eq.~\eqref{eqpamulR}. 
As before, the exponential growth and decay that could result from the gain and loss terms are mapped onto a unitary squeezing operations in the driven quantum model (in this case two-mode squeezing operations).

With this explicit non-degenerate parametric-amplifier (NDPA) Hamiltonian $ {\hat H}_{\mathrm{NDPA}}$ in hand, it is interesting to return to the simple $\mathcal{PT}$ dimer discussed in Sec.~\ref{sec:DPAMapping}.  This corresponds to the case $N=1$ of the 1D $\mathcal{PT}$ chain considered in this section.  In this case, the matrix $\Omega$ becomes an overall constant in the non-Hermtian Hamiltonian $\HHtb$ and can be ignored, and our mapping shows that the dynamics is equivalent to a simple two-mode NDPA in Eq.~\eqref{eqpamulR}.  We stress that this is a distinct mapping from that in Sec.~\ref{sec:DPAMapping}, which involves a single-mode DPA.  By having two modes here, there is no constraint on the phases of mode amplitudes, as the number of complex degrees of freedom is the same as the original non-Hermitian coupled-mode problem.  As we will show in the next section, this lack of constraints remains the same in the general multi-mode version of the problem as well.




\subsection{Generalized non-Hermitian $\mathcal{PT}$-symmetric tight-binding chain}
\label{subsec:GeneralizedPTChain}

We now consider more general $\mathcal{PT}$-symmetric non-Hermitian lattice models, which could be in higher dimensions, have long range hopping terms, and have non-local non-Hermitian terms.  The unitary mapping  ${\mathcal{U}_\mathrm{tb}}$ derived above is also valid for a wide class of these generalized models.  Note first that a generic $\mathcal{PT}$-symmetric non-Hermitian Hamiltonian (in arbitrary dimensions, with $2 N$ sites) can always be written in the form given in Eq.~(\ref{eqPTmul}), where $\mathcal{PT}$ symmetry requires that the coefficient matrices $\Omega,\Gamma$ are real, as well as $\tilde J = J^*$.
Our mapping to a Hermitian parametric amplifier problem (as per Eq.~(\ref{eqpamulR})) remains valid as long as the coefficient matrices $\Omega$, $\Gamma$ and $J$ are all real, symmetric matrices.  This encompasses a much broader class of models than the 1D nearest neighbour, imaginary potential model described by Eqs.~(\ref{eq:1DtbCoeffMat}).  

Among the extra kinds of terms that can be accommodated in the starting non-Hermitian $\mathcal{PT}$ lattice model are:
\begin{itemize}
    \item real detunings of $a_j$ and $b_j$ modes, given by real, diagonal matrix elements of $\Omega$;
    \item real, coherent (i.e.~Hermitian) coupling between any two $a_j$ and $a_{j'}$ (or $b_j$ and $b_{j'}$) modes with a completely real coupling strength, represented by off-diagonal matrix elements of $\Omega$;
    \item real, coherent coupling between any two $a_j$ and $b_{j'}$ modes with a completely real coupling strength, represented by corresponding matrix elements of $J$;
    \item imaginary, dissipative (i.e.~non-Hermitian) couplings between any two different modes, represented by off-diagonal matrix elements of $\Gamma$.
\end{itemize}

As an example, our mapping to a NDPA system remains valid for a 2D tight-binding $\mathcal{PT}$ lattice, as long as the coherent couplings are purely real, and the dissipative couplings are purely imaginary.
Conversely, for non-Hermitian tight-binding models where hopping phases encode non-trivial fluxes, we may construct an example where the mapping does not work.  Necessary conditions for such a mapping to exist are presented in  Appendix~\ref{AppSecConv}, while simple four mode systems where the correspondence fails are discussed in Appendix~\ref{appsec:MapCEx}.

\subsection{Mapping for arbitrary multi-mode non-Hermitian Hamiltonians}
\label{sec:MultiModeQMFS}

In Sec.~\ref{subsec:StandardPTChain} and \ref{subsec:GeneralizedPTChain}, we described a general mapping between a wide class of non-Hermitian, $\mathcal{PT}$ symmetric lattice models and the dynamical matrix of a Hermitian, parametrically driven bosonic system.  Crucially, this mapping preserved the number of modes.  As discussed, it cannot be applied to {\it all} possible $\mathcal{PT}$ lattice models, nor can it be used for systems with broken $\mathcal{PT}$.

In this section, we show how the general QMFS strategy introduced in Sec.~\ref{sec:TwoModeQMFS} can be generalized to map an arbitrary non-Hermitian lattice model to a Hermitian, parametrically-driven bosonic problem.  While more general, this strategy comes with a price:  the driven bosonic system will have twice the number of modes as in the original non-Hermitian Hamiltonian.  

The approach is to generalize the construction presented in Eq.~\eqref{EqHQMFS2m} of Sec.~\ref{sec:TwoModeQMFS} to a general $N$-mode non-Hermitian Hamiltonian $\HH_N $. 
We will use a $2N$-mode bosonic system, with canonical quadrature operators $ {\hat x_{  \pm,j} }$ and ${\hat p_{  \pm,j} }$ for $j =1,2,\ldots, N$.  The only nonzero commutators between the quadratures are  
\begin{equation}
    \left[ { {\hat x_{   \pm,j} },{\hat p_{  \pm,j'} }} \right] =i{\delta _{jj'}},
\end{equation}
for $j,j'=1,2,\ldots, N$. 

To implement the general QMFS strategy, we wish to construct a Hamiltonian where a set of fully commuting collective quadratures has a linear dynamics corresponding to $\HH_N$.  Following the convention for the two-mode case in Eqs.~(\ref{eq:PseudoMode2m},\ref{eq:PseudoModeConj2m}), we first introduce two sets of pseudo-modes $\hat{  {z}}_{\pm,j}~(j=1,2,\ldots,N)$ as
\begin{equation}\label{eq:PseudoModesMulti}
    \hat{  {z}}_{\pm,j} =\hat{x}_{\pm ,j} + i \hat{p}_{\mp ,j}.
\end{equation}

Mirroring the strategy of Sec.~\ref{sec:TwoModeQMFS}, we want a Hermitian bosonic Hamiltonian that yields the equations of motion:
\begin{subequations}
\begin{align} 
i{\partial _t } {\vec{\hat{ z }}}_{+ } &= \HH_N  {\vec{\hat{ z }}}_{+ },\label{eq:EOMPseudoMode}\\
i{\partial _t } {\vec{\hat{ z }}}_{- } &= \HH^\dag_N  {\vec{\hat{ z }}}_{- },\label{eq:EOMPseudoModeConj}
\end{align}
\end{subequations}
where we define $N$-vectors  ${\vec{\hat{ z }}}_{\pm }$ consisting of the pseudo-mode operators ${ {\hat{ z }}}_{\pm,j }$, respectively, for notational convenience. As before, the desired dynamics will only couple mutually commuting quadratures.  It is straightforward to prove that the two equations above generate a dynamics that preserve all canonical commutation relations, i.e. they generate a symplectic transformation of the bosonic system (see Appendix~\ref{appsec:QMFSgen} for details).  Further, one can show that this dynamics is generated by the Hermitian $2N$-mode Hamiltonian
\begin{align}
    {\hat H}_{\mathrm{QMFS}, \mathrm{multi.}} &=\frac{1}{2} \sum\limits^N_{j,j'=1}  \left[ \left({  {\HH_N}+{ \HH^\dag_N } }\right)_{jj'}  \left({\hat a_j^\dag {{\hat a}_{j'}} }-{\hat b_j {{\hat b}_{j'}^\dag} }   \right)\right.\nonumber\\
    & \left.+   { \left({  {\HH_N} -{ \HH^\dag_N } }\right)_{jj'}}  \left({\hat a_j^\dag {{\hat b}_{j'}^\dag} }- {\hat a_{j'}  {{\hat b}_j } }\right) \right],
    \label{eq:HQMFSMulti}
\end{align}
where we define the bosonic mode operators in parallel to Eq.~\eqref{eq:PseudoMode2m} as
\begin{equation} \label{eq:BosonicPseudoModeMulti}
 {{\hat a}_j} \pm {\hat b_j^\dag }=  {{\hat x_{  \pm ,j} } + i{\hat p_{ \mp ,j} }} =\hat{  {z}}_{\pm,j}.
\end{equation}

The approach here is of course directly applicable to the case where the non-Hermitian $\mathcal{H}_N$ describes a lattice model in real space.  Our mapping doubles the number of modes: for every lattice site in the original model, there are now two bosonic modes $\hat{a}_j, \hat{b}_j$.  Note however from Eq.~(\ref{eq:HQMFSMulti}) that our mapping is fully local.  For every band $E_n(\vec{k})$ of $\HH_N$, the closed-form dynamics of the ${\hat a}_j$ and ${\hat b_j^\dag }$ operators will correspondingly contribute two independent bands $E_n(\vec{k})$ and $E_n^*(\vec{k})$ in the bosonic problem; this follows directly from Eqs.~(\ref{eq:EOMPseudoMode})-(\ref{eq:EOMPseudoModeConj}). We stress that this doubled band structure only solves half of the entire BdG problem of the bosonic Hamiltonian; the full band structure will also include contributions from dynamics of the ${\hat b}_j$ and ${\hat a_j^\dag }$ operators, which can also be obtained from Eqs.~(\ref{eq:EOMPseudoMode})-(\ref{eq:EOMPseudoModeConj}) as the $-E_n(\vec{k})$ and $-E_n^*(\vec{k})$ bands.


\section{Applications of dissipation-free non-Hermitian quantum dynamics}

In this section, we discuss how the mappings introduced in the previous sections can be used to realize various well-known non-Hermitian effects in dissipation-free, quantum settings.

\subsection{Exceptional point sensing}
\label{EPmodesplit}

We first consider sensing methods that exploit the strong sensitivity of mode eigenvalues of a $\mathcal{PT}$-symmetric non-Hermitian system that is tuned to the vicinity of an exceptional point (EP) \cite{Wiersig2014,Yang2017}.  The most common version of this scheme involves a simple gain-loss $\mathcal{PT}$ dimer (c.f.~Sec.~\ref{sec:PTDimerReview}) with an effective non-Hermitian Hamiltonian
\begin{equation} 
    \mathcal{H} \left[ \epsilon \right]=\mathcal{H}_\mathcal{PT}= i\frac{\gamma }{2}{\sigma _z} + (g_0 + \epsilon) {\sigma _x}. 
    \label{eq:Heps}
\end{equation}
The goal is to estimate the small parameter $\epsilon$.
If the unperturbed Hamiltonian $\mathcal{H} \left[ 0 \right]$ 
is tuned to the EP by choosing $g_0 = \gamma/2 = g_c$, then the
perturbation $\epsilon$ induces an eigenvalue splitting that scales as $\sqrt{\epsilon}$, i.e. from Eq.~(\ref{eqPT2mspec}), we have:
\begin{equation}
    \left| {\lambda_+ - \lambda_-} \right| \simeq 2 \sqrt{2 g_0 \epsilon}.
    \label{eq:DeltaEP}
\end{equation}
For small $\epsilon \ll g_0$, this is parametrically larger than a conventional mode splitting in a Hermitian system, which would be proportional to $\epsilon$.

To exploit this eigenvalue sensitivity for measurement, it was suggested in Refs.~\cite{Wiersig2014,Yang2017} to look at the reflection of a probe tone applied to the system at frequency $\omega_{\rm p}$.  The frequency-dependent reflection coefficient $R[\omega_{\rm p}]$ would then reflect the parametric mode-splitting of the eigenvalues.  While the advantage of this approach seems obvious, recent studies have shown that the unavoidable noise associated with incoherent gain and loss in the quantum regime can limit any enhancement of signal-to-noise ratio \cite{HKLau2018,LJiang2018}.

Here, we show an analogous EP sensing scheme can be implemented in a parametric amplifier setup, without having to introduce any incoherent gain and loss, and corresponding noise.    While there are many ways to proceed, the simplest is to use the unitary mapping introduced in Sec.~\ref{sec:DPAMapping} that maps the $\PT$ dimer in Eq.~(\ref{eq:Heps}) to a single-mode, degenerate parametric amplifier (DPA).  Letting $\delta = g_0$ and $\nu = \gamma/2$, the Hermitian DPA Hamiltonian corresponding to $\HH[\epsilon]$ is then given by:
\begin{equation}
    {{\hat H}_\mathrm{DPA }} \left[ \epsilon \right] =\left( {\delta+\epsilon} \right) {{\hat a}^\dag }\hat a  +\frac{\nu }{2} \left( { i  {{\hat a}^{\dag 2} }  + h.c.} \right).
    \label{eq:HDPAEP}
\end{equation}
As usual, the tunneling in $\HH[\epsilon]$ becomes a detuning term, and the gain/loss terms in $\HH[\epsilon]$ become a two-photon drive.  We stress that the dynamical matrix of this Hermitian Hamiltonian is unitarily equivalent to $\HH[\epsilon]$, and has the same eigenvalues.  Note that the perturbation $\epsilon$ is now a standard dispersive coupling, something that arises in many measurement contexts.  In the case where $\epsilon$ corresponds to the state of a qubit, this exact setup was realized in a recent superconducting quantum circuit experiment (though operated in a different regime) \cite{Siddiqi2018}.  

\begin{figure}
  \includegraphics[width=80mm]{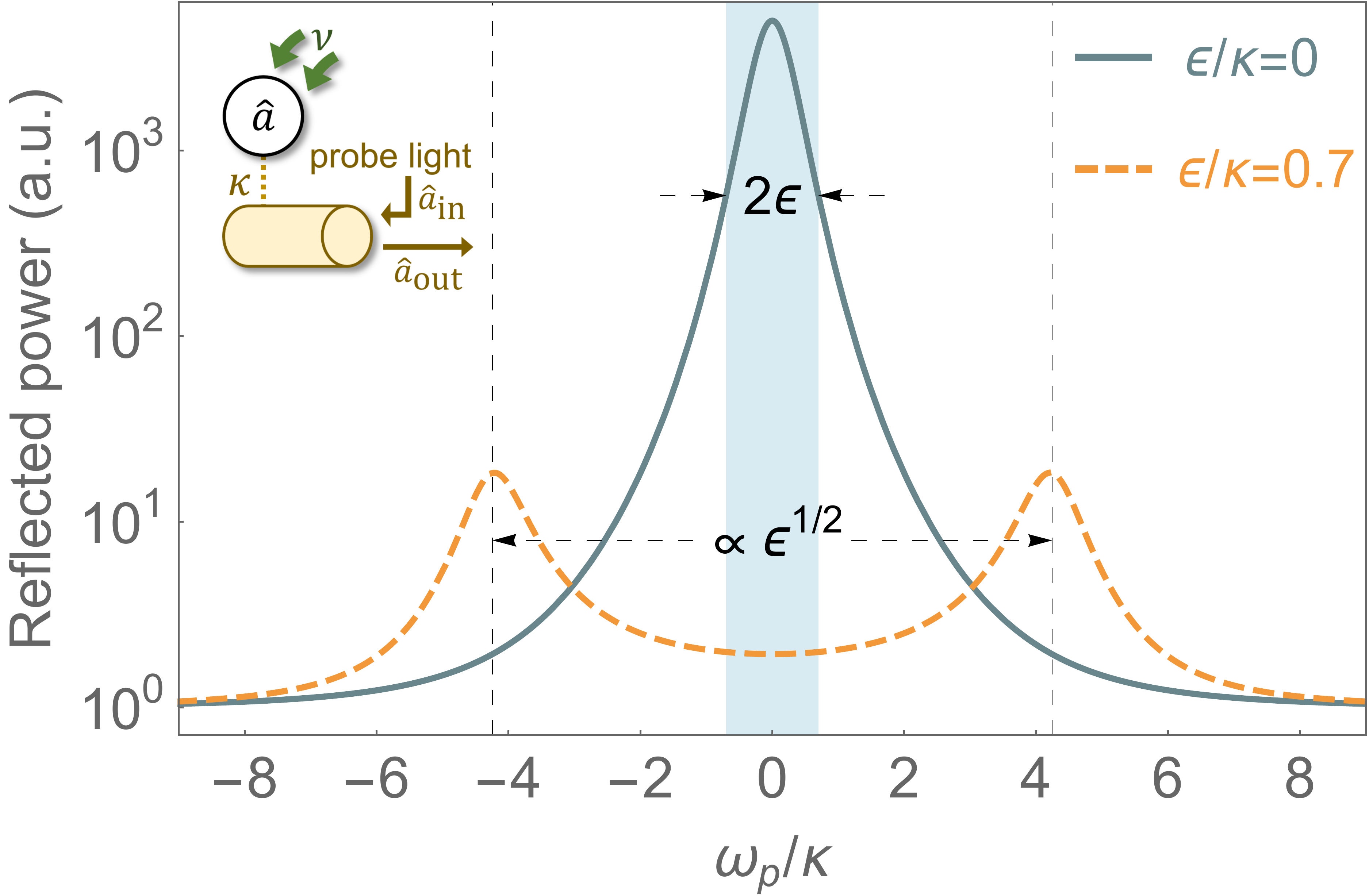}
  \caption{Exceptional-point sensing with a degenerate parametric amplifier (DPA).  A detuned DPA is tuned to an EP by matching the pump detuning and parametric drive amplitudes.  Probe light of frequency $\omega_p$ is sent into the cavity via waveguide (coupling rate $\kappa$).    Plotted is the frequency-dependent reflected flux of the probe tone.  Dark cyan solid line: reflected flux for  the unperturbed system, showing a single peak.  Orange dashed line: reflected flux where the system is perturbed by shifting the cavity frequency an amount $\epsilon =0.7 \kappa$ (c.f.~Eq.~(\ref{eq:HDPAEP})).  One now has two distinct peaks, with a splitting that scales as $\sqrt{\epsilon}$.   Parameters correspond to a parametric drive amplitude $\nu  /\kappa=12.5$, and pump detuning $\delta = \nu$ originally set to the EP. 
}
  \label{RefSpecref}
\end{figure}

We now mimic the EP sensing protocol, by first tuning $\delta = \nu$ so that the unperturbed system is at the EP.  To probe the $\epsilon$-induced mode splitting, we will again look at the reflection of a probe tone applied at frequency $\omega_p$.  We couple the cavity $\hat{a}$ to an input-output waveguide (or transmission line), with a coupling rate $\kappa$.  Using standard input-output theory \cite{Gardiner2004}, the Heisenberg equation of motion of $\hat{a}$ is 
\begin{subequations}
\begin{align}
\frac{d}{dt} {\hat a} =& -i \delta {\hat a} +  \nu {{\hat a}^\dag } - \frac{\kappa }{2}{\hat a} - \sqrt \kappa  {\left( {{{\alpha}_\mathrm{in}}{e^{ - i\omega_p t}} + {{\hat \zeta }_\mathrm{in}}} (t)\right)},
\end{align}
\end{subequations}
where $\alpha_{\rm in}$ is the amplitude of the probe tone, and ${{\hat \zeta }_\mathrm{in}}(t)$ describes vacuum noise entering through the waveguide.  Note that we are working in a rotating frame determined by the frequency of the pump field used to realize the parametric interaction.  

The introduction of the waveguide shifts the eigenvalues of the dynamical matrix by a constant, but the system still possesses an EP.  We pick the pump detuning $\delta = \nu \equiv \delta_c$ so that the unperturbed system is tuned to this EP.  We then calculate the total output flux $P_\mathrm{out} \left(  \omega_p   \right)$ (including both the reflected signal and idler beams), as a function of the probe frequency $\omega_p$, to see how the $\epsilon$-induced mode splitting impacts the light leaving the cavity.
Using the standard input-output relation ${{\hat a}_\mathrm{out}}= {{\hat a}_\mathrm{in}}+ {\sqrt \kappa}{\hat a}$ \cite{Gardiner2004}, the output flux is
\begin{equation} 
\frac{P_\mathrm{out} \left(  \omega_p   \right)}{\left|\alpha_\mathrm{in} \right|^2} =1 +   \frac{ 2\kappa^2   \nu^2}{{{{\left( {f\left[  \omega_p   \right]} \right)}^2} + {\kappa ^2}\left( {{\delta ^2} - {{ \nu }^2}} \right)}},
\end{equation}
where $f\left[  \omega_p   \right] = { \omega_p  ^2} + {\left( {\kappa /2} \right)^2} - {\delta ^2} + {  \nu^2}$.
Note that we do not include the contribution from amplified vacuum fluctuations here, as this yields a background that is independent of both $\omega_p$ and $\alpha_{\rm in}$.
In the limit of a weak coupling to the waveguide, we will observe narrow peak(s) in $P_\mathrm{out}$ that correspond to the dynamical matrix eigenvalues $\lambda_{\pm}$, see Fig.~\ref{RefSpecref}.  For $\epsilon = 0$, there is just a single peak, whereas for non-zero $\epsilon$ there are two peaks, with the expected splitting  $\left| {\lambda_+ - \lambda_-} \right|  \simeq {\rm{2}}\sqrt {{\rm{2}}\epsilon \delta_c } \gg \epsilon$ (see also Eq.~\eqref{eq:DeltaEP}).

We thus see that the EP sensing scheme of Refs.~\cite{Wiersig2014,Yang2017} can be directly implemented in a parametric-amplifier setup, without any need for incoherent gain and loss.  We leave a full analysis of the noise properties and ultimate sensitivity of this scheme (both in the linear and nonlinear response regimes) to a future work.  Note that the general analysis in Ref.~\cite{HKLau2018} of linear-response EP sensing assumed a Hamiltonian that conserves particle number, and thus does not apply directly to the DPA setup described here.  
Also note that higher-order exceptional points have been discussed in the context of sensing; these too can be realized without dissipation using parametrically-driven bosonic modes (see Appendix \ref{app:HigherOrderEP}).

\subsection{Quasi-adiabatic evolution and chiral mode switching}
\label{subsec:EPEncircle}

\begin{figure}
  \includegraphics[width=70mm]{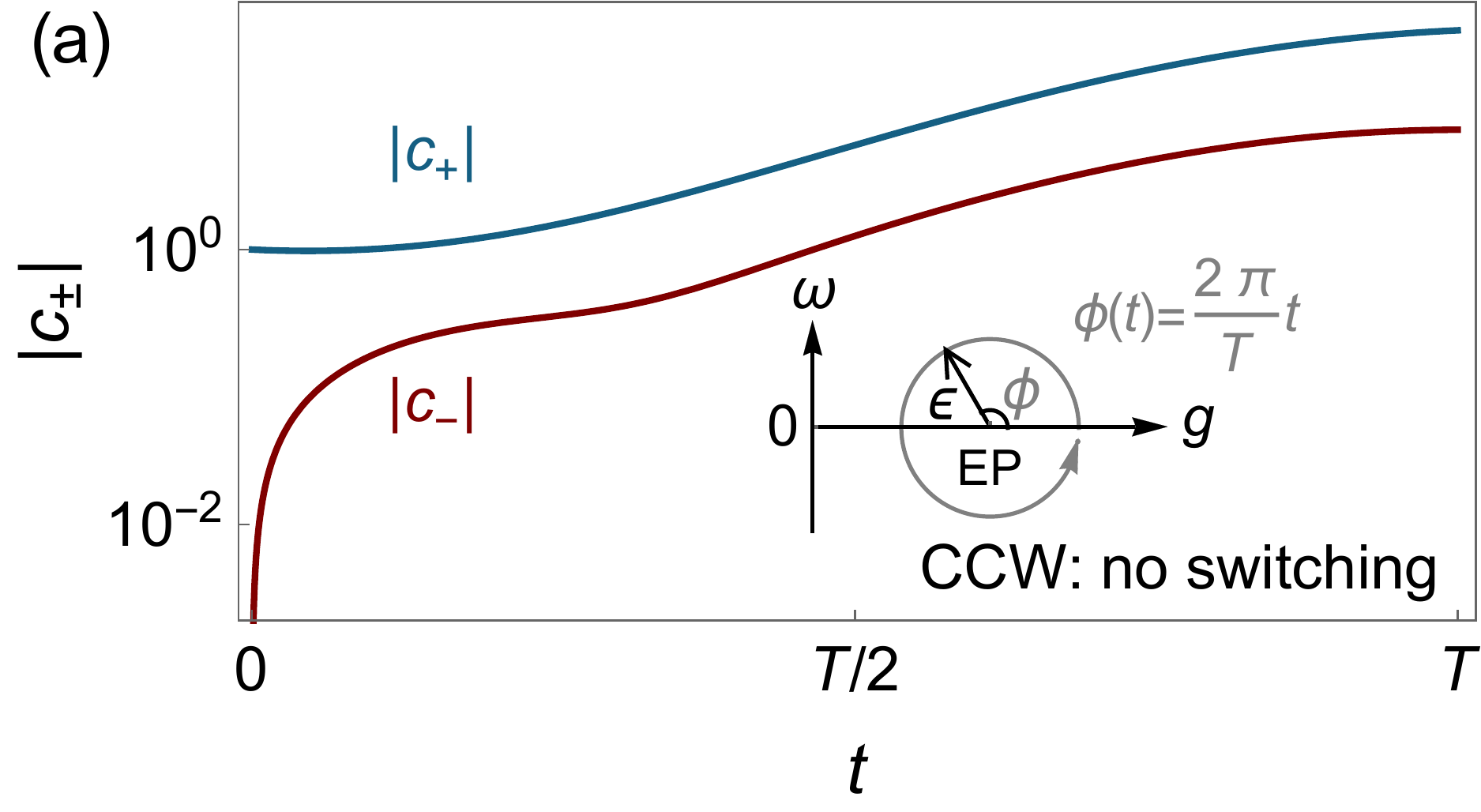}
  \includegraphics[width=70mm]{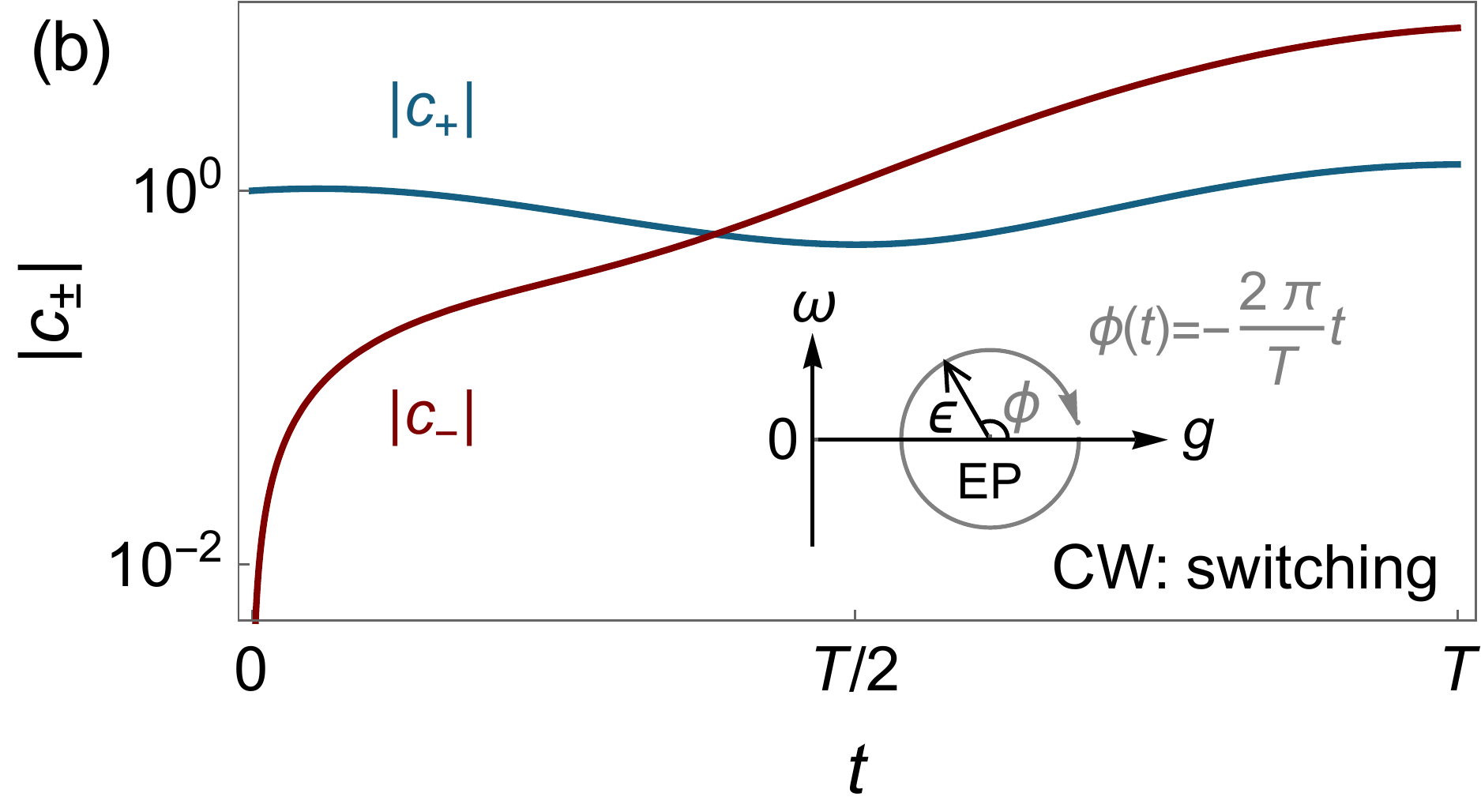}
  \caption{Chiral nature of quasi-adiabatic dynamics in a four-mode Hermitian bosonic system 
  (c.f.~Eq.~(\ref{eqnH2mpa})) whose dynamics mimics the gain-loss dimer in Eq.~(\ref{eq:Homegat}).  In each plot, $g(t)$ and $\omega(t)$ are varied along a circle
  in parameter space (see insets).  (a) Evolution of instantaneous eigenmode amplitudes 
  $\left| \langle \hat{c}_\pm \left (t \right) \rangle \right |$ (c.f.~Eq.~(\ref{eq:chatdefn})), for  
  a counter-clockwise (CCW) parameter variation.  (b) Same, but now for a clockwise (CW) variation.  In both cases, 
  the initial state is a coherent state with $\langle \hat{c}_+(0) \rangle = 1$, 
  $\langle \hat{c}_{-}(0) \rangle= 0$.  For the CCW evolution, one sees an adiabatic evolution (the $+$ mode remains dominant), whereas for CW evolution, there is a non-adiabatic switching, and the $-$ mode is dominant at the end of the protocol.    
   For both plots $\gamma T=20$ and $\epsilon/\gamma=0.1$.}
\label{fig:QADfref}
\end{figure}

Another striking effect associated with exceptional points involves the chirality of non-adiabatic effects in non-Hermitian systems whose parameters are cyclically varied \cite{Moiseyev2011,Uzdin2011,Moiseyev2013,Killingbeck2013,Moiseyev2013b,Viennot2014,Milburn2015, Harris2016, Rotter2016}.  The paradigmatic system is the detuned gain-loss dimer $\HH_\omega$ in Eq.~\eqref{eqnH2m}, where now the tunneling $g$ and detuning $\omega$ are made time-dependent:
\begin{equation}
   {\mathcal{H}}_\omega(t) 
    = \left( {\omega(t)   +i \frac{ \gamma}{2} } \right) 
    {\sigma _z} + g(t)   {\sigma _x}.
    \label{eq:Homegat}
\end{equation}
Consider a cyclic time-variation of parameters, where $\left( g\left( t\right),\omega\left( t\right)  \right)$ follow a closed path in parameter space that encloses one of the two EPs at $\left( g_c =\pm \gamma/2, \omega =0  \right)$ (see inset in Fig.\ref{fig:QADfref}a).  Non-adiabatic effects in such a setup depend crucially on the direction one traverses the path in parameter space:  for one direction, there is no switching between adiabatic eigenmodes,
whereas for the other direction, there is appreciable switching.  Appendix~\ref{appsec:RevQAD} gives a basic introduction to this phenomena; see Ref.~\cite{Milburn2015} for a more comprehensive discussion.

Recent experiments have probed this EP encircling physics in classical settings \cite{Harris2016, Rotter2016}, and it has been suggested that such effects could be useful in quantum settings \cite{Moiseyev2013}.  As usual though, the unavoidable noise associated with incoherent gain and loss in quantum systems would be problematic.  We show here how the mapping introduced in 
Sec.~\ref{sec:TwoModeQMFS} to a dissipation-free driven bosonic system allows one to realize this chiral switching behaviour without any dissipation or noise.  
As a concrete quantum application of our mapping, we show how the chiral switching behaviour impacts the evolution of entanglement in our system.

As discussed in Sec.~\ref{sec:TwoModeQMFS}, we can realize the dynamics of $\HH_{\omega}(t)$ in Eq.~(\ref{eq:Homegat}) without dissipation using a four-mode, parametrically-driven bosonic system with a Hamiltonian $\hat{H}_{\omega \mathrm{PA}}(t)$ given by 
Eq.~\eqref{eqnH2mpa}.  First, consider the non-Hermitian system described by  $\HH_{\omega}(t)$.  The time-evolution matrix $\mathcal{U}_{\omega}(t)$ corresponding to this Hamiltonian relates final and initial mode amplitudes, and is determined by 
\begin{equation}
    i{\partial _t }\mathcal{U}_{\omega}\left( t  \right) = {\mathcal{H}_{\omega}} \left( t \right)\mathcal{U}_{\omega}\left( t  \right), \quad  \mathcal{U}_{\omega}\left( t=0  \right)=1.
    \label{eq:UomegaDefined}
\end{equation}

Our Hermitian, bosonic four mode system has been constructed so that the quasi-mode operators $\hat{z}_1, \hat{z}_2$ defined in Eq.~\eqref{eq:PseudoMode2m} evolve exactly like amplitudes in the non-Hermitian system.  This implies that
\begin{equation}
        \left( \begin{array}{c}
            \hat{z}_1(t) \\
            \hat{z}_2(t)
            \end{array}
        \right) = 
   \mathcal{U}_{\omega}\left( t  \right) \cdot
            \left( \begin{array}{c}
            \hat{z}_1(0) \\
            \hat{z}_2(0)
                    \end{array}
        \right),
        \label{eq:BrokenEOMSolu}
\end{equation}
where we stress that these are operator equations.  Thus, the chiral switching behaviour encoded in $\mathcal{U}_\omega(t)$ will directly manifest itself in the quantum bosonic system, without any need to inject noise to preserve commutation relations.

The chiral switching behaviour is best understood by analyzing the dynamics in terms of the instantaneous eigemodes $\vec{r}_{\pm}(t)$ of $\HH_\omega(t)$.  These are defined via
\begin{equation}
    \HH_{\omega} \left( t\right) {{\vec r}_ \pm \left (t \right)} = \lambda_\pm \left (t \right)   {{\vec r}_ \pm \left (t \right)},
\label{eq:RevDef}
\end{equation}
where explicit forms for the eigenmodes and eigenvalues $\lambda_{\pm}(t)$ are given Eq.~\eqref{appeq:QAD2mevs} of Appendix~\ref{appsec:RevQAD}.
Classically, we could describe the instantaneous state of our system in terms of the amplitudes $c_{\pm}(t)$ of the two eigenmodes.  In our quantum parametric amplifier analogue, these amplitudes become operators:
\begin{equation}
    \hat{c}_\pm (t) =
        {{\vec r}^T_ \pm (t)} \cdot  \vec {\hat z }(t),
        \label{eq:chatdefn}
\end{equation}

Not surprisingly, the average values of these operators behave exactly as the corresponding amplitudes in the classical setup.  
Preparing a particular initial condition would involve displacing the four bosonic modes appropriately.  In Fig.~\ref{fig:QADfref}, we show the evolution of the average instantaneous mode amplitudes $\left| \langle \hat{c}_\pm \left (t \right) \rangle \right |$, for evolution along a circular path in the $(g,\omega)$ parameter space that encircles an EP. In both cases, the initial state is chosen so that only the $+$ eigenmode is initially excited, i.e.~$\left\langle \right.\vec {\hat z }(t=0)\left. \right\rangle={{\vec r}_ + (t=0)}$.   As can be seen from the figure, for evolution corresponding to a counter-clockwise (CCW) encircling, the 
amplitudes of the pseudo-modes correspond to predominantly exciting the instantaneous $+$ eigenmode.  In contrast, for a clockwise encircling, one sees that there is a switching:  at the final time $T$, the pseudo-mode amplitudes correspond to predominantly exciting the $-$ instantaneous eigenmode.
Note that because of the EP structure in our system, the instantaneous eigenmodes at the final time $t=T$ are flipped versions of those at $t=0$, i.e. $\vec{r}_\pm(t=T) = \vec{r}_\mp(t=0)$ \cite{Milburn2015}.

\begin{figure}[h]
  \includegraphics[width=70mm]{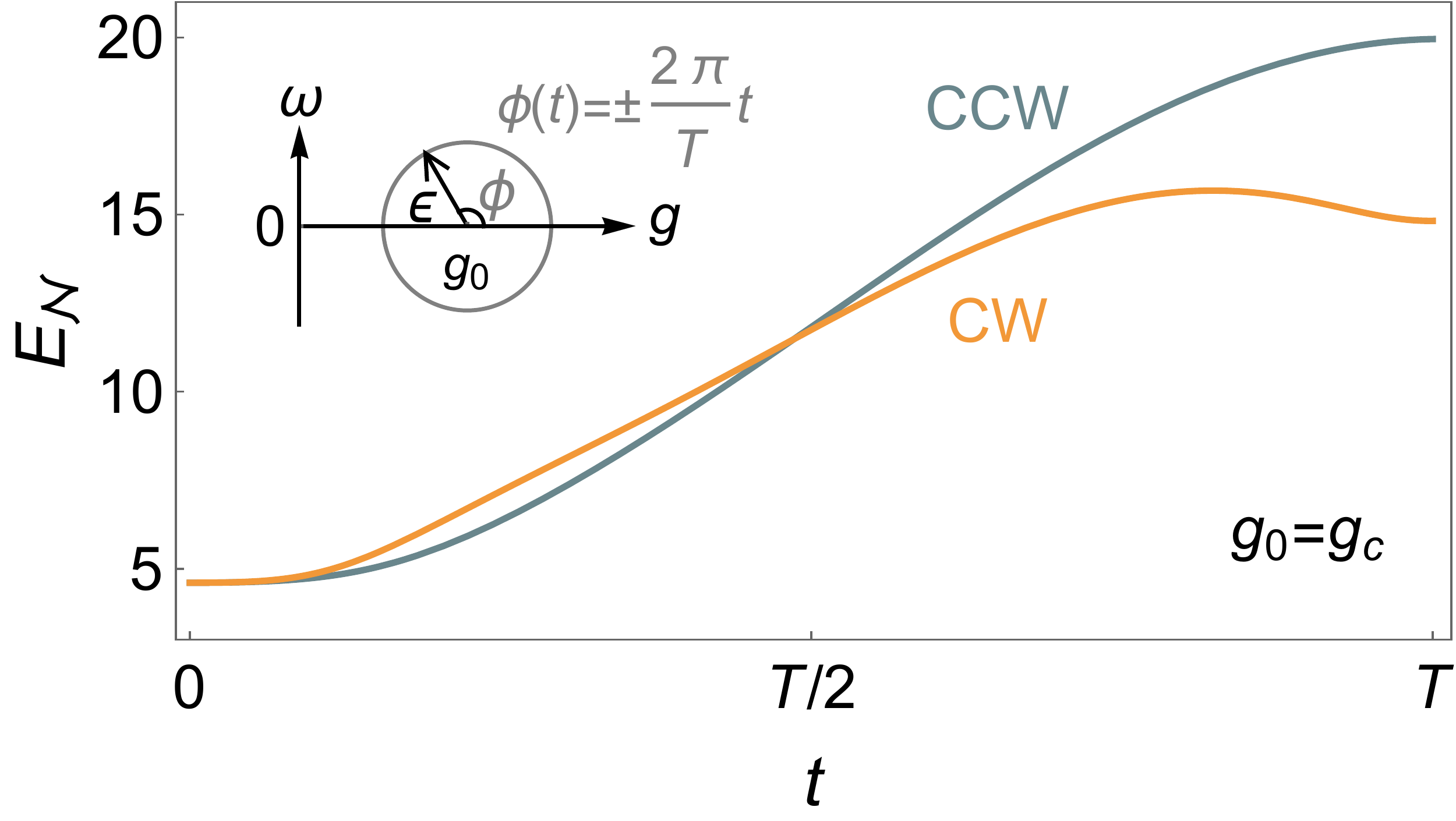}
  \includegraphics[width=70mm]{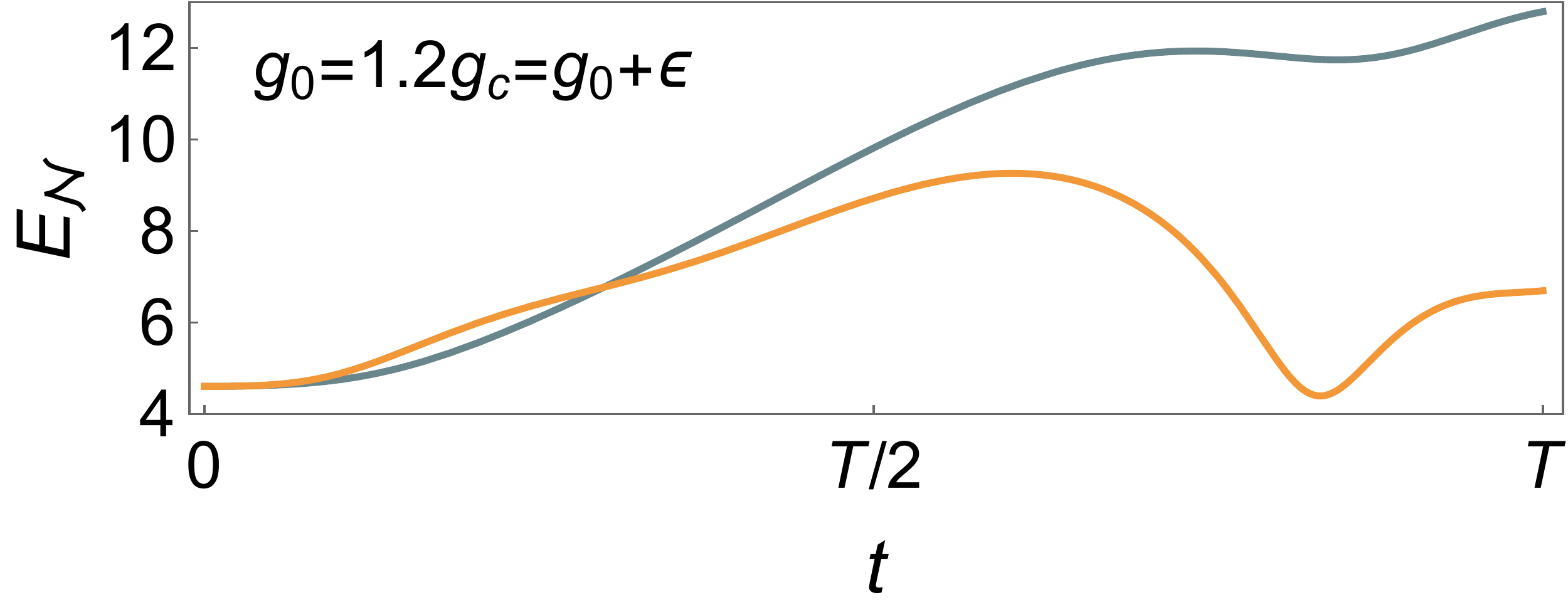}
  \includegraphics[width=70mm]{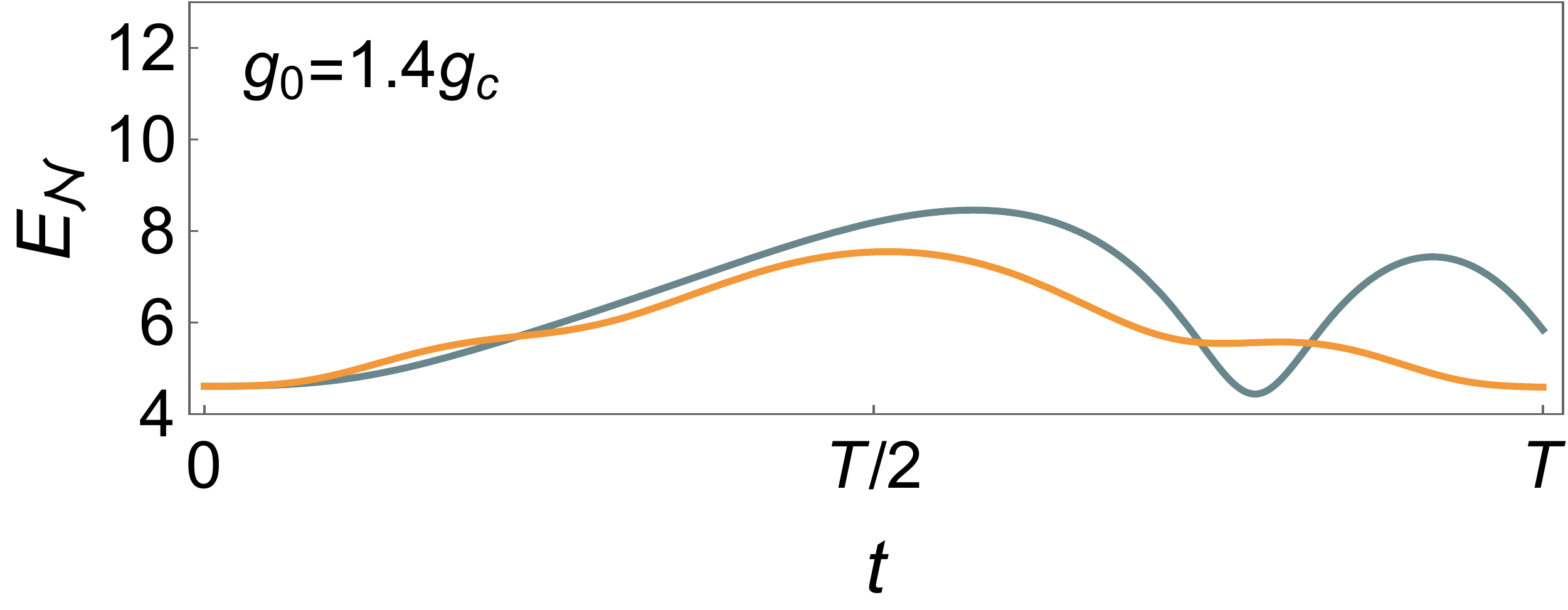}
  \includegraphics[width=70mm]{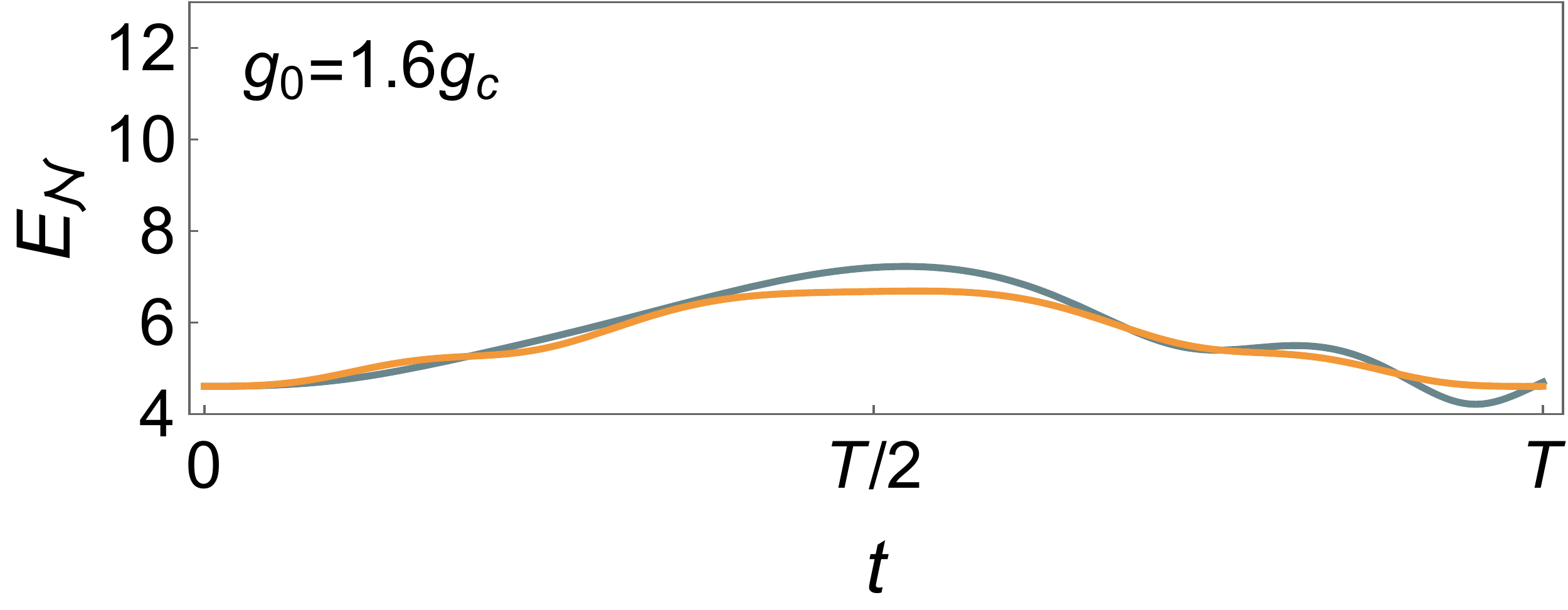}
  \caption{Entanglement evolution during EP encircling.  We consider  
  bipartite entanglement in a Hermitian four-mode bosonic system whose dynamics corresponds to the detuned gain-loss dimer described by $\HH_\omega(t)$ (c.f.~Eq.~(\ref{eq:Homegat})).  The parameters $(g,\omega)$ follow a complete circle in parameter space.  Entanglement (as quantified by the logarithmic negativity $E_N$) between the $a$ modes and $b$ modes is plotted as a function of time; the two curves in each panel are for a clockwise (CW) versus counterclockwise (CCW) parameter variation.  The center of the circular trajectory $(g=g_0,\omega)$ is different for each panel (as indicated).  The top panel corresponds to a trajectory centered on the exceptional point at $g = g_c \equiv \gamma/2$, whereas in the subsequent panels, the trajectory is displaced to the right.  One sees that entanglement generation is manifestly chiral when the trajectory encircles the EP, while this is lost when the trajectory is far from the EP.  The initial state is chosen to asymmetrically populate the $+$ eigenmode (see main text), and $\gamma T = 20$, $\epsilon / \gamma = 0.1$.}  
\label{fig:En}
\end{figure}

A more interesting situation is to consider the evolution of a general quantum state for either a CW or CCW parameter evolution.  In Appendix \ref{app:EPEncirclingDetails}, we derive the quantum unitary transformations describing both these cases, and discuss them using the Bloch-Messiah decomposition \cite{Braunstein2005}.  This allows us to view each transformation as the product of two beam-splitter operations, interspersed with a squeezing operation.  Interestingly, we find that both the CW and CCW complete encirclings are described by the same squeezing operation; the chirality only appears in the initial and final beam-splitter operations.

To see a direct consequence of this, imagine a quantum state with non-zero photon number, but where $\langle \hat{z}_1 \rangle = \langle \hat{z}_2 \rangle = 0$.  Classically, we could imagine at $t=0$ stochastically preparing the system in the $+$ eigenmode with a random phase; the simplest choice would be to take $c_{+}(0)$ to be a Gaussian random variable (while $c_{-}(0)$ is set to zero).  Using our equivalent quantum parametric amplifier setup  $\hat{H}_{\omega \mathrm{PA}}(t)$, we could consider an analogous initial condition.  In particular, we start the quantum four-mode system in a zero-mean Gaussian pure state  whose covariance matrix at $t=0$ predominantly populates the $+$ eigenmode. By this, we mean an initial state where $\langle \hat{c}_+ \hat{c}_+ \rangle \gg \langle \hat{c}_- \hat{c}_- \rangle $.
This state will necessarily have entanglement correlations between the $a$ subsystem (formed by modes $a_1$, $a_2$) and the $b$ subsystem (formed by modes $b_1$, $b_2$).  We can now ask how this entanglement evolves in time as we cyclically vary $g$ and $\omega$ as before.

In Fig.~\ref{fig:En}, we show the evolution of the $a$-$b$ subsystem entanglement (as quantified by the logarithmic negativity \cite{Vidal2002,Plenio2005}), for various circular parameter variations.
 In each case, we start with zero-mean Gaussian states of our four bosonic modes with the same amount of initial entanglement.  This initial state is chosen to have an average total photon number of approximately $100$, and an asymmetry quantified by
\begin{equation}
   \frac{\langle \hat{c}^\dag_+ \hat{c}_+ \rangle }{\langle \hat{c}^\dag_- \hat{c}_- \rangle }
   \simeq  \frac{10^4 }{\left|{\vec r}_+ \left (t=0 \right) \right|^2 \cdot \left|{\vec r}_- \left (t=0 \right) \right|^2} \gg 1.
\end{equation}
Further details and motivation for this choice are given in App.~\ref{app:EPEncirclingStates}; our chosen state corresponds to $e^{ \lambda_0} = 10$ in Eq.~(\ref{eq:ClassicalCovariance}).
The two curves in each panel correspond to CW and CCW traversal of the same circular path in the $(g,\omega)$ parameter space.  The first panel corresponds to the same path as in Fig.~\ref{fig:QADfref}; in the remaining panels, we displace the path so that it eventually no longer encloses the EP.  For paths enclosing the EP, we see that the entanglement evolution exhibits a marked chiral behaviour:  the amount of entanglement depends crucially on the direction that the path is traversed.          This asymmetry gradually becomes negligible as we displace the circular path away from the EP.  The results here show that the chirality associated with EP encircling can indeed have impliciations for quantum dynamics (e.g.~in determining the generation of entanglement).

\subsection{Connecting topology in non-Hermitian systems to Hermitian driven bosonic systems}


As a final application, we discuss how our mappings can be applied
to non-Hermitian lattice models with topologically non-trivial bands.  There has been considerable recent interest in studying such models, see e.g.~\cite{Lee2016,Lieu2018,Lieu2018a,Ueda2018,Sato2018}.  Our mappings provide a route for realizing these topological bands in fully Hermitian bosonic systems, without any need to couple to external dissipation.  More specifically, in Ref.~\cite{Fu2018}, a non-Hermitian Chern number was introduced to characterize bands in 2D non-Hermitian systems.  Using our mapping, it is straightforward to show that these Chern numbers are equivalent to topological invariants that were introduced earlier to characterize bands in Hermitian bosonic systems with pairing terms \cite{Shindou2013,Peano2016}; this is demonstrated in Appendix \ref{app:ChernNumbers}.  Our discussion here complements recent studies showing that the symmetry-based classification of non-Hermitian Hamiltonians can also be applied to anomalous bosonic systems \cite{Lieu2018,Sato2018}.

Despite the immense interest in non-Hermitian topology, most work has focused on models that are topological even if the non-Hermitian terms are set to zero (i.e.~in the absence of gain and loss).  As discussed below, our approach allows us to  construct a model where this is no longer true: non-trivial band topology {\it only} emerges in the presence of non-zero gain and loss.  We accomplish this by constructing the non-Hermitian equivalent of a recently-studied bosonic model where parametric driving induces topology \cite{Peano2016}.


\subsubsection{Non-trivial topology induced by gain and loss}
\label{Sec:PTKagome}

\begin{figure}
\includegraphics[width=44mm]{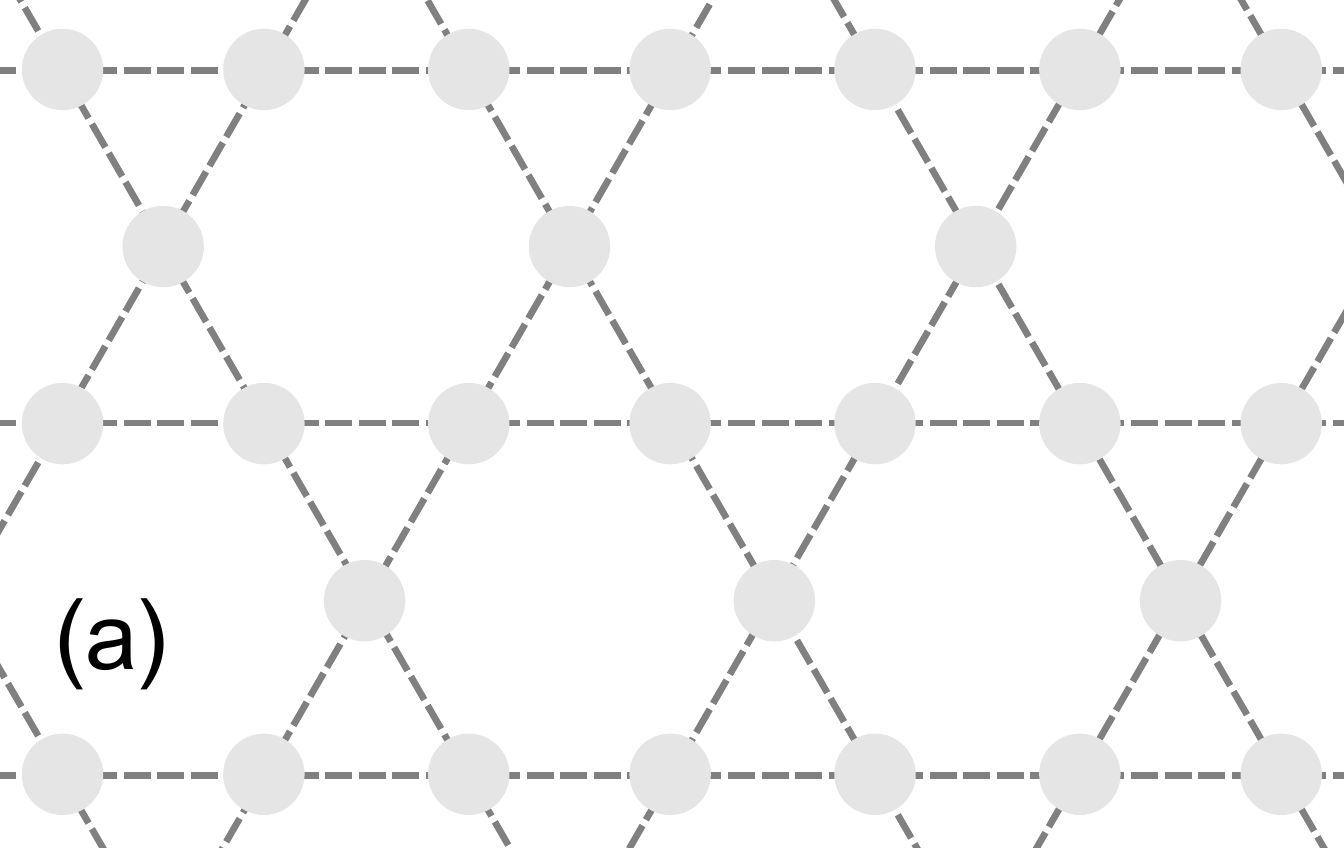}
  \includegraphics[width=36mm]{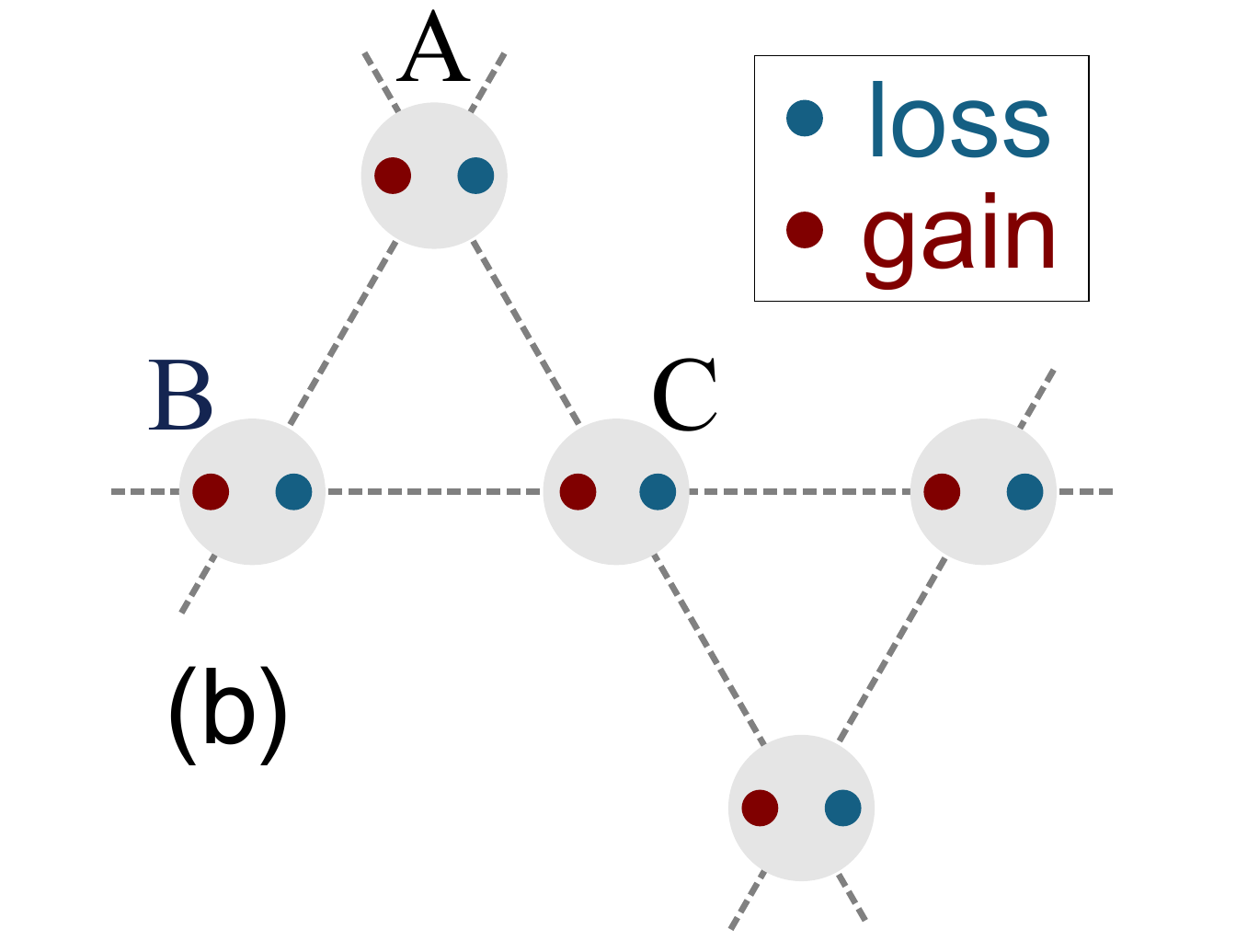}
  \caption{Schematic of the $\PT$ Kagome model (c.f. Eq.~(\ref{eq:HKagomeTotal})), where the system is topologically non-trivial only with the introduction of non-zero gain and loss.  (a) Kagome lattice, where each site (grey circle) is a gain-loss dimer.  (b) Unit cell.}
  \label{fig:Kagome}
\end{figure}

We consider a 2D Kagome lattice, 
where on each lattice site we have a two-cavity $\mathcal{PT}$-symmetric gain-loss dimer (see Fig.~\ref{fig:Kagome}).
The system Hamiltonian will consist of a purely Hermitian hopping terms coupling nearest neighbour lattices, and purely local term which includes non-Hermitian effects:
\begin{equation}
    {{\hat {\mathcal{H}}}_\mathrm{Kagome}} = 
         \hat{\HH}_{\rm hopping} + \hat{\HH}_{\rm local}.
        \label{eq:HKagomeTotal}
    \end{equation}
We will use the composite index ${\mathfrak{j}} = \left( {\mathbf{j},s} \right) = \left( {{j_1},{j_2},s} \right)$ to label both the unit cell $(j_1,j_2)$ and basis element $s = A, B, C$ of each lattice site.  Further, we will use a pseudospin $\uparrow, \downarrow$ to index each element of the cavity dimer located at a given lattice site.   

Letting ${\hat{\psi} _{{\mathfrak{j}} }} = {\left( {\begin{array}{*{20}{c}}
  {{{\hat a}_{{\mathfrak{j}} , \uparrow }}}&{{{\hat a}_{{\mathfrak{j}} , \downarrow }}} 
\end{array}} \right)^T}$
the onsite terms are 
\begin{align}
 {{\hat {\mathcal{H}}}_\mathrm{local}} &=
    \sum\limits_{\mathfrak{j}} {\hat{\psi} _{\mathfrak{j}}^\dag  
    \left( i{{\nu }}{\sigma _z} + {\omega _0}{\sigma _x} \right)
    {\hat{\psi} _{\mathfrak{j}}}}.
    \label{eq:HKagomeLocal}
\end{align}
These local terms describe a $\PT$ dimer at each lattice site, with tunneling amplitude $\omega_0$ and gain/loss rate $\nu$.

The tunneling terms between cavities on nearest neighbour lattice sites is described by the Hermitian Hamiltonian
\begin{align}
{{\hat {\mathcal{H}}}_\mathrm{hopping}} & =\sum\limits_{\left\langle \mathfrak{j}, \mathfrak{j}' \right\rangle } { \psi  _{{\mathfrak{j}}}^\dag {\mathcal{J} \left[{ss'}\right]}{\psi _{{\mathfrak{j'}}}} }
\end{align}
where the hopping matrix elements depend on both sublattice index and pseudo-spin:
\begin{align}
    {\mathcal{J}  \left[{ss'}\right]} & =  \frac{J}{2}\left( {{{e^{i{\varphi _{ss'}}}}}\sqrt 3 {\sigma _0} + {\sigma _x}} \right) \\
{\varphi _{ss'}} & =\left\{{\begin{array}{*{20}{c}}
  {+\frac{\pi }{2}, \quad   ss' = AB,BC,CA,} \\ 
  {-\frac{\pi }{2}, \quad   ss' = BA,CB,AC,} 
\end{array}} \right.
\end{align}
$J$ is the overall hopping amplitude.  We see that there are hopping terms that both preserve and flip the pseudo spin (i.e.~a gain cavity on a given site can tunnel to either a gain or loss cavity on a neighbouring site).  Further, the spin-conserving tunneling is complex, and thus encodes a synthetic gauge field.
The tunneling here can be viewed as a generalized kind of synthetic spin-orbit coupling.  

Consider first the properties of our system in the case where there are no gain/loss terms (i.e.~$\nu=0$), and the Hamiltonian is Hermitian.  In this case, the system has no topologically non-trivial bands, as it is possible to completely gauge away the hopping phases.  To see this, note that in this case $\sigma_x$ on each lattice site commutes with the Hamiltonian.  It thus useful to use a local basis of $\sigma_x$ eigenstates:
 \begin{equation}
    {{\hat a}  _{\mathfrak{j}, \pm}} = \frac{1}{\sqrt 2} \left( {{\hat a}_{{\mathfrak{j}} , \uparrow }} \pm {{\hat a}_{{\mathfrak{j}} , \downarrow }} \right),
 \end{equation}
In this basis, the Hamiltonian decouples into two independent tight-binding models
 \begin{equation}
    {{\HH}_{\mathrm{Kagome},\pm }} = \pm \left(\sum\limits_{\mathfrak{j}} {\omega _0} {{\hat a}  _{\mathfrak{j}, \pm}^\dag {{\hat a}  _{\mathfrak{j}, \pm}}}  +\sum\limits_{\left\langle \mathfrak{j},\mathfrak{j}' \right\rangle } J {e^{ \pm i\frac{{2{\varphi _{ss'}}}}{3}}} { {\hat a} _{{\mathfrak{j}, \pm}}^\dag  {{\hat a}_{{\mathfrak{j'}, \pm}}} } \right) , 
\end{equation}
with uniform onsite energies $\pm {\omega _0}$ and nearest-neighbor couplings $\pm J \exp \left( { \mp i2{\varphi _{ss'}}/3} \right)$. 
We thus have two decoupled Kagome lattices, with the $+$ ($-$) lattice have a synthetic Aharonov-Bohm flux $\pi$ ($-\pi$) in each triangular plaquette.  These fluxes do not break time-reversal symmetry, and can be eliminated by a local gauge transformation: 
 \begin{subequations}
\begin{align}
{{\hat a}_{\mathbf{j},B, \pm}} &\to {{\hat a'}_{\mathbf{j},B, \pm}} = {e^{ \mp i\frac{{2\pi }}{3}}}{{\hat a}_{\mathbf{j},B, \pm}},\\
{{\hat a}_{\mathbf{j},C, \pm}} &\to {{\hat a'}_{\mathbf{j},C, \pm}} = {e^{ \mp i\frac{{4\pi }}{3}}}{{\hat a}_{\mathbf{j},C, \pm}}. 
\end{align}
\end{subequations}
This results in a decoupled pair of time-reversal invariant, topologically trivial Kagome models 
 \begin{equation}
    {{\hat H'}_{\mathrm{Kagome}, \pm }} = \pm \left(\sum\limits_{\mathfrak{j}} {\omega _0} {{\hat a}  _{\mathfrak{j},\pm}^\dag {{\hat a}_{\mathfrak{j},\pm}}}  -
    \sum \limits_{\left\langle \mathfrak{j} \mathfrak{j}' \right\rangle } 
    J   { {\hat a} _{{\mathfrak{j},\pm}}^\dag  {{\hat a}_{{\mathfrak{j'},\pm}}} } \right) . 
 \label{eq:DecoupleKagomes}
 \end{equation}
 
If we now turn on the gain/loss parts of the Hamiltonian (i.e.~make $\nu$ non-zero in Eq.~(\ref{eq:HKagomeLocal})), it is no longer possible to gauge away the hopping phases.  At an intuitive level, the non-Hermitian terms are off-diagonal in the $+/-$ basis used to write Eq.~(\ref{eq:DecoupleKagomes}), and hence can enable hopping processes that pick up non-trivial fluxes.  

The topological properties of the resulting model can be completely understood by mapping the system to a Hermitian, parametrically-driven bosonic model having a single bosonic mode on each lattice site (see Appendix~\ref{AppSecKc}).  As usual, the non-Hermitian gain/loss terms are mapped to parametric driving terms, and the dimer structure is mapped to the particle-hole structure of the bosonic theory.  The resulting bosonic theory is equivalent to the parametrically driven Kagome lattice mode studied by Peano et al in Ref.~\cite{Peano2016}.  The model exhibits topological bands and protected edge states whenever $\nu$ is non-zero.  Since the mapping is fully local in real space (two bosonic modes, with balanced gain and loss, per lattice site), it thus follows that our non-Hermitian $\mathcal{PT}$ model exhibits topological bands (with non-zero Chern number) and edge states if and only if there is non-zero gain loss.  We thus have, as desired, a model where topology is induced by gain/loss.  

\section{Conclusions}

In this paper we have given a thorough discussion of how one can realize non-Hermitian dynamics without the need to couple to external dissipation.  We make use of a simple but surprising fact:  a Hermitian, quadratic bosonic Hamiltonian that breaks particle number conservation necessarily gives rise to a non-Hermitian dynamical matrix.  We have discussed three generic strategies for using this correspondence to realize a given non-Hermitian (linear) Hamiltonian using a quadratic bosonic system.  Given an initial $N$ mode non-Hermitian problem, one can always accomplish this mapping using a $2 N$ mode bosonic system and the quantum-mechanics free subsystem (QMFS) approach discussed in Sec.~\ref{sec:TwoModeQMFS}.  In other more constrained cases, it is possible to mimic the desired dynamics using $N$ modes (Sec.~\ref{sec:Lattices}) or even $N/2$ modes (Sec.~\ref{sec:DPAMapping}).

Our work has considered just a few possible applications and implications of this mapping.  In future work, it will be interesting to use this mapping to explore a wider class of non-Hermitian dynamical phenomena (such as the recently observed non-Hermitian analogue of Fermi arcs \cite{Zhen2018}), and to develop new kinds of quantum control protocols in parametrically driven systems.  It will also be extremely interesting to extend our approach to describe systems with true nonlinearities.

\section*{Acknowledgements}

This work was supported by the University of Chicago
Materials Research Science and Engineering Center, which
is funded by the National Science Foundation under Grant
No. DMR-1420709.


\appendix

\section{Comment on pseudo-Hermiticity}\label{AppSecpsH}
In general, the $2N \times 2N$ dynamical matrix ${\mathcal{H}_{\mathrm{eff  }, N}} $ of a $N$-mode bosonic parametrically-driven system is related to a Hermitian Bogoliubov de Gennes Hamiltonian $H_\mathrm{BdG}$ by
\begin{equation}
{\mathcal{H}_{\mathrm{eff  }, N}} ={ \sigma}_{N,z} H_\mathrm{BdG},
\end{equation}
where ${ \sigma}_{N,z} =\mathrm{diag}\left(\mathbb{I}_N,-\mathbb{I}_N\right)$ is the diagonal matrix incorporating bosonic commutation relations, so that ${\mathcal{H}_{\mathrm{eff  }, N}} $ is always ${ \sigma}_{N,z} $-pseudo-Hermitian as \cite{JMP20021,Lieu2018}
\begin{equation}
{\mathcal{H}^\dag_{\mathrm{eff  }, N}}={ \sigma}_{N,z} {\mathcal{H}_{\mathrm{eff  }, N}}{ \sigma}_{N,z}.\end{equation}
Previous works have shown that $\mathcal{PT}$-symmetric Hamiltonians are always $\eta$-pseudo-Hermitian, where $\eta$ is an invertible Hermitian operator \cite{JMP20021,JMP20022,JMP20023}. However, two pseudo-Hermitian matrices need not be  unitarily equivalent, even though they may be isospectral. 
Hence, the fact that $\mathcal{PT}$-symmetric Hamiltonians are pseudo-Hermitian does not guarantee that they are unitarily equivalent to the dynamical matrix of some Hermitian, bosonic problem.

\section{Canonical form for a general class of $\PT$-symmetric non-Hermitian Hamiltonian}
\label{appsec:PTCanonicalForm}

Consider the most general $2N$-mode $\mathcal{PT}$-symmetric non-Hermitian Hamiltonian.  Using the same conventions as Sec.~\ref{subsec:StandardPTChain}, the Hamiltonian has the form
\begin{equation}
\label{eq:GeneralPT}
{\mathcal{H}_\mathcal{PT}} = \left( {\begin{array}{*{20}{c}}
  {\mathcal{E}}&{\mathcal{F}} \\ 
  {{{\mathcal{F}}^*}}&{{{\mathcal{E}}^*}} 
\end{array}} \right),\end{equation}
where ${\mathcal{E}}$ and ${\mathcal{F}} $ are arbitrary $N \times N$ matrices.
As always, we define time reversal operation as complex conjugation, and the parity operation is an exchange of modes described by 
\begin{equation}
{ { \sigma}_{N,x}} = \left( {\begin{array}{*{20}{c}}
  0&{{\mathbb{I}_N}} \\ 
  {{\mathbb{I}_N}}&0 
\end{array}} \right).\end{equation}

We will show that 
as long as the anti-Hermitian part of $\HHPT$ is full rank, it is always possible to unitarily transform $\HHPT$ to a form $\mathcal{H}_2$ where the non-Hermitian part of $\HH$ is diagonal, namely:
\begin{equation}
\label{eq:CanonicalPTHamiltonian}
{\mathcal{H}_2} = \left( {\begin{array}{*{20}{c}}
  {\tilde \Sigma  + i\Gamma_N }&{\tilde J} \\ 
  {{{\tilde J}^*}}&{{{\tilde \Sigma }^*} - i\Gamma_N } 
\end{array}} \right).\end{equation}
Here $\tilde \Sigma $ is a Hermitian $N \times N$ matrix, $\Gamma_N$ is a real, diagonal, non-negative $N \times N$ matrix, and the $N \times N$ matrix $\tilde{J}$ is symmetric.  Note that 
$\mathcal{H}_2$ still explicitly retains $\PT$ symmetry as defined before, as ${\mathcal{H} ^*_2}={ { \sigma}_{N,x}}{\mathcal{H}_2}{ { \sigma}_{N,x}}$.

We can always write $\HHPT$ in terms of its Hermitian and anti-Hermitian parts as ${\mathcal{H}_\mathcal{PT}}={ {H}_\mathcal{PT}}+i { {\Gamma}_\mathcal{PT}}$. The ${\mathcal{PT}}$ symmetry of ${\mathcal{H}_\mathcal{PT}}$ then implies
\begin{subequations} \label{APeqPTd}
\begin{align}
{ {H}^*_\mathcal{PT}}&  = { { \sigma}_{N,x}}{ {H}_\mathcal{PT} }{ { \sigma}_{N,x}},\\
{ {\Gamma}^*_\mathcal{PT}} & =- { { \sigma}_{N,x}} { {\Gamma}_\mathcal{PT}}{ { \sigma}_{N,x}}.
\label{eq:APTEqn}
\end{align}
\end{subequations}

Eq.~(\ref{eq:APTEqn}) implies that the eigenvalues of ${ {\Gamma}_\mathcal{PT}}$ are real and come in pairs of opposite signs.  It can thus be diagonalized as 
\begin{equation}
\mathcal{U}_{\Gamma}  { {\Gamma}_\mathcal{PT}} \mathcal{U}^\dag_{\Gamma}={\Gamma_D} ,\end{equation}
where ${\Gamma_D}= \mathrm{diag} \left( {{\Gamma_N}, - {\Gamma _N}} \right)$, and ${\Gamma _N}$ is a diagonal $N \times N$ matrix with non-negative entries.  

Using $\mathcal{U}_{\Gamma}$ to transform $\HHPT$, i.e. 
${\mathcal{H}}_1 =\mathcal{U}_{\Gamma}  { \mathcal{H}_\mathcal{PT} } \mathcal{U}^\dag_{\Gamma} $, we obtain
\begin{equation}
{\mathcal{H}}_1= H_1 + i{\Gamma_D} = \left( {\begin{array}{*{20}{c}}
  {{\Sigma _1} + i\Gamma _N }&J \\ 
  {{J^\dag }}&{{\Sigma _2} - i\Gamma _N } 
\end{array}} \right) ,\end{equation}
$\PT$ symmetry implies that the Hermitian matrix $H_1$ and the non-negative matrix ${\Gamma_D}$ must satisfy 
\begin{subequations}
\begin{align}
{H_1} &= {  \tilde{\mathcal{U}}_{\Gamma} } H_1^* {{  \tilde{\mathcal{U}}^\dag_{\Gamma} }},\label{APeqptH1}\\
 {\Gamma_D} &=- {  \tilde{\mathcal{U}}_{\Gamma} } {\Gamma_D} {{  \tilde{\mathcal{U}}^\dag_{\Gamma} }}, \label{APeqptad}
\end{align}
\end{subequations}
where we have introduced the symmetric unitary matrix ${{  \tilde{\mathcal{U}}_{\Gamma} }} = {\mathcal{U}}_{\Gamma} { { \sigma}_{N,x}}{ {\mathcal{U}}^T_{\Gamma}}$.

Eq.~\eqref{APeqptad} can be written explicitly as 
\begin{equation}
  \left( {\begin{array}{*{20}{c}}
  { + \Gamma_N }&0 \\ 
  0&{ - \Gamma_N } 
\end{array}} \right){{  \tilde{\mathcal{U}}_{\Gamma} }} +{{  \tilde{\mathcal{U}}_{\Gamma} }} \left( {\begin{array}{*{20}{c}}
  { + \Gamma_N }&0 \\ 
  0&{ - \Gamma_N } 
\end{array}} \right) = 0,\end{equation}
In what follows, we assume that ${ {\Gamma}_\mathcal{PT}}$ is full rank; physically, this implies that all modes in the system are coupled to the dissipation.  As a consequence,  $\Gamma_N$ has no zeros on the diagonal.   The above equation then provides a constraint on the form of $\tilde{\mathcal{U}}_{\Gamma}$:  its
diagonal blocks must be identically zero.  
We can thus write it as
\begin{equation}
{{  \tilde{\mathcal{U}}_{\Gamma} }} = \left( {\begin{array}{*{20}{c}}
  0&{{{\tilde u}_{12}}} \\ 
  {\tilde u_{12}^T}&0 
\end{array}} \right),\end{equation}
with $\left[ {{{{\tilde u}_{12}}},\Gamma_N } \right] = 0$. 

The remaining $\PT$ condition on ${H_1}$ in  Eq.~\eqref{APeqptH1} now reads
\begin{equation}
\left( {\begin{array}{*{20}{c}}
  {{\Sigma _1}}&J \\ 
  {{J^\dag }}&{{\Sigma _2}} 
\end{array}} \right) \left( {\begin{array}{*{20}{c}}
  0&{{{\tilde u}_{12}}} \\ 
  {\tilde u_{12}^T}&0 
\end{array}} \right)= \left( {\begin{array}{*{20}{c}}
  0&{{{\tilde u}_{12}}} \\ 
  {\tilde u_{12}^T}&0 
\end{array}} \right)\left( {\begin{array}{*{20}{c}}
  {\Sigma _1^*}&{{J^*}} \\ 
  {{J^T}}&{\Sigma _2^*} 
\end{array}} \right),\end{equation}
or equivalently
\begin{subequations}
\begin{align}
&{J^\dag }  {{\tilde u}_{12}}= \tilde u_{12}^T {J^*} \Leftrightarrow J\tilde u_{12}^T = {{\tilde u}_{12}}{J^T},\\
&{\Sigma _1}{{\tilde u}_{12}} ={{\tilde u}_{12}} \Sigma _2^* \Leftrightarrow {\Sigma _2} \tilde u_{12}^T = \tilde u_{12}^T \Sigma _1^*.
\end{align}
\end{subequations}
It follows that there exists a unitary matrix ${\tilde u_{12}}$ that commutes with $\Gamma_N$ and satisfy the equalities
\begin{subequations}
\begin{align}
\Sigma _1^* &= {{\tilde u}^*_{12}}{\Sigma _2 }\tilde u_{12}^T ,\\
J\tilde u_{12}^T = {{\tilde u}_{12}}{J^T} &= {\left( {J\tilde u_{12}^T } \right)^T}= {\left( { \tilde u_{12}^* J^\dag } \right)^*} .
\end{align}
\end{subequations}

We can now finally use this unitary matrix to transform our non-Hermitian $\PT$ Hamiltonian into a simpler, final form $\mathcal{H}_2$:
\begin{align}
{\mathcal{H}_2} & \equiv \left( {\begin{array}{*{20}{c}}
  {\mathbb{I}_N}&0 \\ 
  0&{{{\tilde u}^*_{12}}} 
\end{array}} \right)\left( {\begin{array}{*{20}{c}}
  {{\Sigma _1} + i\Gamma_N }&J \\ 
  {{J^\dag }}&{{\Sigma _2} - i\Gamma_N } 
\end{array}} \right)\left( {\begin{array}{*{20}{c}}
  {\mathbb{I}_N}&0 \\ 
  0&{{\tilde u} _{12}^T } 
\end{array}} \right) \nonumber\\
&= \left( {\begin{array}{*{20}{c}}
  {{\Sigma _1} + i\Gamma_N }&{J\tilde u_{12}^T } \\ 
  {{{\left( {J\tilde u_{12}^T } \right)}^*}}&{\Sigma _1^* - i\Gamma_N } 
\end{array}} \right),\end{align}
Defining $\tilde \Sigma  = {\Sigma _1}$ and $\tilde J = J\tilde u_{12}^\dag$ and 
$\tilde{J} = J \tilde{u}_{12}^T$, this is exactly the form given in Eq.~(\ref{eq:CanonicalPTHamiltonian}).

\section{Constraints on representing $\PT$ non-Hermitian Hamiltonians with Hermitian bosonic Hamiltonians }
\label{AppSecConv}


In this appendix we will derive necessary and sufficient conditions for determining whether a given $2N$-mode $\PT$-symmetric non-Hermitian Hamiltonian is unitarily equivalent to 
the dynamical matrix of a Hermitian, $2N$ mode parametric amplifier system.
The most general $2N$ mode $\mathcal{PT}$-symmetric non-Hermitian Hamiltonian $\HHPT$ was given in Eq.~(\ref{eq:GeneralPT}).  As shown above, as long as its non-Hermitian part is full rank, it can be transformed to the canonical form $\HH_2$ in 
Eq.~(\ref{eq:CanonicalPTHamiltonian}), where the anti-Hermitian terms are diagonal.  We will work with this form in what follows.

The question now is whether it is possible to find a unitary matrix $\mathcal{U}$ that transforms the generic $\PT$ Hamiltonian $\mathcal{H}_2$ in Eq.~(\ref{eq:CanonicalPTHamiltonian}) to a physical bosonic dynamical matrix $\mathcal{M}_N$, i.e.
\begin{equation}
    \mathcal{U} {\mathcal{H}_2} \mathcal{U}^\dag= \mathcal{M}_{N}.
    \label{eq:UnitaryMapping}
\end{equation}
We will consider the least constrained mapping, where $\mathcal{M}_N$ is a $2N \times 2N$ matrix describing {\it non-degenerate} parametric driving.  This is the same situation as in Sec.~\ref{sec:Lattices}: the bosonic theory has $N$ ``$a$" modes and $N$ ``$b$" modes, and the parametric driving conserves the total number of $a$ minus $b$ bosons.   As discussed in the main text (c.f.~Eq.~(\ref{efftoBdGg})), in this case the dynamical matrix will take the form  
\begin{equation}
{\mathcal{M}_{N}}=\left( {\begin{array}{*{20}{c}}
  {{\mu _a}}&{  \nu } \\ 
  {{-\nu ^\dag }}&{{- \mu _b^T}} 
\end{array}} \right),\label{appeq:DynMatform}
\end{equation}
where ${{\mu _{a,b}}}$ are arbitrary Hermitian $N \times N$ matrices, and $\nu$ can be any $N \times N$ matrix.


For a given $\mathcal{PT}$ Hamiltonian $\mathcal{H}_2$, it is not always possible to find a $\mathcal{U}$ and $\mathcal{M}_N$ satisfying Eq.~(\ref{eq:UnitaryMapping}).  This is because the bosonic dynamical matrix
$\mathcal{M}_N$ is pseudo-Hermitian in a constrained fashion.  Recall (see Appendix~\ref{AppSecpsH}) that any physical $\mathcal{M}_N$ must satisfy:
\begin{equation}
{{\mathcal{M}^\dag_{ N}} } = { { \sigma}_{N,z}} \, {\mathcal{M} _{ N}}
\, { { \sigma}_{N,z}},
\label{eq:MPseudoH}
\end{equation}
where as always, ${\sigma}_{N,z}$ is a $z$ Pauli matrix in particle-hole space.

It thus follows that {\it any} $\mathcal{PT}$ Hamiltonian that is unitarily equivalent to a bosonic dynamical matrix $\mathcal{M}_N$ must satisfy 
\begin{equation}
    \mathcal{H}_2^\dagger = 
    \mathcal{W} \mathcal{H}_2 \mathcal{W}^\dagger,
        \label{eq:WDefinition}
\end{equation}
for some $2N \times 2N$ matrix $\mathcal{W}$ satisfying 
\begin{subequations}
\begin{eqnarray}
    \mathcal{W} & = & \mathcal{W}^\dagger = \mathcal{W}^{-1}, \\
    \mathrm{Tr } \, \mathcal{W} &= & 0,
\end{eqnarray}
    \label{eq:WConstraint}
\end{subequations}
(i.e.~a Hermitian unitary matrix with $N$ eigenvalues $+1$ and $N$ eigenvalues $-1$).
If a unitary equivalence existed as per Eq.~(\ref{eq:UnitaryMapping}), we could explicitly construct $\mathcal{W}$ as 
$  \mathcal{W} = \mathcal{U} \sigma_{N,z} \mathcal{U}^\dagger$.
Eqs.~(\ref{eq:WDefinition}) and (\ref{eq:WConstraint}) thus represent a necessary condition for the existence of a $\mathcal{M}_N$ that is unitarily equivalent to a given $\mathcal{H}_2$.  Note this is a tighter constraint than simply requiring $\mathcal{H}_2$ to be pseudo-Hermitian (something that is always true).

To show that this is also a {\it sufficient} condition, suppose one can find a $\mathcal{W}$ satisfying Eqs.~(\ref{eq:WDefinition}) and (\ref{eq:WConstraint}).  $\mathcal{W}$ could be then diagonalized as  $\mathcal{W}=\mathcal{U}_\mathcal{W} { { \sigma}_{N,z}}\mathcal{U}^\dag_\mathcal{W}$ for some unitary $\mathcal{U}_\mathcal{W}$. 
It then easily follows that the matrix $\tilde{\mathcal{M}} \equiv \mathcal{U}_\mathcal{W}^\dagger \mathcal{H}_2 \mathcal{U}_\mathcal{W}$ satisfies the pseudo-Hermiticity condition in Eq.~(\ref{eq:MPseudoH}), and thus represents a valid bosonic dynamical matrix.

We can derive more explicit conditions in the case where the anti-Hermitian part of $\mathcal{H}_2$ is full-rank (i.e.~$\Gamma_N$ is positive).  
In this case, Eq.~(\ref{eq:WDefinition}) can only be satisfied if $\mathcal{W}$ has the form
\begin{equation}
\mathcal{W} = \left( {\begin{array}{*{20}{c}}
  0&{{w_{12}}} \\ 
  {w_{12}^\dag }&0 
\end{array}} \right),\end{equation}
where the unitary matrix ${{w_{12}}}$ commutes with $\Gamma_N$.  This form of $\mathcal{W}$ is explicitly Hermitian, unitary and traceless, so it fulfills all the conditions in Eq.~(\ref{eq:WConstraint}).

From Eq.~(\ref{eq:WConstraint}), we can now derive:
\begin{equation}
\left( {\begin{array}{*{20}{c}}
  0&{{w_{12}}} \\ 
  {w_{12}^\dag }&0 
\end{array}} \right)\left( {\begin{array}{*{20}{c}}
  {\tilde \Sigma }&{\tilde {J}} \\ 
  {{{\tilde {J}}^*}}&{{{\tilde \Sigma }^*}} 
\end{array}} \right) = \left( {\begin{array}{*{20}{c}}
  {\tilde \Sigma }&{\tilde {J}} \\ 
  {{{\tilde {J}}^*}}&{{{\tilde \Sigma }^*}} 
\end{array}} \right)\left( {\begin{array}{*{20}{c}}
  0&{{w_{12}}} \\ 
  {w_{12}^\dag }&0 
\end{array}} \right).\end{equation}
Hence, a necesary condition for $\mathcal{H}_2$ to be unitarily equivalent to a bosonic dynamical matrix is the existence of an $N \times N$ unitary matrix $w_{12}$ that satisfies the equations
\begin{subequations}
\label{appeq:GenPTtoPAthm}
\begin{eqnarray}
{w_{12}}{{\tilde {J}}^*} & =& \tilde {J} w_{12}^\dag    ,\\
{w_{12}}{{\tilde \Sigma }^*} &= & \tilde \Sigma {w_{12}}  .
\end{eqnarray}
\end{subequations}


\section{Constraints on representing Hermitian bosonic Hamiltonians with $\PT$ non-Hermitian Hamiltonians}

We now ask the converse of the question discussed in the previous appendix. Given a generic $2N$-mode non-degenerate parametric amplifier (NDPA) with Hermitian Hamiltonian ${{\hat H}_\mathrm{NDPA,multi.}}$, whose non-Hermitian dynamical matrix ${{\mathcal{M}_{ N}} } $ takes the form (see Eq.~\eqref{appeq:DynMatform})
\begin{equation}
{{\mathcal{M}_{ N}} }  = \left( {\begin{array}{*{20}{c}}
{{\mu _a}}&\nu \\
{ - {\nu ^\dag }}&{ - \mu _b^T}
\end{array}} \right)  ,\end{equation}
we would like to know if there exists unitary matrix $\mathcal{U}_{\mathcal{M}}^\dag$ that transforms ${{\mathcal{M}_{ N}} } $ to an effective non-Hermitian Hamiltonian matrix $\mathcal{H}_{\mathcal{PT}}$ with explicit $\mathcal{PT}$ symmetry
\begin{equation}
\mathcal{U}_{\mathcal{M}} {{\mathcal{M}_{ N}} } {\mathcal{U}^\dag_{\mathcal{M}} } = { { \sigma}_{N,x}}{\left( \mathcal{U}_{\mathcal{M}}  {{\mathcal{M}_{ N}} } {\mathcal{U}^\dag_{\mathcal{M}} }  \right)^*}{ { \sigma}_{N,x}},\end{equation}
or equivalently
\begin{align}
{{\mathcal{M}_{ N}} } &= {\mathcal{U}^\dag_{\mathcal{M}}}{ { \sigma}_{N,x}}{\mathcal{U}_{\mathcal{M}}^*}{\mathcal{M}_{ N}^*}{\mathcal{U}_{\mathcal{M}}^T}{ { \sigma}_{N,x}}{\mathcal{U}_{\mathcal{M}} },\\
\Rightarrow &{\mathcal{M}_{ N}} = {\mathcal{W}_{\mathcal{M}} }{\mathcal{M}_{ N}^*}{\mathcal{W}_{\mathcal{M}}^\dag }, \label{appeq:GenAntiUniDM}
\end{align}
where $\mathcal{W}_{\mathcal{M}} =  {\mathcal{U}^\dag_{\mathcal{M}}}{ { \sigma}_{N,x}}{\mathcal{U}_{\mathcal{M}}^*}$. Thus to determine the existence of a unitarily equivalent $\mathcal{H}_{\mathcal{PT}}$ for dynamical matrix ${{\mathcal{M}_{ N}} } $, we can equivalently ask if there exists a symmetric, unitary matrix $\mathcal{W}_{\mathcal{M}}$ that can be written in the form 
\begin{equation}
\mathcal{W}_{\mathcal{M}} =  {\mathcal{U}^\dag_{\mathcal{M}}}{ { \sigma}_{N,x}}{\mathcal{U}_{\mathcal{M}}^*},
\end{equation}
such that Eq.~\eqref{appeq:GenAntiUniDM} is satisfied. Physically, being unitarily equivalent to any $\mathcal{PT}$-symmetric Hamiltonian can thus be viewed as a generalized $\mathcal{W}_{\mathcal{M}}$-anti-unitary symmetry for the dynamical matrix considered, with specific constraints imposed on the unitary matrix $\mathcal{W}_{\mathcal{M}}$.

In analogy to the derivation of canonical form for $\mathcal{PT}$-symmetric Hamiltonians $\HHPT$, we now transform ${{\mathcal{M}_{ N}} } $ to a more tractable form. Noting that the $N \times N$ off-diagonal block matrix $\nu$ can be written as singular value decomposition $\nu  = \mathcal{V}_a^\dag D_{\nu} {\mathcal{V}_b}$, or equivalently
\begin{equation}
{\mathcal{V}_a}\nu \mathcal{V}_b^\dag  = D_{\nu}={\mathcal{V}_b}{\nu ^\dag }\mathcal{V}_a^\dag ,\end{equation}
we can transform the off-diagonal blocks into non-negative diagonal matrix $D_{\nu}$
\begin{align}{\mathcal{H}_1} &= \left( {\begin{array}{*{20}{c}}
{{\mathcal{V}_a}}&0\\
0&{{\mathcal{V}_b}}
\end{array}} \right)\left( {\begin{array}{*{20}{c}}
{{\mu _a}}&\nu \\
{ - {\nu ^\dag }}&{ - \mu _b^T}
\end{array}} \right)\left( {\begin{array}{*{20}{c}}
{\mathcal{V}_a^\dag }&0\\
0&{\mathcal{V}_b^\dag }
\end{array}} \right) \nonumber\\
&= \left( {\begin{array}{*{20}{c}}
{{\mathcal{V}_a}{\mu _a}\mathcal{V}_a^\dag }&{D_{\nu}}\\
{ - D_{\nu}}&{ - {\mathcal{V}_b}\mu _b^T \mathcal{V}_b^\dag }
\end{array}} \right),\end{align}
which can be rewritten in terms of $N \times N$ Hermitian matrices $\Sigma ,\Delta $ as
\begin{equation}
\label{eq:CanonicalPADM}
{\mathcal{H}_1} = \left( {\begin{array}{*{20}{c}}
{\Sigma  + \Delta }&{D_{\nu}}\\
{ - D_{\nu}}&{\Delta  - \Sigma }
\end{array}} \right).
\end{equation}

The next step is to rotate the anti-Hermitian part to diagonal blocks via a unitary transformation, where we obtain
\begin{equation}
{\mathcal{H}_2} = \left( {\begin{array}{*{20}{c}}
{\Delta  + i {D_{\nu}}}&\Sigma \\
\Sigma &{\Delta  - i {D_{\nu}}}
\end{array}} \right)= H_1 + i{\Gamma_{\nu}},\end{equation}
so that the equality in Eq.~\eqref{appeq:GenAntiUniDM} can be equivalently written as the conditions on the $2N \times 2N$ Hermitian matrix $H_1$ and the $2N \times 2N$  non-negative diagonal matrix ${\Gamma_{\nu}}$
\begin{subequations}
\begin{align}
&H_1 = {\tilde{\mathcal{ W}}_{\mathcal{M}} }{H_1^*} {\tilde{\mathcal{ W}}_{\mathcal{M}}^\dag } ,\label{appeq:GenAntiUniH1} \\
&{\Gamma_{\nu}} = - {\tilde{\mathcal{ W}}_{\mathcal{M}} }{{\Gamma_{\nu}^*}} {\tilde{\mathcal{ W}}_{\mathcal{M}}^\dag } , \label{appeq:GenAntiUniGamma}
\end{align}
\end{subequations}
where ${\tilde{\mathcal{ W}}_{\mathcal{M}} } = {\tilde{\mathcal{U}}^\dag_{\mathcal{M}}}{ { \sigma}_{N,x}}{\tilde{\mathcal{U}}_{\mathcal{M}}^*}$ should again be symmetric and unitary for the unitary equivalence between ${{\mathcal{M}_{ N}} } $ and any $\mathcal{PT}$-symmetric Hamiltonian matrix $\HHPT$ to exist. To proceed and obtain necessary and sufficient conditions for the existence of such unitary equivalence, we now assume that ${D_{\nu}}$ is positive definite, so that Eq.~\eqref{appeq:GenAntiUniGamma} requires that ${\tilde{\mathcal{ W}}_{\mathcal{M}} }$ must take the form
\begin{equation}
{\tilde{\mathcal{ W}}_{\mathcal{M}} }= \left( {\begin{array}{*{20}{c}}
0&{{w_{12}}}\\
{w_{12}^T}&0
\end{array}} \right) ,\end{equation}
where the off-diagonal blocks must be unitary and commute with the diagonal matrix $\left[ {{w_{12}},{D_{\nu}}} \right] = 0$. We note that the criterion ${\tilde{\mathcal{ W}}_{\mathcal{M}} } = {\tilde{\mathcal{U}}^\dag_{\mathcal{M}}}{ { \sigma}_{N,x}}{\tilde{\mathcal{U}}_{\mathcal{M}}^*}$ is automatically satisfied as
\begin{equation}
{\tilde{\mathcal{ W}}_{\mathcal{M}} }  = \left( {\begin{array}{*{20}{c}}
{\mathbb{I}_N}&0\\
0&{w_{12}^T}
\end{array}} \right)\left( {\begin{array}{*{20}{c}}
0&{\mathbb{I}_N}\\
{\mathbb{I}_N}&0
\end{array}} \right)\left( {\begin{array}{*{20}{c}}
{\mathbb{I}_N}&0\\
0&{{w_{12}}}
\end{array}} \right).\end{equation}
Substituting the form of ${\tilde{\mathcal{ W}}_{\mathcal{M}} }$ into Eq.~\eqref{appeq:GenAntiUniH1}, we obtain
 \begin{equation}
\left( {\begin{array}{*{20}{c}}
\Delta &\Sigma \\
\Sigma &\Delta 
\end{array}} \right) \left( {\begin{array}{*{20}{c}}
0&{{w_{12}}}\\
{w_{12}^T}&0
\end{array}} \right)=\left( {\begin{array}{*{20}{c}}
0&{{w_{12}}}\\
{w_{12}^T}&0
\end{array}} \right) \left( {\begin{array}{*{20}{c}}
{{\Delta ^*}}&{{\Sigma ^*}}\\
{{\Sigma ^*}}&{{\Delta ^*}}
\end{array}} \right) .\end{equation}
We thus obtain a set of sufficient conditions for a given dynamical matrix ${{\mathcal{M}_{ N}} } $ to be unitarily equivalent to a $\mathcal{PT}$-symmetric Hamiltonian $\HHPT$, which is the existence of a unitary matrix ${w_{12}}$ such that
\begin{subequations}
\begin{eqnarray}
{w_{12}}\Sigma  &= & {\Sigma ^*} w_{12}^T  ,\\
{w_{12}}\Delta  &= & {\Delta ^*}{w_{12}}  ,
\end{eqnarray}
\end{subequations}
which also commutes with the diagonal matrix $\left[ {{w_{12}},{D_{\nu}}} \right] = 0$. For a positive definite diagonal matrix ${D_{\nu}}$ corresponding to the parametric drivings, the conditions above will also be necessary conditions.

\section{Examples of four-mode $\mathcal{PT}$ and PA models where the correspondence fails}
\label{appsec:MapCEx}
In this section, we will present a four-mode $\mathcal{PT}$-symmetric non-Hermitian Hamiltonian ${{\hat {\HH}}_{\mathcal{PT},4}}$, for which there does not exist any unitarily equivalent Hermitian parametric Hamiltonian of four (or less) modes. According to results in Appendix~\ref{appsec:PTCanonicalForm}, it suffices to only consider $\mathcal{PT}$-symmetric Hamiltonians in the canonical form of $\HH_2$ in Eq.~\eqref{eq:CanonicalPTHamiltonian}. We start with a tight-binding four-mode Hamiltonian in the canonical form
\begin{align}{{\hat {\HH}}_{\mathrm{tb},4}} &= i \frac{\gamma}{2} \left( {\hat a_1^\dag {{\hat a}_1} + \hat a_2^\dag {{\hat a}_2} - \hat b_1^\dag {{\hat b}_1} - \hat b_2^\dag {{\hat b}_2}} \right)  \nonumber\\
&+ g\left( {\hat a_1^\dag {{\hat a}_2} + \hat b_1^\dag {{\hat b}_2} + \hat a_1^\dag {{\hat b}_1} + \hat a_2^\dag {{\hat b}_2} + h.c.} \right), \end{align} 
where the coefficients $\gamma$ and $g$ are both real, so that it takes the form of $\HHtb$ in Eq.~\eqref{eqPTmulc} the main text. Now we add a Hermitian perturbation ${{\hat V}_1}$ as
\begin{equation}
{{\hat V}_1} = \delta \left( {i\hat a_2^\dag {{\hat b}_2} + h.c.} \right),  
\end{equation} 
so that the total Hamiltonian is now given by ${{\hat {\HH}}'_{\mathrm{tb},4}}= {{\hat {\HH}}_{\mathrm{tb},4}}+ {{\hat V}_1}$ . We note that the perturbation $\left( {\delta  > 0} \right)$ breaks time reversal symmetry of the coherent part in the Hamiltonian ${{\hat {\HH}}'_{\mathrm{tb},4}}$, which still takes the canonical form in Eq.~\eqref{eq:CanonicalPTHamiltonian} with component coefficient $2 \times 2$ matrices now given by
\begin{subequations}
\begin{align}
&\tilde \Sigma = g \sigma_x,\\
&{\tilde J} =  \left(  t + \frac{i\delta}{2}\right)\sigma_0-  \frac{i\delta}{2} \sigma_z,\\
&\Gamma_N = \frac{\gamma}{2} \sigma_0,
\end{align}
\end{subequations}
for which we can check the conditions in Eqs.~\eqref{appeq:GenPTtoPAthm} algebraically. It is straightforward but tedious to check these conditins for all 
possible $2 \times 2$ unitary matrices ${w_{12}}$.  We find that it is impossible to construct a unitary ${w_{12}}$ such that all the conditions are satisfied. Thus, we have constructed a four-mode $\mathcal{PT}$-symmetric to which there does not exist any unitarily equivalent PA system
having four or less modes.

Conversely, there also exists parametric model whose dynamical matrix is not unitarily equivalent to any $\mathcal{PT}$-symmetric system with equal number of modes. If we have a parametric model that (up to a local gauge transformation) fits the form in Eq.~\eqref{eqpamulR}
\begin{equation}{{\hat H}_{\mathrm{p.a.},4}} = g \left( {\hat a_1^\dag {{\hat a}_2} - \hat b_1^\dag {{\hat b}_2}} \right) + {\nu _1}\hat a_1^\dag \hat b_1^\dag  + {\nu _2}\hat a_2^\dag \hat b_2^\dag  + h.c.,  
\end{equation} 
where $g$ and $\nu_{1} \ne \nu_{2}$ are real positive parameters, then the correspondence can be broken by adding a Hermitian perturbation as beam-splitter interactions with completely imaginary phase ${\hat V}_2=\delta \left( {i\hat a_1^\dag {{\hat a}_2} - i\hat b_1^\dag {{\hat b}_2}  + h.c.} \right)$, where we assume ${\delta  > 0} $ without lack of generality. We note that the perturbation introduces a nontrivial phase into the parametric Hamiltonian, and the dynamics of the system can still be described by a $4 \times 4$ non-Hermitian dynamical matrix, which is automatically in the form in Eq.~\eqref{eq:CanonicalPADM}. The corresponding canonical form in Eq.~\eqref{eq:CanonicalPADM} now has coefficient $2 \times 2$ matrices as
\begin{subequations}
\begin{align}
& \Delta = g \sigma_x,\\
& \Sigma = - \delta \sigma_y,\\
& D_\nu = \frac{{\nu _1}+{\nu _2}}{2} \sigma_0+ \frac{{\nu _1}-{\nu _2}}{2} \sigma_z ,
\end{align}
\end{subequations}
but now the dynamics is not unitarily equivalent to any four-mode $\mathcal{PT}$-symmetric Hamiltonian.

\section{Higher-order exceptional point in PA systems}
\label{app:HigherOrderEP}

In Sec.~\ref{EPmodesplit} of the main text, we have presented EP enhanced mode splitting based on the $\sqrt{\epsilon}$ scaling of the splitting of eigenvalues in a $\mathcal{PT}$-symmetric non-Hermitian system. With some minor twists on the multimode mapping in Sec.~\ref{subsec:GeneralizedPTChain}, unitary mappings from $\mathcal{PT}$-symmetric system with odd number of modes to non-degenerate parametric amplifiers with equal number of modes could also be constructed. The idea is to leave the single $\mathcal{PT}$-symmetric mode unchanged, perform mapping for the rest of the modes as before, and assign coherent, particle-number conserving interaction terms to realize dynamics of the remaining bosonic mode. As proposed in Refs.~\cite{Demange2011,Lin2016,Jing2017}, such NDPA could exhibit higher order exceptional point, with mode splitting scaling as $ {\epsilon^{1/3}}$.

We first describe the basics ingredients of higher order exceptional point and the corresponding enhanced mode splitting in a $\mathcal{PT}$-symmetric three-mode system. Although the higher order exceptional points can also be found in systems with more number of modes, we focus on the $\mathcal{PT}$ trimer setup for demonstrating purpose.  The sensing scheme now consists of an unperturbed three-mode system with the Hamiltonian given by
\begin{equation} 
\mathcal{H}_\mathrm{HOEP} \left[ 0 \right] = \left( {\begin{array}{*{20}{cccc}}
{+i\frac{\gamma }{2}}&g&0\\
g&0&g\\
0&g&{ - i\frac{\gamma }{2}}
\end{array}} \right),
\end{equation} 
 and we intend to estimate the small parameter $ {\epsilon}$ by probing the output power spectrum of the perturbed Hamiltonian
 \begin{equation} 
\mathcal{H} \left[ \epsilon \right]=\left( {\begin{array}{*{20}{cccc}}
{+i\frac{\gamma }{2}}&g&0\\
g&{\epsilon}&g\\
0&g&{ - i\frac{\gamma }{2}}
\end{array}} \right). 
\end{equation}
If the unperturbed Hamiltonian $\mathcal{H} \left[ 0 \right]$ is set to the third-order EP $(g_c=\sqrt{2}\gamma/4)$, then the power spectrum has a single resonance peak. In this case, the small perturbation $\epsilon$ in the mode detuning will induce mode splitting in the output spectrum, which scales as $ {\epsilon^{1/3}}$ and may be even more sensitive than the splitting scaled as $\sqrt{\epsilon}$ in $\mathcal{PT}$ dimer settings. The $ {\epsilon^{1/3}}$ scaling of mode splitting with respect to small $\epsilon$ perturbations close to the third-order exceptional point has recently been verified in optical experiments~\cite{Hodaei2017}.

Again we aim to achieve the same $ {\epsilon^{1/3}}$ scaling of mode splitting in a PA setup, without having to introduce any external bath (i.e., noise source). Applying the unitary mapping in Sec.~\ref{sec:DPAMapping} to the two-mode subsystem with gain and loss, the corresponding $\epsilon$-dependent NDPA Hamiltonian can be obtained as
\begin{equation}
{{\hat H}_\mathrm{NDPA, 3 }} \left[ \epsilon \right] =   \epsilon {{\hat b}^\dag }\hat b  + \left(\sqrt{2}g {{\hat a}^\dag_2 } \hat b+ i\frac{\nu }{2}    {{\hat a}^{\dag }_1 }{{\hat a}^{\dag }_2}  + h.c. \right)  ,
\end{equation}
where the detuning term $ \epsilon$ becomes the detuning term of the third bosonic mode $b$, and the gain and loss at rate $\gamma$ are transformed into the parametric drive with strength $\nu =\gamma $.
Note that this bosonic Hamiltonian has the general structure of a driven three-mode optomechanical system, where a mechanical resonator $\hat{a}_2$ interacts with two electromagnetic modes $\hat{b}, \hat{a}_1$ via radiation-pressure interactions.  This setup has been previously studied for entanglement generation \cite{Yingdan2013,Tian2013,Yingdan2015,Liang2019}.

It is also interesting to consider the form of the Hamiltonian when $\epsilon= 0$, $g = \sqrt{2} \gamma / 4$ and we are exactly tuned to the EP.  As discussed in Sec.~\ref{sganaSEC}, EP's in bosonic Hamiltonians coincide with conserved Hermitian quadrature variables.  The same is true in our system.  Making the gauge change $\hat{b} \rightarrow \hat{b}' =-i \hat{b}$, the Hamiltonian at the exceptional point can be written
\begin{equation}
   {{\hat H}_\mathrm{NDPA, 3 }} = \frac{\gamma }{{2}} \left( \hat{X}_2 (\hat{P}_1 - \hat{P}_{b'} ) + \hat{P}_2 (\hat{X}_1 + \hat{X}_{b'} ) \right) ,
\end{equation}
where we have introduced standard quadrature operators for each mode.  It follows immediately that there are two conserved quadratures $\hat{Q}_\pm$ in this system:
\begin{align}
    \hat{Q}_{-} &=
        \frac{1}{\sqrt{2}} \left( \hat{P}_1 - \hat{P}_{b'} \right), \\
    \hat{Q}_+ &=
        \frac{1}{\sqrt{2}} \left( \hat{X}_1 + \hat{X}_{b'} \right). 
\end{align}


\section{QMFS for multi-mode systems: construction and corresponding symplectic transformations}
\label{appsec:QMFSMulti}

In Sec.~\ref{sec:MultiModeQMFS}, we discussed how the dynamics of an arbitrary $N$ mode non-Hermitian Hamiltonian could be realized using a QMFS embedded in a $2N$ mode bosonic system.  We provide more details here as to how one constructs such multi-mode QMFS, and also discuss properties of the corresponding symplectic transformation generated by this dynamics.  


\subsection{Constraints on a general QMFS}

Consider an $N$ mode linear and Hermitian bosonic system where the dynamics does not couple canonically conjugate quadratures.  We can write the equations of motion as
\begin{subequations}\label{EqAppsymgenuv}
\begin{align}
{\partial _t }{{\vec {\hat q}}}\left(  t  \right) = \mathcal{A}\left(  t  \right){{\vec {\hat q}}}\left( t \right),\\
{\partial _t }{{\vec {\hat \pi}}}\left(  t  \right) = \mathcal{B}\left(  t  \right){{\vec {\hat \pi}}}\left(  t  \right),
\end{align}
\end{subequations}
where $\mathcal{A}\left( t  \right)$ and $\mathcal{B}\left( t  \right)$ are generic real dynamical matrices, and ${{\vec {\hat q}}}$ and ${{\vec {\hat \pi}}}$ are both column vectors formed by $N$ quadrature operatures satisfying the canonical commutation relations as
\begin{equation}\label{AppEquvcc}
\left[ {{{\hat q}_j},{{\hat \pi}_{j'}}} \right] = i{\delta _{jj'}},
\end{equation}
with all other commutators between the quadratures vanishing $\left[ {{{\hat q}_j},{{\hat q}_{j'}}} \right] =\left[ {{{\hat \pi}_j},{{\hat \pi}_{j'}}} \right] =0$. 


We require the dynamics to preserve the canonical commutation relations at all times.  It is straightforward to show that a necessary and sufficient condition to ensure this is that at all times:
\begin{equation}
      {\mathcal{B}\left(   t  \right) } = - \mathcal{A}^T\left(  t  \right).
      \label{eq:QMFSABConstraint}
\end{equation}
for all time $ t$. 

Similarly, if integration of the equations of motion yields 
\begin{subequations}\label{}
\begin{align}
{{\vec {\hat q}}}\left(  t  \right) &= {\mathcal{U}_\mathcal{A}}\left(  t  \right){{\vec {\hat q}}}\left( 0 \right),\\
{{\vec {\hat \pi}}}\left(  t  \right) &= {\mathcal{U}_\mathcal{B}}\left(  t  \right){{\vec {\hat \pi}}}\left( 0 \right),
\end{align}
\end{subequations}
then the preservation of canonical commutation relations holds if and only if
\begin{equation}\label{AppEqSyevocr}
 {\mathcal{U}_\mathcal{A}}\left(  t  \right)\mathcal{U}_\mathcal{B}^T\left(  t  \right) = {\mathbb{I}_N}.
\end{equation}

\subsection{QMFS for realizing arbitrary multi-mode non-Hermtian dynamics}
\label{appsec:QMFSgen}

We can use the results above to verify the QMFS dynamics presented in Sec.~\ref{sec:MultiModeQMFS} does indeed correspond to a Hermitian bosonic Hamiltonian. in Sec.~\ref{sec:MultiModeQMFS}, an arbitrary non-Hermitian Hamiltonian $\mathcal{H}_N$ was encoded in a QMFS via Eq.~\eqref{eq:EOMPseudoMode}, i.e.
\begin{equation}
    i{\partial _t } {\vec{\hat{ z }}}_{+ } = \HH_N  {\vec{\hat{ z }}}_{+ },
\end{equation}
with pseudo-modes $\vec{z}_{\pm}$  defined 
in Eq.~\eqref{eq:PseudoModesMulti}. 

In terms of the column vectors formed by quadrature operators $ {{\vec {\hat x }}_{+}}$ and ${{\vec {\hat p} _- }} $, we have
\begin{equation}
\label{eq:EOMGenQMFSquad}
{\partial _t }\left( {\begin{array}{*{20}{c}}
  {{\vec {\hat x }}_+} \\ 
  {{\vec {\hat p} _- }} 
\end{array}} \right) = \left( {\begin{array}{*{20}{c}}
  {{\operatorname{Im} \HH_N}}&{\operatorname{Re} \HH_N} \\ 
  { -\operatorname{Re} \HH_N}&{{\operatorname{Im} \HH_N}} 
\end{array}} \right)\left( {\begin{array}{*{20}{c}}
  {{\vec {\hat x }}_+} \\ 
  {{\vec {\hat p} _- }} 
\end{array}} \right).
\end{equation}


Now, using Eq.~(\ref{eq:QMFSABConstraint}) to ensure conjugate quadratures evolve appropriately, 
we obtain
\begin{equation}
\label{eq:EOMGenQMFSquadConj}
{\partial _t }\left( {\begin{array}{*{20}{c}}
  {{\vec {\hat p }}_+} \\ 
  {-{\vec {\hat x} } _-} 
\end{array}} \right) = \left( {\begin{array}{*{20}{c}}
  {-{\operatorname{Im} \HH^T_N}}&{\operatorname{Re} \HH^T_N} \\ 
  { -\operatorname{Re} \HH^T_N}&{-{\operatorname{Im} \HH^T_N}} 
\end{array}} \right)\left( {\begin{array}{*{20}{c}}
 {{\vec {\hat p }}_+} \\ 
  {-{\vec {\hat x} } _-} 
\end{array}} \right).
\end{equation}
The above equation is equivalent to
\begin{equation}
i{\partial _t } {\vec{\hat{ z }}}_{- }  = \HH^\dag_N  {\vec{\hat{ z }}}_{- },
\end{equation}
as given in Sec.~\ref{sec:MultiModeQMFS}.  It thus follows that the QMFS dynamics given in Sec.~\ref{sec:MultiModeQMFS} does indeed preserve canonical commutation relations.  

The above approach is also valid for an  arbitrary time-dependent non-Hermitian dynamical matrix $ \HH_N \left(  t \right)$.  Note first that the classical amplitude evolution is controlled by the $N \times N$ complex matrix $\mathcal{U}_N \left(  t \right)$.  It satisfies:
\begin{equation} 
i{\partial _t }\mathcal{U}_N\left(   t  \right) = \HH_N \left( t \right)\mathcal{U}_N\left(  t  \right).
    \label{eq:SymplecticU}
\end{equation}
In terms of the quadratures ${{  {\hat x }}_{ \pm, j}}$ and ${  {\hat p} } _{ \pm, j}$, we have
\begin{subequations}
\label{appeq:SympTransformGen}
\begin{align}
\left( {\begin{array}{*{20}{c}}
  {{\vec {\hat x }}_+ \left(  t  \right)} \\ 
  {{\vec {\hat p} _- }\left( t  \right)} 
\end{array}} \right) &= \left( {\begin{array}{*{20}{c}}
  {{\operatorname{Re} \mathcal{U}_N}\left( t  \right)}&{-\operatorname{Im} \mathcal{U}_N \left( t  \right)} \\ 
  { \operatorname{Im} \mathcal{U}_N \left( t  \right)}&{{\operatorname{Re} \mathcal{U}_N}\left( t  \right)} 
\end{array}} \right)\left( {\begin{array}{*{20}{c}}
  {{\vec {\hat x }}_+ \left( 0  \right)} \\ 
  {{\vec {\hat p} _- }\left( 0  \right)} 
\end{array}} \right),\\
 \left( {\begin{array}{*{20}{c}}
  {{\vec {\hat p }}_+ \left( t  \right)} \\ 
  {-{\vec {\hat x} } _- \left( t  \right)} 
\end{array}} \right) &= \left( {\begin{array}{*{20}{c}}
  {\operatorname{Re} {\mathcal{V}_N}\left( t  \right)}&{  \operatorname{Im} {\mathcal{V}_N}\left( t \right)} \\ 
  {-\operatorname{Im}{\mathcal{V}_N}\left( t  \right)}&{\operatorname{Re} {\mathcal{V}_N}\left( t  \right)} 
\end{array}} \right)\left( {\begin{array}{*{20}{c}}
 {{\vec {\hat p }}_+ \left( 0 \right)} \\ 
  {-{\vec {\hat x} } _- \left( 0 \right)} 
\end{array}} \right),
\end{align}
\end{subequations}
where the coefficient matrix ${\mathcal{V}_N}\left( t  \right)$ is defined as 
\begin{equation}\label{appeq:scriptVGen}
{\mathcal{V}_N}\left( t  \right) \equiv \left[{{\mathcal{U} }_N^{T}\left( t  \right)}\right]^{-1}.
\end{equation}
It follows immediately that the constraint in  Eq.~\eqref{AppEqSyevocr} is obeyed.

Finally, the above transformation can be equivalently described in terms of the bosonic annihilation and creation operators as
\begin{equation}\label{appeq:SympTFMultiMat}
 \left( {\begin{array}{*{20}{c}}
  { {\vec {\hat a }} }\left(  t  \right) \\ 
  { {\vec {\hat b }} }\left(  t  \right)
\end{array}} \right) = A \left(  t  \right) \cdot \left( {\begin{array}{*{20}{c}}
  { {\vec {\hat a }} }\left( 0  \right) \\ 
  { {\vec {\hat b }} }\left( 0  \right)
\end{array}} \right)+B \left(  t  \right) \cdot \left( {\begin{array}{*{20}{c}}
  { {\vec {\hat {a } }^\dag} }\left( 0 \right) \\ 
  { {\vec {\hat {b } }^\dag}}\left( 0 \right)
\end{array}} \right),
\end{equation}
where
\begin{subequations}\label{appeq:SympTFMulti}
\begin{align}
&A \left(  t  \right)={ \frac{1}{2}} \left( {\begin{array}{*{20}{c}}
{ {{  \mathcal{U}_N}\left(  t  \right)+\left[{{\mathcal{U} }_N^{\dag}\left( t \right)}\right]^{-1}} }&0\\
0&{ {{  \mathcal{U}^*_N}\left(  t  \right)+\left[{{\mathcal{U} }_N^{T}\left( t \right)}\right]^{-1}}  }
\end{array}} \right),\\
&B \left(  t  \right)= { \frac{1}{2}} \left( {\begin{array}{*{20}{c}}
0&{ {{  \mathcal{U}_N}\left(  t  \right)-\left[{{\mathcal{U} }_N^{\dag}\left( t \right)}\right]^{-1}}  }\\
{ {{  \mathcal{U}^*_N}\left( t  \right)-\left[{{\mathcal{U} }_N^{T}\left( t  \right)}\right]^{-1}} }&0
\end{array}} \right).
\end{align}
\end{subequations}

\subsection{Bloch-Messiah representation of the multi-mode QMFS evolution}
\label{appsec:BMRSympGen}

As discussed above, the non-Hermitian Hamiltonian $\HH_N(t)$ induces a symplectic (i.e.~unitary) transformation in the corresponding bosonic QMFS system.  To understand its nature better, it is helpful to use the Bloch-Messiah (BM) decomposition \cite{Braunstein2005}.  This reduces an arbitrary Gaussian unitary to a sequence of three simple operations: a beam-splitter operation, a product of single-mode squeezing operations, then another beam-splitter operation. 

In terms of Eq.~(\ref{appeq:SympTFMultiMat}), the Bloch-Messiah decomposition corresponds to
\begin{subequations}
\begin{align}
&A =  U_{\mathrm{BM}} {D_A}{{ V_{\mathrm{BM}}^\dag} },\\
&B =   U_{\mathrm{BM}} {D_B}{{ V_{\mathrm{BM}}^T}} ,
\end{align}
\label{eqs:BM}
\end{subequations}
where $U_{\mathrm{BM}}$ and $V_{\mathrm{BM}}$ are unitary, and ${D_A},{D_B}$ are non-negative diagonal matrices with the constraint ${D_A^2}={D^2_B}+\mathbb{I}_{2N}$.

The Bloch-Messiah matrices can be explicitly computed.  We first write a singular value decomposition for the time evolution matrix 
${  \mathcal{U}_N}\left( t \right)$ in Eq.~(\ref{eq:SymplecticU}):
\begin{equation}
    {  \mathcal{U}_N}\left(  t  \right)= {W_1}D_{\mathcal{U}}W_2^\dag ,
\end{equation}
and then use this to define the diagonal unitary matrix $W_{\mathcal{U}}$:
\begin{equation}
    W_{\mathcal{U}} ={ {  \sqrt{\left({D_{\mathcal{U}} -D^{-1}_{\mathcal{U}} }\right)\cdot {{\left| {D_{\mathcal{U}} -D^{-1}_{\mathcal{U}} } \right|}^{-1}}} } }.
\end{equation}

With these definitions, the unitary matrices in the BM decomposition (describing initial and final beam-splitter operations) are given by:
\begin{subequations}\label{appeq:BMRQMFSGen}
    \begin{align}
    U_{\mathrm{BM}}&=\left( {\begin{array}{*{20}{c}}
    {{W_1}}&0\\
    0&{ {W^*_1}  }
    \end{array}} \right)\cdot  \frac{1}{\sqrt{2}}\left( \begin{matrix}
   1 & -1  \\
   1 & 1  \\
\end{matrix} \right) \cdot \left( {\begin{array}{*{20}{c}}
{W_{\mathcal{U}} }&0\\
0&{ W^*_{\mathcal{U}}}
\end{array}} \right) ,\\
{ V_{\mathrm{BM}}^\dag} &=\left( {\begin{array}{*{20}{c}}
{W^*_{\mathcal{U}} }&0\\
0&{ W_{\mathcal{U}}}
\end{array}} \right) \cdot \frac{1}{\sqrt{2}}\left( \begin{matrix}
   1 & 1  \\
   -1 & 1  \\
\end{matrix} \right) \cdot \left( {\begin{array}{*{20}{c}}
{ W_2^\dag }&0\\
0&{ W_2^{T}  }
\end{array}} \right).
\end{align}
\end{subequations}
Correspondingly, the diagonal matrices describing the squeezing operations in the BM decomposition can be computed as:
\begin{subequations}
\label{appeq:QMFSBMRdiag}
\begin{align}
    {D_A}&= \left( {\begin{array}{*{20}{c}}
{ { \frac{D_{\mathcal{U}} +D^{-1}_{\mathcal{U}} }{2}} }&0\\
0&{  { \frac{D_{\mathcal{U}} +D^{-1}_{\mathcal{U}} }{2}} }
\end{array}} \right),\\
    {D_B}&=\left( {\begin{array}{*{20}{c}}
{ { \frac{{\left| {D^{-1}_{\mathcal{U}} -D_{\mathcal{U}} } \right|} }{2}} }&0\\
0&{  { \frac{{\left| {D^{-1}_{\mathcal{U}} -D_{\mathcal{U}} } \right|}}{2}} }
\end{array}} \right).
\end{align}
\end{subequations}


\section{Review of the EP encircling}
\label{appsec:RevQAD}
In this section, we briefly review the quasi-adiabatic dynamical phenomena by encircling an EP in two-mode non-Hermitian systems. We use the convention in Ref.~\cite{Milburn2015} for clarity. We consider a system evolving according to the time-dependent non-Hermitian Hamiltonian ${\mathcal{H}}_\omega \left( t \right)$
given in Eq.~(\ref{eq:Homegat})
whose instantaneous eigenvalues 
are given by 
\begin{equation}{\lambda _ \pm(t) }= \pm  \sqrt{{{{\left( {\omega(t)  + i {\gamma }/{2} } \right)}^2} + {g(t)^2}}} .\end{equation}
We assume that the parameters are varied at a rate much slower than the  eigenvalue gap $\left| {{\lambda _ + }-{\lambda _ -}} \right|$. 
We choose the left (right) instantaneous eigenvectors $\vec{l}_\pm(t)$ ($\vec{r}_\pm(t)$) of $\mathcal{H}_{\omega}(t)$ to be biorthonormal and satisfy ${\vec l_ \pm } ={\vec r_ \pm } $, implying:
\begin{subequations}
\label{appeq:QAD2mevs}
\begin{align}
{\vec r_ \pm }  &= {\left(  {1+\rho _{\pm }^{2}} \right)}^{-\frac{1}{2}} \cdot \left( {\begin{array}{*{20}{c}}
{\rho_\pm}\\
1
\end{array}} \right),\\
{\rho_\pm}&=\frac{\omega  + i {\gamma }/{2} + {\lambda _ \pm } }{g}.
\end{align}
\end{subequations}

We consider varying parameters $g(t)$ and $\omega(t)$ along a circle that encloses the EP:
\begin{subequations}
\label{eqQADparam}
\begin{align}
&g\left(t \right)= g_0+\epsilon \cos \phi\left( t \right),\\
 &\omega\left( t \right) = \epsilon \sin \phi\left( t \right),
\end{align}
\end{subequations}
where the center of the circle is taken to be the exceptional point $g_0 =  \gamma /2$, and $\epsilon <\gamma/2$ is a small positive parameter characterizing the encircling radius. The circling phase is chosen such that EP is encircled once, with $\phi \left( {{t_i }} \right) =\phi \left( {{t_f }} \right) $ during the time duration that we consider, and the evolution time is chosen such that $t_f - t_i =T \gg 1/\left| {{\lambda _ + }-{\lambda _ -}} \right|$.

The system is prepared in one of the instantaneous eigenmodes (e.g. the ${\lambda _ + }$ branch) ${{\vec z}  \left( { t_i} \right)}={{\vec r}_+   \left( { t_i} \right)}$ at the beginning time $t_i$, and we consider solution to the equation of motion $i{\partial _t }{\vec z} ={\mathcal{H}_\omega}{\vec z}$. For the Hamiltonian matrix ${\mathcal{H}_{\omega}} \left( { t } \right)$ at given time $t$, we expand the state vector ${\vec z} \left( { t } \right)$ at time $t$ in terms of instantaneous eigenvectors as
\begin{equation}
\label{appeq:QADeigmod}
{\vec z} \left( { t } \right)={c_+} \left( { t } \right) {\vec r_ + } \left( { t } \right)+ {c_-}  \left( { t } \right) {\vec r _-} \left( { t } \right).
\end{equation}

If the dynamics were adiabatic, we would expect the amplitude ${c_-}  \left( { t } \right) $ to be negligible for the entire protocol. Note that as the EP is encircled once during the time evolution, the instantaneous eigenmodes undergo a switch by the end of the evolution, i.e.~$\lambda_\pm \left( { t_f } \right)=\lambda_\mp \left( { t_i } \right)$. Surprisingly, for the path that encircles the EP once, the adiabatic prediction holds for parametric encircling path along only one direction, whereas a non-adiabatic transition will occur in the opposite direction, as depicted in Fig.~\ref{fig:QADfref} in the main text. This chiral mode switching phenomenon could be interpreted physically as a consequence of stability loss delay~\cite{Milburn2015}.

\section{Exceptional point encircling: details}
\label{app:EPEncirclingDetails}

We present here additional details for the quasi-adiabatic exceptional point encircling dynamics presented in Sec.~\ref{subsec:EPEncircle}; the focus is on how the dynamics of the non-Hermitian gain-loss dimer Hamiltonian $\mathcal{H}_{\omega}(t)$ (c.f.~Eq.~(\ref{eq:Homegat})) directly determines the quantum evolution in our Hermitian, four-mode bosonic system (Hamiltonian $\hat{H}_{\omega \textrm{PA}}(t)$, c.f.~Eq.~(\ref{eqnH2mpa})).  We will make explicit use of the Bloch-Messiah (BM) reduction of the symplectic transformation generated by $\mathcal{H}_{\omega}(t)$ (as introduced in App.~\ref{appsec:BMRSympGen}).

\subsection{Symmetry constraints}
\label{appsec:SymmUomega}
Note first that, by construction, ${\mathcal{H}}_\omega(t)$ is a symmetric matrix, and satisfies the chiral symmetry condition $\left\{ {\HH_{\omega} \left( t\right),{\sigma _y}} \right\} = 0$.  As a result, the amplitude-evolution matrix $\mathcal{U}_{\omega}\left( t  \right)$ generated by $\mathcal{H}_\omega(t)$ 
(c.f.~Eq.~(\ref{eq:UomegaDefined})) obeys the constraint
\begin{equation}
    \label{appeq:UomegaCond}
    {\left[ {{\mathcal{U}^T_{\omega} \left( t  \right)}} \right]^{ - 1}} = {\sigma _y}\mathcal{U}_{\omega} \left( t  \right){\sigma _y}.
\end{equation}
It also follows that $\det\mathcal{U}_{\omega} \left( t  \right)=1$.

These conditions can be used to constrain the form of the unitary operator ${\hat U}_{\omega}\left( t  \right)$ which describes evolution in the four-mode Hermitian bosonic system corresponding to $\mathcal{H}_\omega(t)$ (constructed using the QMFS approach).
One finds that the diagonal matrices ${D_A}\left( t  \right)$ and ${D_B} \left( t  \right)$ in the BM reduction of ${\hat U}_{\omega}\left( t  \right)$ (c.f.~Eqs.~(\ref{appeq:SympTFMultiMat}) and (\ref{eqs:BM})) 
are both proportional to the unit matrix, i.e.
\begin{subequations}
\begin{align}
 {D_A}\left( t  \right)  &=\cosh {{\lambda }_{s} \left( t  \right)} \cdot  \mathbb{I}_4,\\
 {D_B}\left( t  \right) &= \sinh {{\lambda }_{s} \left( t  \right)} \cdot \mathbb{I}_4,
\end{align}
\end{subequations}
with the squeezing parameter $\lambda_s(t)$ given by
\begin{equation}
\cosh {{\lambda }_{s}\left( t  \right)} = \frac{1}{2} {\sqrt{\mathrm{tr} \left[{  \mathcal{U}^\dag _{\omega}}\left(  t  \right) { \mathcal{U}_{\omega}}\left(  t  \right)\right]+2 }}  .
\end{equation}
It follows that the squeezing part of the BM decomposition necessarily corresponds to four identical single-mode squeezing operations.

\subsection{Symplectic transformation for the four-mode QMFS Hermitian bosonic system}

\label{SecappQMFS2m}
We now derive the explicit form of the symplectic transformation for the collective quadratures ${{  {\hat x }}_{ \pm, j}}$ and ${  {\hat q} } _{ \pm, j}$ ($j =1,2$) under the time evolution generated by the time-dependent parametric Hamiltonian ${{\hat H}_{\omega \mathrm{PA}} \left( t  \right)}$ (c.f.~Eq.~(\ref{eqnH2mpa})). 
Using Eq.~\eqref{appeq:SympTransformGen}, we immediately have
\begin{subequations}\label{appeq:SympomegaGen}
\begin{align}
\left( {\begin{array}{*{20}{c}}
  {{{\hat x}_{1, + }}\left( t  \right)} \\ 
  {{{\hat x}_{2, + }}\left( t  \right)} \\ 
  {{{\hat p}_{1, - }}\left( t  \right)} \\ 
  {{{\hat p}_{2, - }}\left( t  \right)} 
\end{array}} \right) &= \left( {\begin{array}{*{20}{c}}
  {\operatorname{Re} \mathcal{U}_{\omega} \left( t  \right)}&{ - \operatorname{Im} \mathcal{U}_{\omega} \left( t  \right) } \\ 
  {\operatorname{Im} \mathcal{U}_{\omega} \left( t  \right)}&{\operatorname{Re} \mathcal{U}_{\omega} \left( t  \right)} 
\end{array}} \right)\left( {\begin{array}{*{20}{c}}
  {{{\hat x}_{1, + }}\left( 0 \right)} \\ 
  {{{\hat x}_{2, + }}\left( 0 \right)} \\ 
  {{{\hat p}_{1, - }}\left( 0 \right)} \\ 
  {{{\hat p}_{2, - }}\left( 0 \right)} 
\end{array}} \right),\\
\left( {\begin{array}{*{20}{c}}
  {{{\hat p}_{1, + }}\left( t  \right)} \\ 
  {{{\hat p}_{2, + }}\left( t  \right)} \\ 
  {-{{\hat x}_{1, - }}\left( t  \right)} \\ 
  {-{{\hat x}_{2, - }}\left( t  \right)} 
\end{array}} \right) &= \left( {\begin{array}{*{20}{c}}
  {\operatorname{Re} {\mathcal{V}_{\omega} }\left( t  \right)}&{  \operatorname{Im} {\mathcal{V}_{\omega} }\left( t  \right)} \\ 
  {-\operatorname{Im} {\mathcal{V}_{\omega} }\left( t  \right)}&{\operatorname{Re} {\mathcal{V}_{\omega} }\left( t  \right)} 
\end{array}} \right)\left( {\begin{array}{*{20}{c}}
  {{{\hat p}_{1, + }}\left( 0 \right)} \\ 
  {{{\hat p}_{2, + }}\left( 0 \right)} \\ 
  {-{{\hat x}_{1, - }}\left( 0 \right)} \\ 
  {-{{\hat x}_{2, - }}\left( 0 \right)} 
\end{array}} \right),\label{appeq:SympomegaGen2}
\end{align}
\end{subequations}
where  ${\mathcal{V}_{\omega}}\left( t  \right) \equiv {\left[ {{\mathcal{U}^T_{\omega} \left( t  \right)}} \right]^{ - 1}}$ (see Eq.~\eqref{appeq:scriptVGen}). 

As derived in Eq.~\eqref{appeq:UomegaCond}, the chiral symmetry of $\HH_{\omega} \left( t\right)$ ensures the equality ${\mathcal{V}_{\omega}}\left( t  \right)  = {\sigma _y}\mathcal{U}_{\omega} \left( t  \right){\sigma _y}$, so that the equation of motion Eq.~\eqref{appeq:SympomegaGen2} can be equivalently rewritten in terms of the matrix $ \mathcal{U}_{\omega} \left( t  \right)$ as
\begin{equation}
\left( {\begin{array}{*{20}{c}}
  {{{\hat x}_{2, - }}\left(  t  \right)} \\ 
  {{-{\hat x}_{1, - }}\left(  t  \right)} \\ 
  {{{\hat p}_{2, + }}\left(  t  \right)} \\ 
  {{-{\hat p}_{1, + }}\left(  t  \right)} 
\end{array}} \right) = \left( {\begin{array}{*{20}{c}}
  {\operatorname{Re} \mathcal{U}_{\omega} \left( t  \right)}&{  \operatorname{Im} \mathcal{U}_{\omega} \left( t  \right)} \\ 
  {-\operatorname{Im} \mathcal{U}_{\omega} \left( t  \right)}&{\operatorname{Re} \mathcal{U}_{\omega} \left( t  \right)} 
\end{array}} \right)\left( {\begin{array}{*{20}{c}}
  {{{\hat x}_{2, - }}\left( 0  \right)} \\ 
  {{-{\hat x}_{1, - }}\left( 0 \right)} \\ 
  {{{\hat p}_{2, + }}\left( 0 \right)} \\ 
  {{-{\hat p}_{1, + }}\left( 0  \right)}  
\end{array}} \right).
\end{equation}

Note that the symplectic transform presented here in terms of the collective quadratures $ {\hat x_{  \pm,j} }$ and ${\hat p_{  \pm,j} }$ is equivalent to the one discussed in Eq.~\eqref{eq:BrokenEOMSolu} in the main text, which can be compactly written using the pseudo-modes $\hat{  {z}}_{ j}$ and $\hat{ \tilde z}_{ j}$
(c.f.~Eqs.~\eqref{eq:PseudoMode2m} and \eqref{eq:PseudoModeConj2m}) as
\begin{subequations}\label{appeq:SympomegaPM}
\begin{align}
        \left( \begin{array}{c}
            \hat{z}_1(t) \\
            \hat{z}_2(t)
            \end{array}
        \right) &= 
   \mathcal{U}_{\omega}\left( t  \right) \cdot
            \left( \begin{array}{c}
            \hat{z}_1(0) \\
            \hat{z}_2(0)
                    \end{array}
        \right),\\
        \left( \begin{array}{c}
            \hat{ \tilde z}_1( t ) \\
            \hat{ \tilde z}_2( t )
            \end{array}
        \right) &= 
    {\left[ {{\mathcal{U}^\dag_{\omega} \left( t  \right)}} \right]^{ - 1}} \cdot
            \left( \begin{array}{c}
            \hat{ \tilde z}_1(0) \\
            \hat{ \tilde z}_2(0)
                    \end{array}
        \right).
        \label{appeq:SympomegaPM2}
\end{align}
\end{subequations}

\subsection{Comparing symplectic transformations for clockwise and counterclockwise encirclings}
\label{appsec:BMRUomega}

Consider a general case of a multimode Hermitian bosonic system which corresponds (via the QMFS mapping) to a time-dependent non-Hermitian Hamiltonian.  We take this latter Hamiltonian to be symmetric and periodic (period $T_0$), i.e.
\begin{equation}\label{appeq:nHMatSymmPer}
    \HH_N \left( t \right)=\HH^T_N \left( t \right)=\HH_N \left( t +T_0 \right). 
\end{equation}

As discussed in Appendix~\ref{appsec:QMFSMulti}, the corresponding symplectic transformations in the QMFS setup are fully characterized by the non-unitary time-evolution matrices $\mathcal{U}_N\left(   T_0  \right)$ and $ {\mathcal{\tilde  U}}_N\left(   T_0  \right)$, which are $t=T_0$ solutions to the equations of motion
\begin{subequations}
\begin{align}
 i{\partial _t }\mathcal{U}_N\left(   t  \right) &= \HH_N \left( t \right)\mathcal{U}_N\left(  t  \right),\\
 i{\partial _t } {\mathcal{\tilde  U}} _N\left(   t  \right) &= \HH_N \left( -t \right) {\mathcal{\tilde  U}}_N\left(  t  \right).
\end{align}
\end{subequations}

As (by assumption) $\HH_N \left( t \right)$ is a symmetric matrix, one finds:
\begin{equation}
{\mathcal{\tilde  U}} _N\left(   t  \right)=\left[{{\mathcal{U} }_N^{T}\left( - t \right)}\right]^{-1}.
\end{equation}
Furthermore, periodicity of $\HH_N \left( t \right)$ leads to the relation that ${{\mathcal{U} }_N \left( -T_0 \right)}=\left[{{\mathcal{U} }_N \left( T_0 \right)}\right]^{-1}$, so that we have
\begin{equation}\label{appeq:TIscriptUSym}
{\mathcal{\tilde  U}} _N\left(   T_0   \right)=\left[{{\mathcal{U} }_N^{T}\left( - T_0  \right)}\right]^{-1}={{\mathcal{U} }_N^{T}\left(   T_0  \right)}.
\end{equation}

We see that the classical amplitude-evolution matrices associated with forward and backwards evolution are related by a simple transpose operation.  We can use this and the results of App.~\ref{appsec:BMRSympGen} to then directly relate the unitary evolutions in the corresponding four-mode bosonic QMFS systems.  It follows that the BM decompositions for forward and backwards evolution are related via:
\begin{subequations}\label{appeq:BMRQMFSGenTR}
\begin{eqnarray}
\tilde U_{\mathrm{BM}} \left(   T_0  \right) &=&{{ V_{\mathrm{BM}}^*}\left(   T_0  \right)  },\\
\tilde V_{\mathrm{BM}} \left(   T_0  \right) &=& {{ U_{\mathrm{BM}}^*}\left(   T_0  \right)  },\\
{\tilde D_A}\left(   T_0  \right)={D_A}\left(   T_0  \right) &\Leftrightarrow & {\tilde D_B}\left(   T_0  \right) ={D_B}\left(   T_0  \right) .
\end{eqnarray}
\label{eqs:BMConstraints}
\end{subequations}
Here, tildes indicate backwards evolution.  Note that the squeezing aspect of the evolution (as parameterized by the $D$ matrices) is the same irrespective of the direction.  Finally, note that these results apply directly to our two-mode problem of interest (i.e.~$\mathcal{H}_N(t) \rightarrow \mathcal{H}_\omega(t)$, $\mathcal{U}_N(t) \rightarrow \mathcal{U}_\omega(t)$), as $\mathcal{H}_\omega(t)$ satisfies Eq.~(\ref{appeq:nHMatSymmPer}).


\subsection{Evolution of quantum states via EP encircling}
\label{app:EPEncirclingStates}

\subsubsection{Evolution of an initial vacuum state}

Having built up the necessary machinery, we can now study how the switching dynamics encoded in $\mathcal{U}_\omega(t)$ influences the evolution of quantum states in our four-mode bosonic system (lowering operators $\hat{a}_1, \hat{a}_2, \hat{b}_1, \hat{b}_2$).  The unitary evolution operator $\hat{U}_\omega(t)$ of our system is defined by
\begin{equation}\label{eq:SympTwoModeQAD}
    i{\partial _t } {\hat U}_{\omega}\left( t  \right) = {{\hat H}_{\omega \mathrm{PA}}} \left( t \right) {\hat U}_{\omega}\left( t  \right), \quad  {\hat U}_{\omega}\left( t=0  \right)=\mathbb{I},
\end{equation}
where ${\hat H}_{\omega \mathrm{PA}}(t)$ is given by Eq.~(\ref{eqnH2mpa}).  $\hat{U}_\omega(t)$ generates a symplectic (i.e.~commutation-relation preserving) linear transformation of the system's mode operators.  As established in Appendix~\ref{SecappQMFS2m}, the form of this transformation is completely determined by the amplitude-evolution matrix $\mathcal{U}_\omega(t)$ of the original non-Hermitian problem.

As in the main text, we consider a cyclic evolution where $(g(t),\omega(t))$ evolve along a closed circle, starting and ending at the same point in parameter space.  The non-Hermitian system's evolution is different for these two directions, corresponding to two distinct evolution matrices $\mathcal{U}_\circlearrowleft (t)$ and $\mathcal{U}_\circlearrowright (t)$.  When transformed to the instantaneous eigenmode basis of $\HH_{\omega} \left( t\right)$, one of these encodes the switching behaviour seen in Fig.~\ref{fig:QADfref}, the other has no switching behaviour.  

Using our mapping, we have two corresponding unitary transformations $\hat{U}_\circlearrowleft \left( t\right)$ and $\hat{U}_\circlearrowright \left( t\right)$ for our quantum four-mode system; we wish to understand their asymmetry.
This is best accomplished by using the Bloch-Messiah decomposition (see Appendix~\ref{appsec:BMRUomega}), which represents each transformation as a product of two beam-splitter transformations, interspersed with a (diagonal) squeezing transformation.  We find that the squeezing associated with both $\hat{U}_\circlearrowleft \left(  T \right)$ and $\hat{U}_\circlearrowright \left(  T \right)$ are identical, with the asymmetry manifesting itself only in the beamsplitter operations (c.f.~Eqs.(\ref{eqs:BMConstraints}))    

To see the the physical consequences of this asymmetry, consider first the case where all four modes start in vacuum, and parameters are cyclically evolved on the path shown in Fig.~\ref{fig:QADfref}(a).  CW or CCW traversal of this path results in two different final states for our four bosonic modes, 
$\left| \Psi _{ \circlearrowleft } \left(t =T\right) \right\rangle$ versus
$\left| \Psi _{ \circlearrowright } \left(t =T\right) \right\rangle$.  
These final states are necessarily Gaussian and have zero means, and are thus fully characterized by their covariance matrix.  First, consider beam-splitter type correlations between $a$ and $b$ modes.  Due to the block structure of the symplectic transformation in Eqs.~\eqref{appeq:SympTFMulti}, these vanish for all times $t$, i.e. 
\begin{equation}
    \left\langle {\hat a_j^\dag \left( t  \right){{\hat b}_{j'}}\left( t  \right)} \right\rangle =\left\langle {\hat b_j^\dag \left(  t  \right){{\hat a}_{j'}}\left(  t \right)} \right\rangle =0.
\end{equation}

Moreover, the photon numbers are identical for all four modes:
\begin{align}
&\left\langle {\hat a_j^\dag \left( t  \right){{\hat a}_{j'}}\left( t  \right)} \right\rangle  = \left\langle {\hat b_j^\dag \left( t  \right){{\hat b}_{j'}}\left( t  \right)} \right\rangle  =\delta_{jj'}  {\sinh ^2}{\lambda _s}\nonumber\\
&=\frac{\delta_{jj'}}{4}\left( {   {\mathrm{tr} \left[{  \mathcal{U}^\dag _\omega}\left( t  \right) { \mathcal{U}_\omega}\left( t  \right)\right]  }- 2} \right).
\end{align}
As the squeezing parameter $\lambda_s$ is the same at the final time $T$ irrespective of encircling direction, the same is necessarily true for these average photon numbers.

Finally, the only non-zero anomalous (squeezing) correlators are given by
\begin{align}
&\left\langle {\hat a_j  \left( t  \right){{\hat b}_{j'}}\left(  t  \right)} \right\rangle  = \left\langle {\hat b_{j'}  \left(  t  \right){{\hat a}_{j}}\left( t  \right)} \right\rangle  ={D}_{jj'}   \\
&{D}=\frac{1}{4}    { \mathcal{U}_\omega}\left(  t  \right){  \mathcal{U}^\dag _\omega}\left(  t  \right) -\frac{1}{4}   \left[ { \mathcal{U}_\omega}\left(  t  \right){  \mathcal{U}^\dag _\omega}\left(  t  \right)\right]^{-1}  ,
\end{align}
where ${D}$ is a Hermitian matrix.
These correlators (at the final time $t=T$) will depend on the direction of the encircling.

Finally, we could look at bipartite entanglement between different subsystems.  Consider for example the entanglement between the $a$ subsystem (formed by modes $a_1,a_2$) and the $b$ subsystem (formed by modes $b_1, b_2$).  Quantifying the entanglement via the logarithmic negativity $E_N$ \cite{Vidal2002,Plenio2005}, one finds:
\begin{align}
    E_N\left[ {{\rho _{ab}}\left( t  \right)} \right] = &\left( {\cosh {\lambda _s} + 1} \right)\log \left( {\cosh {\lambda _s}} \right) \nonumber\\
&- \left( {\cosh {\lambda _s} - 1} \right)\log \left( {\sinh {\lambda _s}} \right).
\end{align}
The entanglement only depends on the squeezing paramter $\lambda_s$.  As this is identical for both encircling directions, the generated $a-b$ entanglement is thus also insensitive to direction.  
The net result is that if we start with a vacuum state, the asymmetry between the states 
$\left| \Psi _{ \circlearrowleft } \left(t =T\right) \right\rangle$ and
$\left| \Psi _{ \circlearrowright } \left(t =T\right) \right\rangle$ is subtle: both have the same average photon number and entanglement properties, and differ only in the phase of two-mode squeezing correlators between $a$ and $b$ modes.

\subsubsection{Construction of an asymmetric initial quantum state}

We now finally turn to the case presented in Sec.~\ref{subsec:EPEncircle} of the main text, where we consider an initial, pure quantum state that corresponds to selectively populating one of the two eigenmodes of $\mathcal{H}_\omega(0)$.
In the classical case, the chiral mode switching behaviour depends crucially on having such an asymmetric initial state.  The same is true in the quantum case.  

To construct a suitable initial state, we first consider the classical two-mode problem.  Using our convention for the instantaneous eigenvectors $\vec{r}_+(t)$, $\vec{r}_-(t)$ of $\mathcal{H}_\omega(t)$, one finds that the vectors $\vec{r}_+(t)$ and $\left( \vec{r}_-(t) \right)^*$ are orthogonal.  They thus serve as a good basis, and we can write any initial set of amplitudes in the classical problem as:
\begin{equation}
    \vec z  =
        \xi \frac{ {\vec r} _ + }{\left| {\vec r} _ + \right|} +
        \xi_\perp \frac{ \left({\vec r} _ -\right)^* }{\left| {\vec r} _ - \right|} ,
\end{equation}
where $\xi$ and $\xi_\perp$ are complex numbers.  $\xi_\perp$ is proportional to the amplitude $c_{-}$ defined in Eq.~(\ref{eq:chatdefn}), whereas $\xi$ describes the amount of population in the mode ${\vec r} _ +$ (when we make this vector part of an orthonormal basis).

We could now imagine a random classical state which selectively populates the $+$ eigenmode with a random phase.  In particular, take $\xi, \xi_\perp$ to be complex Gaussian random variables with zero mean, and where the only non-zero covariances are:
\begin{subequations}
\begin{eqnarray}
    \overline{ \left( \textrm{Re } \xi \right)^2} = \overline{ \left( \textrm{Im } \xi \right)^2} = 
        \frac{1}{2}e^{2 \lambda_0}, \\
    \overline{ \left( \textrm{Re } \xi_\perp \right)^2} = \overline{ \left( \textrm{Im } \xi_\perp \right)^2} = \frac{1}{2}e^{- 2 \lambda_0}. 
\end{eqnarray}
\label{eq:ClassicalCovariance}
\end{subequations}
The parameter $\lambda_0 > 0$ determines the asymmetry of the initial state.

Turning to our quantum system, the components of $\hat{z}$ become operators as per Eq.~(\ref{eq:BrokenEOM}) and (\ref{eq:PseudoMode2m}): they are linear combinations of the QMFS collective quadrature operators $( \hat{x}_{+ ,1},  \hat{x}_{+ ,2}, \hat{p}_{- ,1}, \hat{p}_{- ,2} )$.  It immediately follows that the amplitudes $\xi$, $\xi_\perp$ become commuting operators that are linear combinations of these QMFS collective quadrature operators; one can easily find the relevant orthogonal transformation.  

We can now construct a quantum Gaussian state where the operators $\hat{\xi}$, $\hat{\xi}_\perp$ have a covariance matrix that coincides with Eqs.~(\ref{eq:ClassicalCovariance}).  This in turn defines the covariance matrix of the QMFS collective quadrature operators.  This of course does not specify the entire state:  we also need to specify covariances
involving collective quadratures conjugate to those in the QMFS, i.e.~$\left( {\hat{p}_{+ ,1},  \hat{p}_{+ ,2}, -\hat{x}_{- ,1}, -\hat{x}_{- ,2}} \right)$.  We do this by insisting on two additional requirements:
\begin{itemize}
    \item The covariance matrix of the entire system describes a physical state compatible with the uncertainty principle \cite{BraunsteinRMP}.
    \item The covariance matrix of the entire system describes a pure state.
    \item  There are no classical (i.e.~symmetrized) correlations between a collective quadrature from the main QMFS, and the secondary QMFS 
\end{itemize}
These conditions allow us to find a pure zero-mean Gaussian state parameterized by $\lambda_0$, where the amplitude-operators $\hat{\xi}, \hat{\xi}_\perp$ have covariances given by Eqs.~(\ref{eq:ClassicalCovariance}).  This is the kind of initial state used for the calculations of entanglement dynamics in Sec.~\ref{subsec:EPEncircle}.

Note that while our state clearly has a strong asymmetry favouring the $+$ instantaneous eigenmode, there is still some population of the $-$ eigenmode, as $\xi_\perp$ is not exactly zero.  One cannot find a physical state where $\langle \hat{\xi}_\perp^\dagger \hat{\xi}_\perp \rangle$ is strictly zero as this would violate the uncertainty principle (i.e.~as this quantity becomes smaller and smaller, the covariances of operators outside of the QMFS would diverge).


\section{Relating non-Hermitian and bosonic topological invariants}
\label{app:ChernNumbers}

In this appendix, we exploit the mappings established earlier to show that non-Hermitian Chern numbers \cite{Fu2018} are equivalent to the Chern numbers for anomalous bosonic problems.

\subsection{Relating Bogoliubov transformations to non-Hermitian eigenvectors}

  As a prerequisite, we will establish the connection between Bogoliubov transformations (in the bosonic problem) to the eigenvectors of the non-Hermitian problem.

Consider first a translationally-invariant Hermitian bosonic $\mathrm{2D}$ lattice model having a  primitive unit cell with $N$ sites.  The Hamiltonian can be written as 
$\hat{H} = \frac{1}{2} \sum_{\mathbf{k}} \hat{H}_{\mathbf{k}}$, where the Bloch Hamiltonian for quasimomentum $\mathbf{k}$ has the general form
\begin{align}
    {{\hat H}_{\mathbf{k}}} = &\sum\limits_{i,j = 1}^N {\left( {{\mu _{{\mathbf{k}},ij}}{\hat a}_{{\mathbf{k}},i}^\dag {{\hat a}_{{\mathbf{k}},j}} + {\mu _{ - {\mathbf{k}},ij}}{\hat a}_{ - {\mathbf{k}},i}^\dag {{\hat a}_{ - {\mathbf{k}},j}}} \right)}    \nonumber\\
    &+ 
        \sum\limits_{i,j = 1}^N {\left( {{\nu _{{\mathbf{k}},ij}}{\hat a}_{{\mathbf{k}},i}^\dag {\hat a}_{ - {\mathbf{k}},j}^\dag  + h.c.} \right)},
\end{align} 
where $\hat{a}_{\mathbf{k},i}$ is the annihilation operator corresponding to quasi-momentum $\mathbf{k}$ and site $i$ in the unit cell.
Hermiticity requires ${\mu _{  {\mathbf{k}}}} = \mu _{  {\mathbf{k}}}^\dag $; further, ${\nu _{\mathbf{k}}} = \nu _{ - {\mathbf{k}}}^T$ as bosonic lowering operators commute. 

The Heisenberg equations of motion now take the compact form
\begin{equation}\label{EqMMgeom}
i{\partial _t}{\left| {\hat a_{\mathbf{k}}} \right\rangle} = \left( {\begin{array}{*{20}{c}}
{{\mu _{\mathbf{k}}}}&{{\nu _{\mathbf{k}}}}\\
{ - \nu _{\mathbf{k}}^\dag }&{ - \mu _{ - {\mathbf{k}}}^T}
\end{array}} \right){\left| {\hat a_{\mathbf{k}}} \right\rangle},\end{equation}
where we define the column vector $\left| {\hat a_{\mathbf{k}}} \right\rangle $ formed by the $2N$ coupled operators as
\begin{equation}\left| {\hat a_{\mathbf{k}}} \right\rangle = \left( {{\hat a}_{{\mathbf{k}},1}},{{\hat a}_{{\mathbf{k}},2}}, \cdots ,{{\hat a}_{{\mathbf{k}},N}},\hat a_{ - {\mathbf{k}},1}^\dag ,\right.\hat a_{ - {\mathbf{k}},2}^\dag , \cdots ,\left.\hat a_{ - {\mathbf{k}},N}^\dag  \right)^T.\end{equation} 
We can now interpret the dynamical matrix of our driven bosonic system as an effective non-Hermitian Bloch Hamiltonian $\mathcal{H}_\mathrm{eff}\left( { {\mathbf{k}}} \right)$ of a lattice with $2N$ sites in the unit cell, 
\begin{equation}\label{efftoBdGg}
{\mathcal{H}_\mathrm{eff}} \left( { {\mathbf{k}}} \right)= \left( {\begin{array}{*{20}{c}}
{{\mu _{\mathbf{k}}}}&{{\nu _{\mathbf{k}}}}\\
{ - \nu _{\mathbf{k}}^\dag }&{ - \mu _{ - {\mathbf{k}}}^T}
\end{array}} \right).
\end{equation}
This non-Hermitian Bloch Hamiltonian is related to the  Hermitian Bogoliubov-de Gennes (BdG) Bloch Hamiltonian ${H_\mathrm{BdG}} \left( { {\mathbf{k}}} \right)$ by ${\mathcal{H}_\mathrm{eff}}\left( { {\mathbf{k}}} \right) =  {{ \sigma} _{N,z}}{H_\mathrm{BdG}}\left( { {\mathbf{k}}} \right)$, where
\begin{equation}
{H_\mathrm{BdG}} \left( { {\mathbf{k}}} \right)= \left( {\begin{array}{*{20}{c}}
{{\mu _{\mathbf{k}}}}&{{\nu _{\mathbf{k}}}}\\
{\nu _{\mathbf{k}}^\dag }&{\mu _{ - {\mathbf{k}}}^T}
\end{array}} \right).
\end{equation}
and $\sigma_{N,z}$ is a $z$ Pauli matrix in particle-hole space.

We next define the  left and right eigenvectors of the matrix ${\mathcal{H}_\mathrm{eff}}\left( { {\mathbf{k}}} \right)$:
\begin{align}&{\mathcal{H}_\mathrm{eff}}\left( {\mathbf{k}} \right){\left| {{\mathbf{k}},j} \right\rangle {}_{\mathrm{R}}} = {E_j}\left( {\mathbf{k}} \right){\left| {{\mathbf{k}},j} \right\rangle {}_{\mathrm{R}}},\\
&  {}_{\mathrm{L}}\left\langle {{\mathbf{k}},j} \right|{\mathcal{H}_\mathrm{eff}}\left( {\mathbf{k}} \right) = {E_j}\left( {\mathbf{k}} \right){}_{\mathrm{L}}\left\langle {{\mathbf{k}},j} \right|.\end{align}
Here $j=1,2,\cdots,2N$, and we choose the eigenvectors to satisfy the biorthonormal condition
\begin{equation}{}_{\mathrm{L}}{\left\langle {{{\mathbf{k}},j}}
 \mathrel{\left | {\vphantom {{{\mathbf{k}},l} {{\mathbf{k}},l'}}}
 \right. \kern-\nulldelimiterspace}
 {{{\mathbf{k}},j'}} \right\rangle {}_{\mathrm{R}}} = {\delta _{j,j'}}. 
 \label{eq:Biorth}
 \end{equation}
 
We will focus exclusively on the regime where the parametric driving is sufficiently weak that our system is stable, and the spectrum  ${E_j}\left( {\mathbf{k}} \right)$ is purely real.  ${{\hat H}_{\mathbf{k}}}$ can then be diagonalized via a Bogoliubov transformation.  Not surprisingly, the quasiparticle operators that diagaonlize the Hamiltonian are directly related to the eigenvectors of ${\mathcal{H}_\mathrm{eff}}\left( { {\mathbf{k}}} \right)$.  To see this explicitly, we
use the fact that all left eigenvectors either have a real, non-zero ``expectation" of $\sigma_{N,z}$ that is either positive or negative (see e.g. \cite{Peano2016PRX}).  They can thus be chosen to obey the symplectic normalization condition 
\begin{equation}{}_{\mathrm{L}}\left\langle {{\mathbf{k}},n, \pm } \right| {{ \sigma} _{N,z}} \left| {{\mathbf{k}},n', \pm } \right\rangle   {}_{\mathrm{L}} =  \pm {\delta _{n,n'}},\label{Eqsigznorm} \end{equation}
All left eigenvectors are now labelled by a sign $\pm$, and the index $n$ runs from $1$ to $N$.  We denote the corresponding eigenvalues $E_{n,\pm}(\mathbf{k})$.
With this convention, it follows from Eq.~(\ref{eq:Biorth}) that the corresponding right eigenvectors are given by 
\begin{equation}
    \left| {{\mathbf{k}},n, \pm } \right\rangle   {}_{\mathrm{R}}=\pm{{ \sigma} _{N,z}}\left| {{\mathbf{k}},n, \pm } \right\rangle   {}_{\mathrm{L}}.
\label{eq:REigVecSyplectic}
\end{equation}
The eigenvectors now let us express the equations of motion in diagonal form; this can be accommplished by using the positive-norm eigenvectors only.  We introduce new bosonic quasiparticles via $${{\hat \beta }_{{\mathbf{k}},n}} = {{}_{\mathrm{L}}}\left\langle {{\mathbf{k}},n,+}
 \mathrel{\left | {\vphantom {l {\hat a}}}
 \right. \kern-\nulldelimiterspace}
 {{\hat a}} \right\rangle.$$ 
 They satisfy
 $$i{\partial _t}{{\hat \beta }_{{\mathbf{k}},n}}={E_{n,+}}\left( {\mathbf{k}} \right){{\hat \beta }_{{\mathbf{k}},n}}.$$
 This represents a canonical Bogoliubov transformation, and the Hamiltonian is diagonal when expressed in terms of these operators.  We thus see the (expected) relation between the Bogoliubov transformation and the eigenvectors of our non-Hermitian Hamiltonian.

\subsection{Equivalence of bosonic and non-Hermitian Chern numbers}

With the above relations in hand, we can now show that the non-Hermitian Chern number introduced in Ref.~\cite{Fu2018} coincides with the previously introduced Chern number for anomalous bosonic problems \cite{Shindou2013}. 

We start with the bosonic system.
The Berry connection $ {A}_{nn}\left( {\mathbf{k}} \right)$ for the $n^\mathrm{th}$ band was introduced in Refs.~\cite{Shindou2013,Barnett2013,Peano2016} as
\begin{equation}
    { {A}_{nn}} \left( {\mathbf{k}} \right)
        =i  \cdot {{}_{\mathrm{L}}\left\langle {{\mathbf{k}},n} \right| }{{ \sigma} _{N,z}}{\nabla _{\mathbf{k}}}{\left| {{\mathbf{k}},n} \right\rangle {}_{\mathrm{L}}}, 
\end{equation}
and the corresponding (quantized) Chern number is given by:
\begin{align}
{C_n} = &\frac{1}{{2\pi }}\int\limits_{\mathrm{BZ}} {\left( {\nabla  \times { {A}_{nn}}} \right) \cdot \hat z \, {d^2}{\mathbf{k}}} 
\end{align}
This serves as a topological invariant to characterize bands in an anomalous, stable bosonic system.  

Now, using Eq.~(\ref{eq:REigVecSyplectic}), we can equivalently write this Chern number in terms of left and right eigenvectors of the non-Hermitian Hamiltonian 
$\mathcal{H}_{\rm eff}(\mathbf{k})$
\begin{align}
{C_n} = &\frac{1}{{2\pi }}\int\limits_{\mathrm{BZ}} {\left( {\nabla  \times { {A}_{nn}}} \right) \cdot \hat z{d^2}{\mathbf{k}}}   \nonumber\\
=& \frac{i}{{2\pi }}\int\limits_{\mathrm{BZ}} {{\epsilon _{3ij}}{\partial _i}\left({}_{\mathrm{L}} {\left\langle {{\mathbf{k}},n,+} \right|{\partial _j}{{\left| {{\mathbf{k}},n,+} \right\rangle }}} {}_{\mathrm{R}}\right){d^2}{\mathbf{k}}}  \nonumber\\
=& \frac{i}{{2\pi }}\int\limits_{\mathrm{BZ}} {{\epsilon _{3ij}}\left( {{\partial _i}\left\langle {{\mathbf{k}},n,+} \right|_{\mathrm{L}}} \right)\left( {{\partial _j}{{\left| {{\mathbf{k}},n,+} \right\rangle }_{\mathrm{R}}}} \right){d^2}{\mathbf{k}}}. \label{eq:BosonicChernNum}
\end{align}

We can now compare this expression against the generalized Chern numbers
$ {N}_n^{\mathrm{\alpha \beta} }$
introduced in Ref.~\cite{Fu2018} for 2D non-Hermitian Hamiltonians.  These are defined as
\begin{align}&  {B}_{n,ij}^{\alpha \beta }\left( {\mathbf{k}} \right) = i\left\langle {{\partial _i}\psi _n^\alpha \left( {\mathbf{k}} \right)} \right|\left. {{\partial _j}\psi _n^\beta \left( {\mathbf{k}} \right)} \right\rangle ,\\
& {N}_n^{\alpha \beta } = \frac{1}{{2\pi }}\int\limits_{BZ} {{\epsilon _{ij}} {B}_{n,ij}^{\alpha \beta }\left( {\mathbf{k}} \right){d^2}{\mathbf{k}}}.
\end{align}
Here, the indices $\alpha, \beta = \mathrm{L,R}$, and
$\left|{\psi _n^\mathrm{L} \left( {\mathbf{k}} \right)} \right\rangle$
($\left|{\psi _n^\mathrm{R} \left( {\mathbf{k}} \right)} \right\rangle$)
denotes the left (right) eigenvector of the given non-Hermitian Bloch Hamiltonian $\HH(\mathbf{k})$.  Ref.~\cite{Fu2018} shows that all four Chern numbers $N^{\alpha \beta}_n$ for a given band $n$ are identical.  

We see now that the bosonic Chern number in Eq.~\eqref{eq:BosonicChernNum} is identical to the generalized non-Hermitian Chern number $ {N}_n^{\mathrm{LR} }$.  Thus, as long as the bosonic Hamiltonian ${{\hat H}_{\mathbf{k}}}$ has well-defined Chern numbers, we can always find the corresponding non-Hermitian lattice model ${\mathcal{H}_\mathrm{eff}} \left( { {\mathbf{k}}} \right)$ whose topological invariants are exactly the same. 

While the correspondence found here here may not seem that surprising, it provides an interesting recipe for constructing non-trivial non-Hermitian topological models: start with a topological bosonic model, and then construct its non-Hermitian analogue.  We pursue this approach in the next section.

\section{Topological $\mathcal{PT}$-symmetric model inspired by the mapping between $\mathcal{PT}$ and PA systems}

\subsection{Correspondence between the dimer Kagome Hamiltonian and the bosonic parametric model}\label{AppSecKc}
Due to the correspondence between the Chern number based on the bosonic symplectic normalization relation and the generalized Chern number for the non-Hermitian dynamical matrix, the analysis in Ref.~\cite{Peano2016} on the topological phases of the system also applies to the equivalent non-Hermitian problem. One interesting and probably exotic feature for the bosonic model is that the nontrivial topological phases are completely due to the parametric drive, without which we would only have a trivial Kagome lattice model with nearest neighbor tunnel couplings. Thus for the non-Hermitian model, nontrivial topological phases can only exist if the effective Hamiltonian has nonzero non-Hermitian components. This is in contrast to some previous work based on a topologically nontrivial coherent Hamiltonian on non-Hermitian topological systems, where the anti-Hermitian part of dynamics is usually introduced as a perturbation \cite{Lee2016,Szameit2017}. Here we combine the correspondence of topological phases and the mapping between some parametric models and $\mathcal{PT}$-symmetric systems to construct a non-Hermitian $\mathcal{PT}$-symmetric lattice model, where nontrivial topological phase emerges from an otherwise topologically trivial Hermitian model when one adds balanced onsite gain and loss terms to the model properly.

In Ref.~\cite{Peano2016}, non-trivial topological states can be created by adding parametric coupling with proper arrangement of phases to a topologically trivial Kagome lattice model that only have identical coherent hopping. We consider a parametric Hamiltonian ${\hat H}_\mathrm{p.a., Kagome}={\hat H}_0+{\hat H}_{\mathrm{L}}$ consisting of a topologically trivial tight-binding Kagome lattice model that conserves particle number
\begin{equation}{{\hat H}_0} = {\omega _0}\sum\limits_{\mathfrak{j}} {{\hat a}_{\mathfrak{j}}^\dag {{\hat a}_{\mathfrak{j}}}}  - J\sum\limits_{\left\langle {{\mathbf{j}},{\mathbf{j'}}} \right\rangle } {{\hat a}_{\mathfrak{j'}}^\dag {{\hat a}_{\mathfrak{j'}}}} ,
\end{equation}
and a local parametric drive term ${H_{\mathrm{L}}}$
\begin{equation}{{\hat H}_{\mathrm{L}}} =  - \frac{1}{2} {{\nu  }\sum\limits_{\mathfrak{j}} {{e^{i{\phi _s}}}{\hat a}_{\mathfrak{j}}^\dag {\hat a}_{\mathfrak{j}}^\dag }   }   + h.c.,
\end{equation}
where the index ${\mathfrak{j}} = \left( {\mathbf{j},s} \right) = \left( {{j_1},{j_2},s} \right)$ incorporates both periodicity in real space and sub-lattices $s=A,B,C$, and ${\nu }$, ${\phi _s}$ denote the parametric drive strength and phase, respectively. Transforming the mode operators to the reciprocal ${\mathbf{k}}$ space, the system dynamics is closed with respect to the set of operators $  \left| {\hat a_{\mathbf{k}}} \right\rangle=  \left( {{{\hat a}_{{\mathbf{k}},A}},{{\hat a}_{{\mathbf{k}},B}},{{\hat a}_{{\mathbf{k}},C}},{\hat a}_{ - {\mathbf{k}},A}^\dag ,{\hat a}_{ - {\mathbf{k}},B}^\dag ,{\hat a}_{ - {\mathbf{k}},C}^\dag } \right)^T$, so that as a special case of Eq.~\eqref{EqMMgeom}, the equations of motion can be written in the compact form
\begin{equation}
i{\partial _t}  \left| {\hat a_{\mathbf{k}}} \right\rangle = {\mathcal{H}_\mathrm{eff,K .}}\left( {\mathbf{k}} \right)  \left| {\hat a_{\mathbf{k}} } \right\rangle,
\end{equation}
where the dynamical matrix is given by
\begin{equation}{\mathcal{H}_\mathrm{eff,K .}}\left( {\mathbf{k}} \right) = \left( {\begin{array}{*{20}{c}}
  {{\omega _0}\mathbb{I}_3 - J\tau \left( {\mathbf{k}} \right)}&{{h }} \\ 
  { - h ^\dag }&{ - {\omega _0}\mathbb{I}_3 + J\tau \left( {\mathbf{k}} \right)} 
\end{array}} \right) .\end{equation}
The matrix $\tau \left( {\mathbf{k}} \right)$ is formed by geometrical factors of the tight-binding Kagome lattice
\begin{equation}\tau \left( {\mathbf{k}} \right) = \left( {\begin{array}{*{20}{c}}
  0&{1 + {e^{ - i{\mathbf{k}} \cdot {{\mathbf{a_1}}}}}}&{1 + {e^{i{\mathbf{k}} \cdot {{\mathbf{a_3}} }}}} \\ 
  {1 + {e^{i{\mathbf{k}} \cdot {{\mathbf{a_1}}}}}}&0&{1 + {e^{ - i{\mathbf{k}} \cdot {{\mathbf{a_2}}}}}} \\ 
  {1 + {e^{ - i{\mathbf{k}} \cdot {{\mathbf{a_3}}}}}}&{1 + {e^{i{\mathbf{k}} \cdot {{\mathbf{a_2}}}}}}&0 
\end{array}} \right),\end{equation}
where ${\mathbf{a_1}} = \left( { - 1, - \sqrt 3 } \right), {\mathbf{a_2}} = \left( {  2,0} \right)$ are the lattice vectors, and ${\mathbf{a_3}} = \left( { - 1,\sqrt 3 } \right)$; the coefficient matrix ${h } =  - {\nu }\exp \left( {i\Phi } \right)$, $\mathbb{I}_3$ is the $3 \times 3$ identity matrix, and $\Phi$ is the diagonal matrix formed by the phases carried by local parametric drives 
\begin{equation}\Phi = \mathrm{diag}\left( {\phi_A ,\phi_B ,\phi_C } \right) = \mathrm{diag}\left( {0,\phi ,2\phi } \right),\quad \phi  = \frac{{2\pi }}{3}.\end{equation}
The off-diagonal tunnelings in $h $ can be rotated to onsite gain and loss via a unitary transformation
\begin{equation}
    \label{Eq:UKagome}
   { \mathcal{U}_\mathrm{K .}}  = \frac{1}{{\sqrt 2 }}\left( {\begin{array}{*{20}{c}}
  {{e^{2i \Phi }}}&{   {e^{2i \Phi }}} \\ 
  {i{e^{ -2i \Phi}}}&{- i{e^{ - 2i \Phi}}} 
\end{array}} \right) .
\end{equation}
so that ${\mathcal{H}_{\mathcal{PT},\mathrm{K.}}}\left( {\mathbf{k}} \right) =  { \mathcal{U}_\mathrm{K .}^\dag} {\mathcal{H}_\mathrm{eff,K .}}\left( {\mathbf{k}} \right){ { \mathcal{U}_\mathrm{K .}}}$ is
\begin{equation}
   {\mathcal{H}_{\mathcal{PT},\mathrm{K.}}}\left( {\mathbf{k}} \right)  =\left( {\begin{array}{*{20}{c}}
  {  {\Sigma} \left( {\mathbf{k}} \right) - i{\nu  }\mathbb{I}_3}&{  {\Delta} \left( {\mathbf{k}} \right)} \\ 
  {   {\Delta} \left( {\mathbf{k}} \right)}&{ {\Sigma} \left( {\mathbf{k}} \right) + i{\nu }\mathbb{I}_3} 
\end{array}} \right) ,
\end{equation}
where $ {\Sigma} \left( {\mathbf{k}} \right)$ and $ {\Delta} \left( {\mathbf{k}} \right)$ are Hermitian matrices with matrix elements given by
\begin{subequations}
\begin{align}
  & { {\Sigma} _{ss'}}\left( {\mathbf{k}} \right) = iJ{\tau _{ss'}}\left( {\mathbf{k}} \right)\sin  \left( {2{\phi _s} - 2{\phi _{s'}}} \right),\\
   &  { {\Delta} _{ss'}}\left( {\mathbf{k}} \right) =  \omega_0{\delta _{ss'}} - J{\tau _{ss'}}\left( {\mathbf{k}} \right)\cos  \left( {2{\phi _s} - 2{\phi _{s'}}} \right) .
\end{align}
\end{subequations}
The corresponding real space Hamiltonian ${{\hat {\mathcal{H}}}_\mathrm{Kagome}} $ for the lattice model $ {\mathcal{H}_{\mathcal{PT},\mathrm{K.}}}\left( {\mathbf{k}} \right)$ is presented in Sec.~\ref{Sec:PTKagome} in the main text.

Before ending this section, we note that the unitary mapping ${ \mathcal{U}_\mathrm{K .}}$ in Eq.~\eqref{Eq:UKagome} is local in real space, so that any topological edge modes of the parametric Kagome lattice model ${\hat H}_\mathrm{p.a., Kagome}$ will also be mapped to topological edge modes of the $\PT$-symmetric model.

\bibliographystyle{naturemag}
\pagestyle{plain}
\bibliography{ref}

\end{document}